\documentclass[12pt,a4paper]{article}
\usepackage{jheppub}
\usepackage[FIGTOPCAP]{subfigure}
\usepackage{amsmath}
\usepackage{amsfonts}
\usepackage{amssymb}
\usepackage{tensor,epigraph}
\usepackage{cancel}
\usepackage{ulem} 
\usepackage{musixtex}
\usepackage{booktabs}
\usepackage{empheq}
\usepackage{amsmath}
\usepackage{amssymb}
\usepackage{amsthm}
\usepackage{comment}
\usepackage{psfrag}
\usepackage{graphicx}
\usepackage{color}
\usepackage{bbm}
\usepackage{color}

\newcommand{\be}{\begin{equation}}
\newcommand{\ee}{\end{equation}}

\definecolor{klgreen}{rgb}{0.0, 0.5, 0.0}

\newcommand{\exclude}[1]{}

\newcommand{\beq}{\begin{equation}}
\newcommand{\eeq}{\end{equation}}
\newcommand{\bea}{\begin{eqnarray}}
\newcommand{\eea}{\end{eqnarray}}
\long\def\/*#1*/{}

\newcommand{\Tr}{{\rm Tr}}

\newcommand{\junk}[1]{}

\title{Probing inside a charged hairy black hole in massive gravity}
\author[a]{Mirmani Mirjalali}
\author[a]{, Seyed Ali Hosseini Mansoori}
\author[b]{, Leila Shahkarami}
\author[a]{, and Morteza Rafiee}

\affiliation[a]{Faculty of Physics, Shahrood University of Technology, P.O. Box 3619995161, Shahrood, Iran\vspace{0.1cm}}
\affiliation[b]{School of Physics, Damghan University, Damghan, 41167-36716, Iran \vspace{0.1cm}}
\vspace{0.1cm}

\emailAdd{mirmanimirjalali4@gmail.com}
\emailAdd{shosseini@shahroodut.ac.ir}
\emailAdd{l.shahkarami@du.ac.ir}
\emailAdd{m.rafiee@shahroodut.ac.ir}
\abstract{In this paper, we investigate the internal structure of a charged hairy black hole solution in the non-linear massive gravity. We first consider the impact of various configurations of massive gravity on the condensate operator and then probe the black hole interior dynamics. Like a standard holographic superconductor system, just below the critical temperature, the interior evolves through several distinct epochs, including a collapse of the Einstein-Rosen bridge, Josephson oscillations of the scalar field, and finally a Kasner (or Kasner inversion) cosmology. However, for the large massive gravity parameter, we see distinguishing features for the interior dynamics. In this regime, at a given temperature, the Einstein-Rosen bridge collapse and subsequent Josephson oscillations epochs completely disappear  from the interior dynamics and the final Kasner cosmology epoch starts exactly after the would-be inner horizon and the system does not experience the Kasner inversion epoch. 
 }

\begin{document}
\maketitle
\section{Introduction}
In recent years there has been growing interest in understanding the internal structure of a black hole. Take for instance, a Reisser-Nordstrom (RN) black hole \cite{reissner1916eigengravitation,nordstrom1918energy} or a Kerr black hole \cite{kerr1963gravitational}. After crossing the event horizon, there is an inner Cauchy horizon before reaching near the singularity where the spacetime curvature becomes infinite. The appearance of Cauchy horizon results in breaking down the predictability of the classical dynamics \cite{geroch1970domain} and appear to violate the strong cosmic censorship (SCC) conjecture \cite{ Penrose:1969pc}.


On the other hand, holographic arguments suggest that Cauchy horizons cannot survive in the full quantum gravity theories \cite{Papadodimas:2019msp, Balasubramanian:2019qwk}. In this respect, recently in Refs.\,\cite{Frenkel:2020ysx, Hartnoll:2020rwq} the authors considered a deformation of a thermal CFT state by a relevant scalar operator, which in the bulk results in a deformation of the black hole singularity, at late interior times, into a more general Kasner geometry from which it can be concluded that there is no Cauchy horizon as supported by the SCC conjecture. The study has been soon extended to the self-interacting $\phi^4$ scalar theory \cite{Wang:2020nkd} and holographic axionic models with a neutral scalar hair \cite{Mansoori:2021wxf}.

A similar situation happens for the asymptotically Anti-de Sitter (AdS) solutions known as holographic superconductors in the presence of a massive scalar and vector hair charged under a Maxwell field \cite{Hartnoll:2020fhc, Cai:2020wrp, Henneaux:2022ijt, Cai:2021obq}. See also Refs.\,\cite{Sword:2021pfm, Liu:2021hap,Devecioglu:2021xug, VandeMoortel:2021gsp,Auzzi:2022bfd,An:2022lvo}  for further theoretical developments in the \textit{no inner-horizon theorem}. Moreover, it has been shown that the Null Energy Condition (NEC) necessarily removes the Cauchy horizon of black holes in the presence of the charged hairs \cite{Yang:2021civ,An:2021plu,Caceres:2022smh}. 

In addition, there is an interesting dynamics inside the horizon of such holographic superconductor models \cite{Hartnoll:2020fhc}. Just below the critical temperature $T_{c}$ when the scalar field is very small, the interior evolves through several dynamical epochs inside the event horizon, including a collapse of the Einstein-Rosen bridge, the Josephson oscillations of the scalar field, and the final Kasner singularity. Notice that the Kasner geometry exponent is extremely sensitive to the black hole temperature near the critical temperature so that at a discrete set of temperatures, an infinite number of Kasner inversions is seen numerically \cite{Hartnoll:2020fhc}. The same behavior has been reported in an asymptotically flat hairy black hole \cite{Dias:2021afz} and a holographic axionic  model including Einstein-Maxwell-scalar (EMS) coupling term \cite{Sword:2021pfm}.

In this paper, we extend the no inner-horizon theorem to charged black holes in the context of the non-linear massive gravity theory and study the interior dynamics of the corresponding hairy black holes.
In fact, here we build our setup on a recent holographic model of non-linear massive gravity in which, by giving a mass to the graviton in the bulk, the momentum is no longer conserved at the boundary \cite{Vegh:2013sk}. Indeed, the bulk graviton mass breaks the diffeomorphism invariance of the gravitational action and therefore, within the holographic framework, the stress-energy tensor of the dual field theory is not conserved and in turn the momentum can be dissipated. One reason that makes this model interesting is that the black brane solutions of this model, in the absence of any scalar hair and even Maxwell tensor, have an inner horizon due to the finite graviton mass. In addition to this fact, we expect the extra bulk degrees of freedom produced by including massive gravity terms introduce distinguishing features for the interior dynamics. The main goal of the present work is to consider this issue.


The remainder of the paper is organized as follows. In Section \ref{Sec2}, we first introduce the gravitational model used in this work, a massive gravity theory coupled to a charged scalar field $\phi$ in analogy with a holographic superconductor. Then we investigate the condensation operator as a feature of the black hole exterior beyond the event horizon. In Section \ref{Sec3}, we discuss dynamical epochs inside the horizon like  the collapse of the Einstein-Rosen bridge associated with the instability of the inner Cauchy horizon triggered by the charged scalar hair. 
Sections \ref{Sec4} and \ref{Sec5} are devoted to study Josephson oscillations epoch and the Kasner form near the spacelike singularity. In Section \ref{largelimi}, we investigate the interior dynamics of the massive gravity black holes at the large massive parameter limit. Finally, we conclude with some discussions in Section \ref{Sec6}.

\section{Model} \label{Sec2}
We work here with a new class of non-linear massive gravity theories presented for the first time in \cite{Vegh:2013sk} in which the ghost field is absent. In such a theory, there is a special coupling between the metric tensor $g_{\mu \nu}$ and a fixed reference metric $f_{\mu \nu}$, giving a mass to $g_{\mu \nu}$ and breaking diffeomorphism invariance. From the holographic point of view, the violation of the diffeomorphism in the bulk gravity side results in the dissipation of the momentum in the boundary field theory.  By including a negative cosmological constant and a minimally coupled Maxwell field $A_{\mu}$, the gravitational action of such a massive gravity is given by
\begin{equation}\label{action1}
S=\int dx^4 \sqrt{-g}\Big[R+6-\frac{1}{4}F_{\mu \nu}F^{\mu \nu}+ \alpha \Tr\mathcal{K}+\beta \Big[\Big(\Tr \mathcal{K}\Big)^2-\Tr \mathcal{K}^2\Big]\Big]\,,
\end{equation} 
where $\alpha$ and $\beta$ are arbitrary dimensionless constants\footnote{Here, we have chosen units in such a way that the gravitational constant $16 \pi G=1$ and the AdS radius $L$ is set to one.}. In addition, the matrix $\mathcal{K}$ is defined as
\begin{equation}
\mathcal{K}^{\mu}_{\rho} \mathcal{K}_{\nu}^{\rho}=g^{\mu \rho} \gamma_{\rho \nu}, \hspace{0.5cm} \text{with} \hspace{0.5cm} \gamma_{\mu \nu}=\text{diag}\Big(0,0,\kappa,\kappa\Big),
\end{equation}
in which $\kappa$ is a constant. Note that all indices are raised and lowered with the dynamical metric $g_{\mu \nu}$ in the coordinate ($t,r,x,y$) where $r$ is the holographic radial direction with the UV boundary at $r=0$. Clearly, this choice of the reference metric explicitly breaks the bulk differmorphism along the spatial directions $x$ and $y$. Correspondingly, the boundary theory has no conserved momentum currents. 
  The action (\ref{action1}) admits black brane solutions corresponding to the following metric ansatz \cite{Vegh:2013sk}
  \begin{eqnarray}\label{normalsol}
  ds^2&=&\frac{1}{r^2} \Big(-f(r) dt^2+\frac{dr^2}{f(r)}+dx^2+dy^2\Big),\\
 \textbf{A}&=&\Phi(r) dt\,,\quad \Phi(r)=\rho\Big(r_{+}-r\Big),\\
 f(r)&=& 1- \frac{r^3}{r_{+}^3}+\frac{\rho^2}{4}\Big(r^4-r^3 r_{+}\Big)+\alpha \frac{\kappa}{2 r_{+} }\Big(r r_{+}-\frac{r^3}{r_{+}}\Big)+\beta  \kappa^2 \Big(r^2-\frac{r^3}{r_{+}}\Big),
  \end{eqnarray}
where $\rho$ is the charge density.
  As long as $\frac{r_{+}^4 \rho^2}{4}- \kappa r_{+}\Big(\alpha+ \beta \kappa r_{+}\Big)\le 3$ (due to the positiveness of the black hole temperature) and $\rho>0$, a smooth Cauchy horizon exists in the interior of such a black hole. Moreover, for stability of both the bulk and boundary theory, one requires \cite{Vegh:2013sk,Davison:2013jba} 
  \begin{equation}\label{stabilitycon}
  M^2(r)=- \beta-\frac{\alpha}{2 \kappa r}\geq 0,
  \end{equation} 
for all $r$. It means that at the IR horizon ($r_{+}$), we also have $M^2(r_{+})\geq 0$.
  Now, we are interested in the black hole solutions that can develop scalar hair through adding a particular coupling of a massive charged scalar field via
  \begin{equation}\label{action2}
  S^{\phi}=\int d^4 x \sqrt{-g} \left[-g^{\mu \nu} \Big(\partial_{\mu} \phi-iq A_{\mu} \phi\Big)\Big(\partial_{\nu} \phi-iq A_{\nu} \phi\Big)-m^2 \phi^2\right],
  \end{equation}
  to the Maxwell term of the action (\ref{action1}). According to the gauge/gravity duality, the Maxwell field $A_{\mu}$ is dual to the
current of a global $U(1)$ symmetry in the boundary field theory. In addition, the mass square $m^2$ of the bulk scalar field $\phi$ determines the scaling dimension $\Delta=3/2+\sqrt{9/2+m^2}$ of the dual operator $ O $ with charge $q$ under this global symmetry\footnote{Note that the negative value of $m^2$ corresponds to a relevant operator with $\Delta<3$ in the boundary field theory.} \cite{Hartnoll:2008vx}. 
 In order to find static black hole solutions in the new setup, we take the following ansatz,
\begin{equation}\label{metric1}
ds^2=\frac{1}{r^2} \Big(-f(r) e^{-\chi (r)} dt^2+\frac{dr^2}{f(r)}+dx^2+dy^2\Big),\quad \phi=\phi(r)\,, \quad \textbf{A}=\Phi(r) dt\,,
\end{equation}
where, in our choice of coordinates, the AdS boundary is at $r = 0$ and the black hole singularity locates at $r \to \infty$. In addition, at the event horizon $r_+$, the blackening function $f(r_+)$ vanishes, while we require $\Phi(r_{+})=0$ in order to get a finite norm of the gauge field at the horizon. Moreover, in terms of the metric functions the temperature is defined as
\begin{equation}\label{tem1}
T=\frac{1}{4 \pi} |f'(r_{+})| e^{-\chi(r_{+})/2}.
\end{equation}
From the actions \eqref{action1} and \eqref{action2}, the independent equations of motion read
\begin{eqnarray}
r^2 e^{-\frac{\chi}{2}} \left(e^{\frac{\chi}{2}} \Phi' \right)'&=&\frac{2 q^2 \phi^2}{f}\Phi\,, \label{eq1}\\
r^4 e^{\frac{\chi}{2}} \left(\frac{e^{-\frac{\chi}{2}} f \phi'}{r^2}\right)'&=&\left(m^2 -\frac{q^2 r^2 e^{\chi}\Phi^2}{f}\right)\phi\,, \label{eq2}\\
\chi'&=&\frac{r q^2e^{\chi}}{f^2} \phi^2 \Phi^2+r(\phi')^2\,,\label{eq4}\\
4 e^{\frac{\chi}{2}} r^4 \Big(\frac{e^{-\frac{\chi}{2}}}{r^3} f\Big)'&=&2 m^2 \phi^2 +r^4 e^{\chi}\Phi'^2-4 \kappa r (\alpha +\beta \kappa r)-12\,,\label{eq5} 
\end{eqnarray}
with a prime denoting the derivative with respect to $r$. In general, these coupled differential equations do not allow analytical solutions, thus we solve them numerically by integrating out from the horizon to the boundary at $r=0$.
It turns out that, by the regularity condition at the horizon, one finds that the above equations are fully determined by three independent parameters at the horizon, $\phi_{+}=\phi(r_{+})$, $\Phi_{+}=\Phi(r_{+})$, and $\chi_{+}=\chi(r_{+})$ for given $r_{+}$, $\alpha$, $\beta$, $q$ and $m$. From the regularity of the Euclidean on-shell action, the temperature  \eqref{tem1} is obtained in terms of these quantities as follows:
\begin{equation}
T=\frac{e^{-\frac{\chi_{+}}{2}}}{16 \pi r_{+}}\Big[12+4 \phi_{+}^2-e^{\chi_{+}} \Phi_{+}^2 r_{+}^4+4\kappa r_{+} \Big(\alpha+\beta \kappa r_{+}\Big)\Big].
\end{equation}
On the other hand, near the AdS boundary $r\to 0$, the equations \eqref{eq1}--\eqref{eq5} are solved by the following series expansions:
\begin{eqnarray}
 \phi&=&\phi_{0}\, r+ \left\langle O \right\rangle r^2+\dots\,,\\
\Phi&=& \mu- \rho r +\dots\, ,
 \\
 \chi&=&\frac{\phi_{0}^{2}}{2}\, r^2+\frac{4 \phi_{0} \left\langle O\right\rangle}{3} r^3+\dots\,,\\ 
f &=& 1+\frac{\alpha \kappa}{2} r+\beta \kappa^2 r^2+\phi_{0}^2 r^2-\left\langle T_{tt}\right\rangle r^3+\dots \, ,
\end{eqnarray}
where we have considered $m^2=-2$ to be concrete, and have taken the normalization of the time coordinate of the metric at the boundary such that $\chi(r=0)=0$ in order for the temperature of the black hole to be the temperature of the
boundary field theory. 
 Here, $\phi_{0}$ is the source of the scalar operator $O$ of the boundary field theory\footnote{The general asymptotic behavior of the scalar field $\phi$ as $r \to 0$ is
 \begin{equation*}
 \phi=\phi_{0}r+\phi_{1} r^2.
 \end{equation*}
 In the framework of holographic superconductors, both of these falloffs are normalizable, so one can choose boundary conditions in which either one vanishes. As a result, the condensate of the scalar operator $O$ in the boundary field theory dual to the field $\phi$ is given by \cite{Hartnoll:2008vx}
 \begin{equation*}
 \left\langle O\right\rangle=\begin{cases}
			\sqrt{2} \phi_{0} & \text{as} \ \phi_{1}=0,\\
             \sqrt{2} \phi_{1} & \text{as} \ \phi_{0}=0.
		 \end{cases}
 \end{equation*}
 Here we are interested in working with the first branch of the above relation.}. Moreover, $\mu$, $\rho$, and $\left\langle T_{tt}\right\rangle$ denote chemical potential, charge density and the energy density in the boundary field theory, respectively.  
After imposing regularity at the horizon $r_{+}$, the boundary quantities $\{\rho, \mu, \left\langle O\right\rangle, \left\langle T_{tt}\right\rangle \}$ can be written in terms of three independent parameters at the horizon, i.e., $\phi_{+}=\phi(r_{+})$, $\Phi_{+}=\Phi(r_{+})$, and $\chi_{+}=\chi(r_{+})$. Nevertheless, the system allows the following scaling symmetry: 
\begin{equation}
r\rightarrow r/\lambda,\quad (\phi_0, T,\kappa, \mu )\rightarrow \lambda(\phi_0, T,\kappa, \mu),\quad (\rho, \left\langle O\right\rangle)\rightarrow \lambda^2 (\rho,\left\langle O\right\rangle),\quad \left\langle T_{tt}\right\rangle\rightarrow \lambda^3 \left\langle T_{tt}\right\rangle,
\end{equation}
with $\lambda$ a constant as we set $q=r_{+}=1$. However, to find the solutions numerically, it is more convenient to use a different set of dimensionless quantities such as $\langle O\rangle/\rho^2$, $T/\sqrt{\rho}$, and $\kappa/\mu$. 

\begin{figure}
 \includegraphics[scale=0.5]{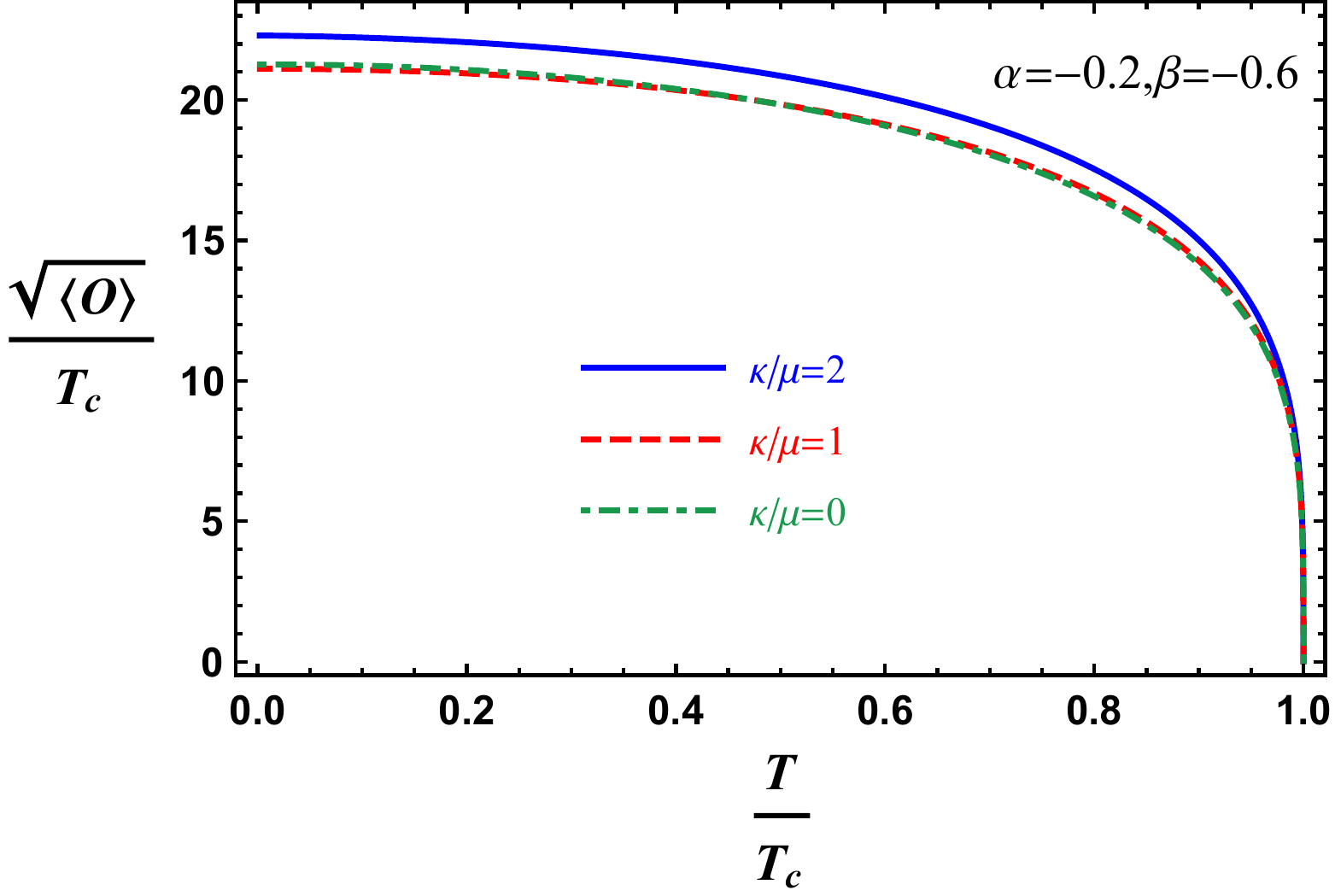}
 \includegraphics[scale=0.5]{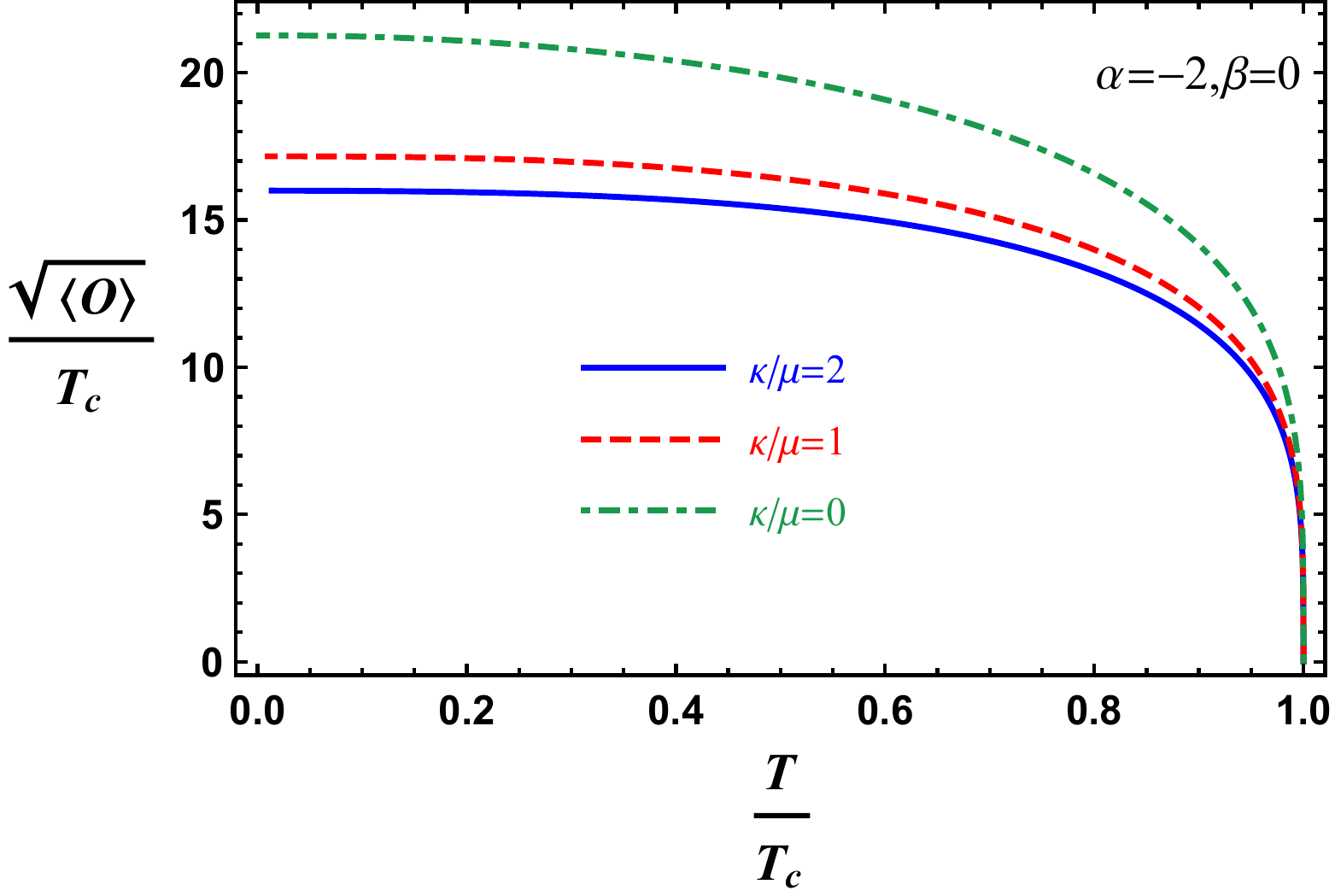}
 \begin{center}
 \includegraphics[scale=0.5]{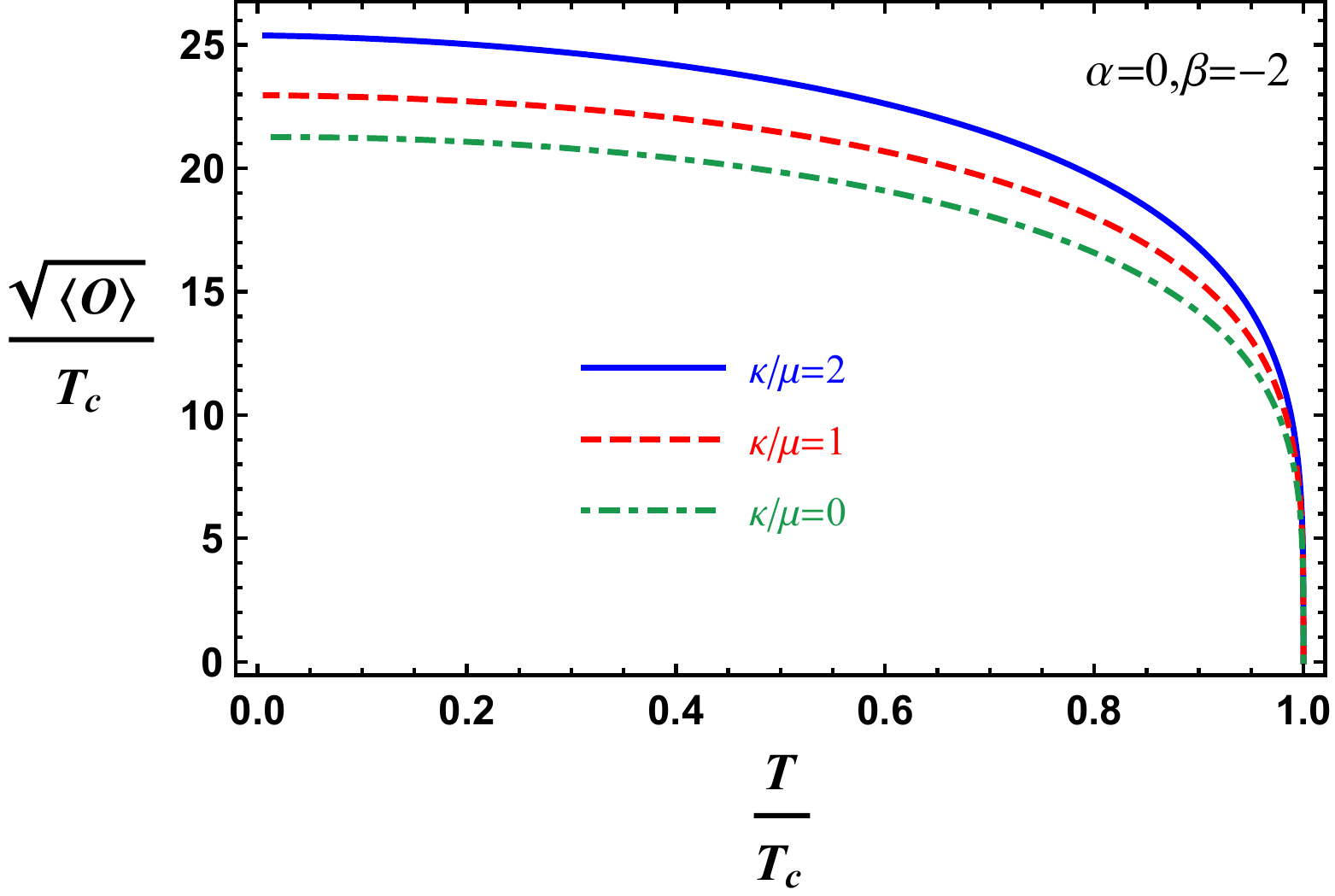}
 \end{center}
 \caption{ The value of the condensate as a function of the temperature in various configurations of the massive gravity.   \label{Fig1}}
\end{figure}

In Fig.\,\ref{Fig1}, we have depicted the condensate for the charged scalar field as a function of the temperature for various values of $\kappa/\mu=\{0,1,2\}$ in different configurations of the massive gravity background. These plots have been achieved by numerically solving the differential equations by using a shooting method.

As shown in Fig.\,\ref{Fig1}, in all cases, below a critical temperature $(T < T_{c})$, a charged condensate forms. In other words, as temperature decreases below a critical value, the system undergoes a phase transition from the normal phase to the superconducting phase. Note that in the normal phase for $T>T_{c}$ where the charged scalar hair vanishes ($\phi=0$), the solution becomes \eqref{normalsol}. It is clear that our setup also possesses a second order transition from the normal state to the superconducting state close to the critical point, which is a square root behavior predicted by the mean field theory, i.e., $\langle O\rangle \propto T_{c}^2 (T-T_{c})^{1/2}$.
Moreover, for the configuration $\beta=3\alpha=-0.6$ and $\alpha=0, \beta=-2$, the condensate increases as $\kappa/\mu$ enhances, while for the case $\alpha=-2, \beta=0$ this behavior is inverse. These results are in agreement with those reported in \cite{Zeng:2014uoa}.

Recently, in \cite{Hartnoll:2020fhc} it has been shown that there is an interesting dynamics inside the horizon of a holographic superconductor. Below the critical temperature, the interior evolves through several distinct epochs, including the collapse of the Einstein-Rosen bridge, Josephson oscillations of the scalar field, and Kasner geometry.
After investigating the condensate as a feature of the black hole exterior, in the next sections we study the interior of the black brane solution associated with our model in order to see whether the same dynamical epochs are observed inside it.

.
\section{Collapse of the Einstein-Rosen bridge}\label{Sec3}


 Just like in recent studies \cite{Hartnoll:2020fhc,Hartnoll:2020rwq,Frenkel:2020ysx,Dias:2021afz}, we expect that the black hole interior ends at a spacelike singularity at $r \to \infty$ after passing the collapse of the Einstein-Rosen (ER) bridge and the Josephson oscillations epochs. Moreover, one can observe a crossover that happens at the location of the would-be inner horizon $r_{\mathcal{I}}$. Particularly, there is typically an inner horizon $r_{\mathcal{I}}$ for the black hole solution \eqref{normalsol} in the absence of the charged scalar hair $\phi=0$. In practice, by increasing the black hole temperature slightly below the critical temperature, the solution resembles \eqref{normalsol} until we go near the would-be inner Cauchy horizon. At this point, the scalar field undergoes an instability very much. This instability is so fast for small values of the scalar field, which indicates the nonlinear nature of the dynamics in this regime. In addition, near the would-be inner horizon, the $tt$ element of metric, i.e., $g_{tt}$ approaches its would-be zero value, it suddenly suffers a very rapid collapse and becomes exponentially small. This fact is associated with the collapse of the Einstein-Rosen bridge \cite{Hartnoll:2020rwq}. Therefore, in this section, we describe ER bridge for our case of study.

 During the ER collapse and oscillatory epochs, the mass of the scalar field in Eqs.\,(\ref{eq2}) and (\ref{eq5}) and the charge term in the Maxwell equation \eqref{eq1} can be dropped \cite{Hartnoll:2020fhc}. By taking these approximations, the equations of motion read
\begin{eqnarray}\label{eqER1}
\Phi'&=&E_0 e^{-\chi/2}\,, \\\label{eqER2}
r^2 e^{-\frac{\chi}{2}} \left(\frac{e^{-\frac{\chi}{2}} f \phi'}{r^2}\right)'&=& -\frac{q^2 \Phi^2}{f}\phi\,, \\\label{eqER3}
\chi'&=&\frac{r q^2e^{\chi}}{f^2} \phi^2 \Phi^2+r(\phi')^2\,,\\\label{eqER4}
4 e^{\frac{\chi}{2}} r^4 \Big(\frac{e^{-\frac{\chi}{2}}}{r^3} f\Big)'&=&r^4 E_0^2-4 \kappa r (\alpha +\beta \kappa r)-12\,, 
\end{eqnarray}
where $E_{0}$ is the constant electric field.

\begin{figure}
 \includegraphics[width=0.45\linewidth]{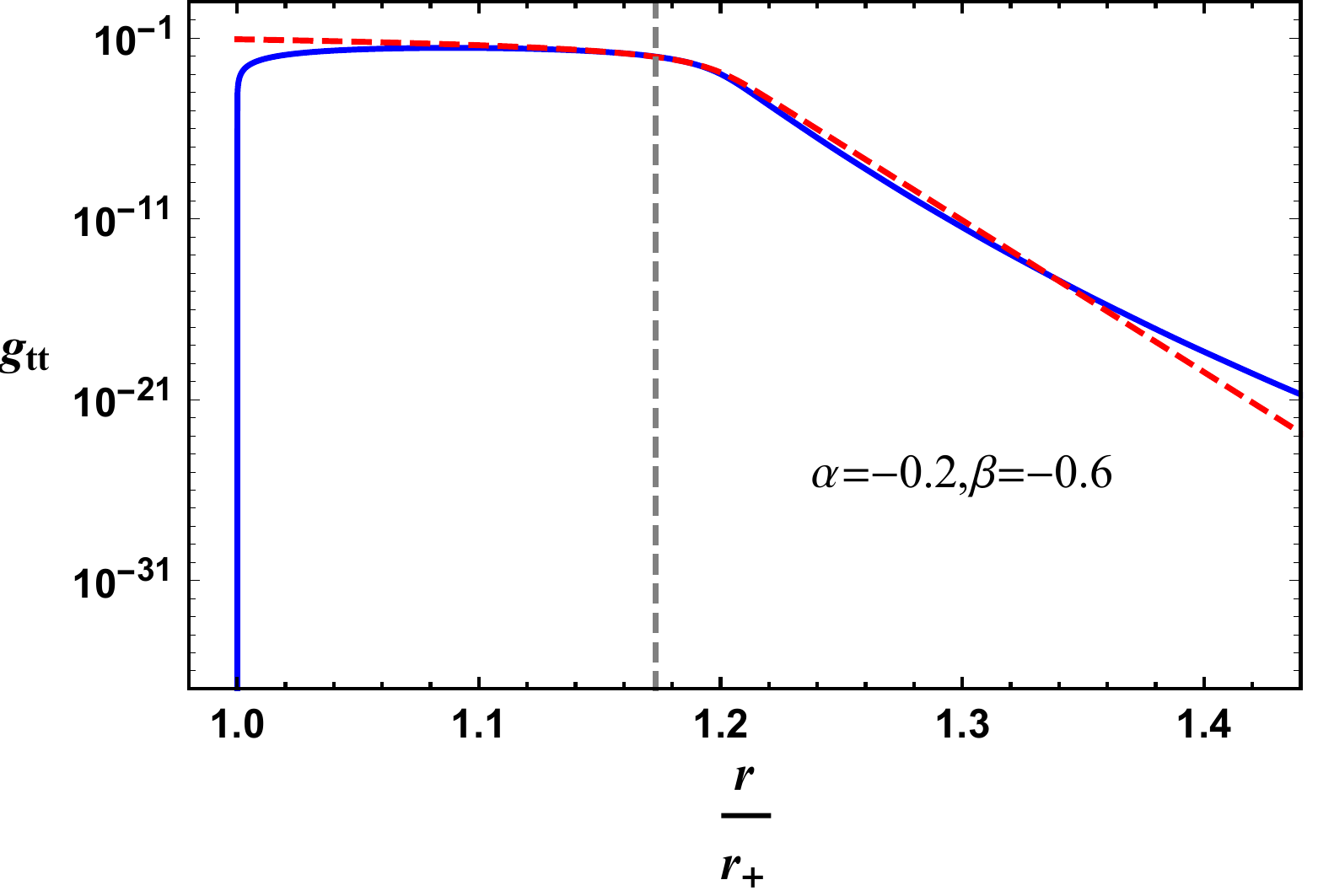}\hspace{.1cm}
 \includegraphics[width=0.45\linewidth]{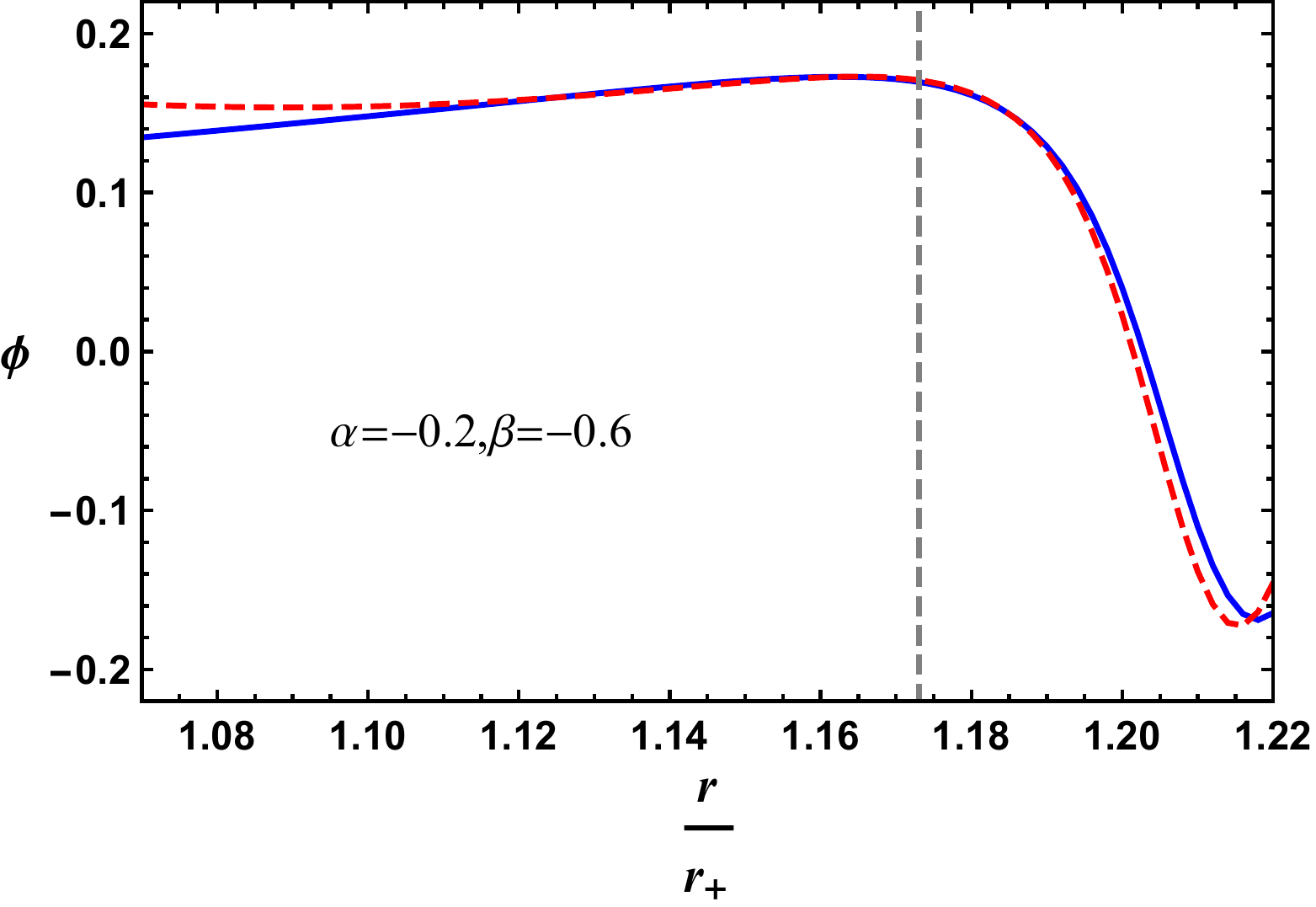}\\
 \includegraphics[width=0.45\linewidth]{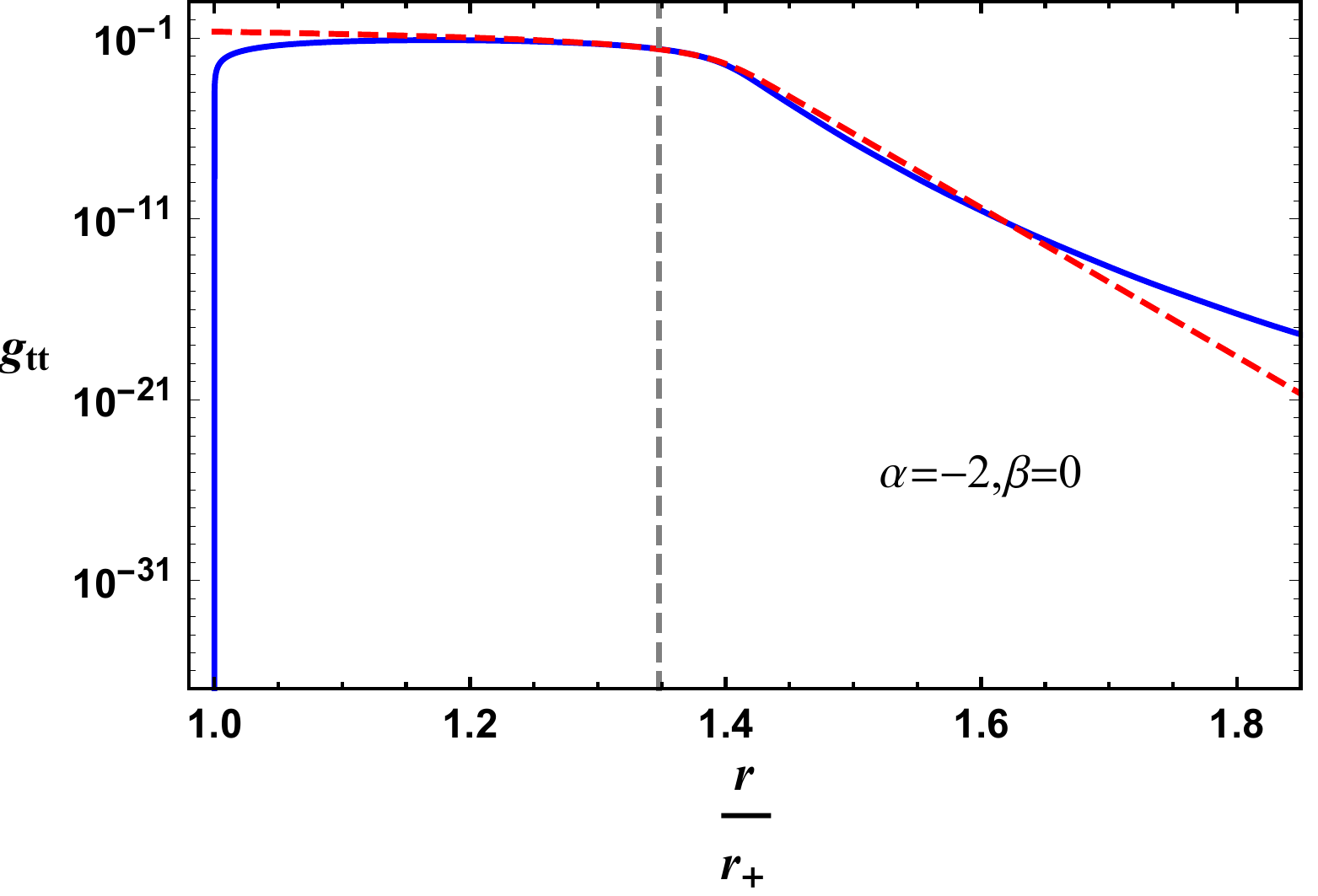}\hspace{.1cm}
 \includegraphics[width=0.45\linewidth]{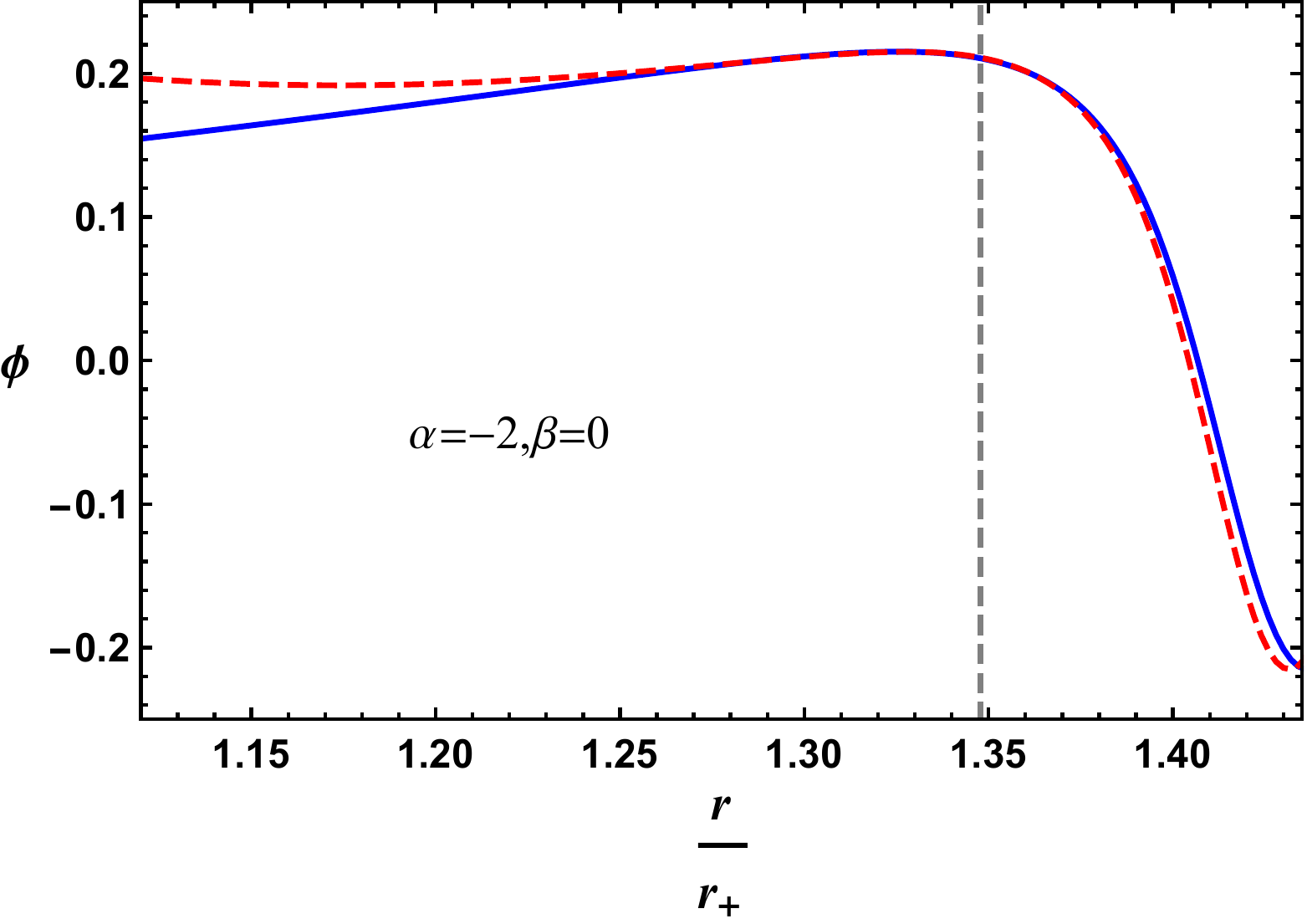}\\\includegraphics[width=0.45\linewidth]{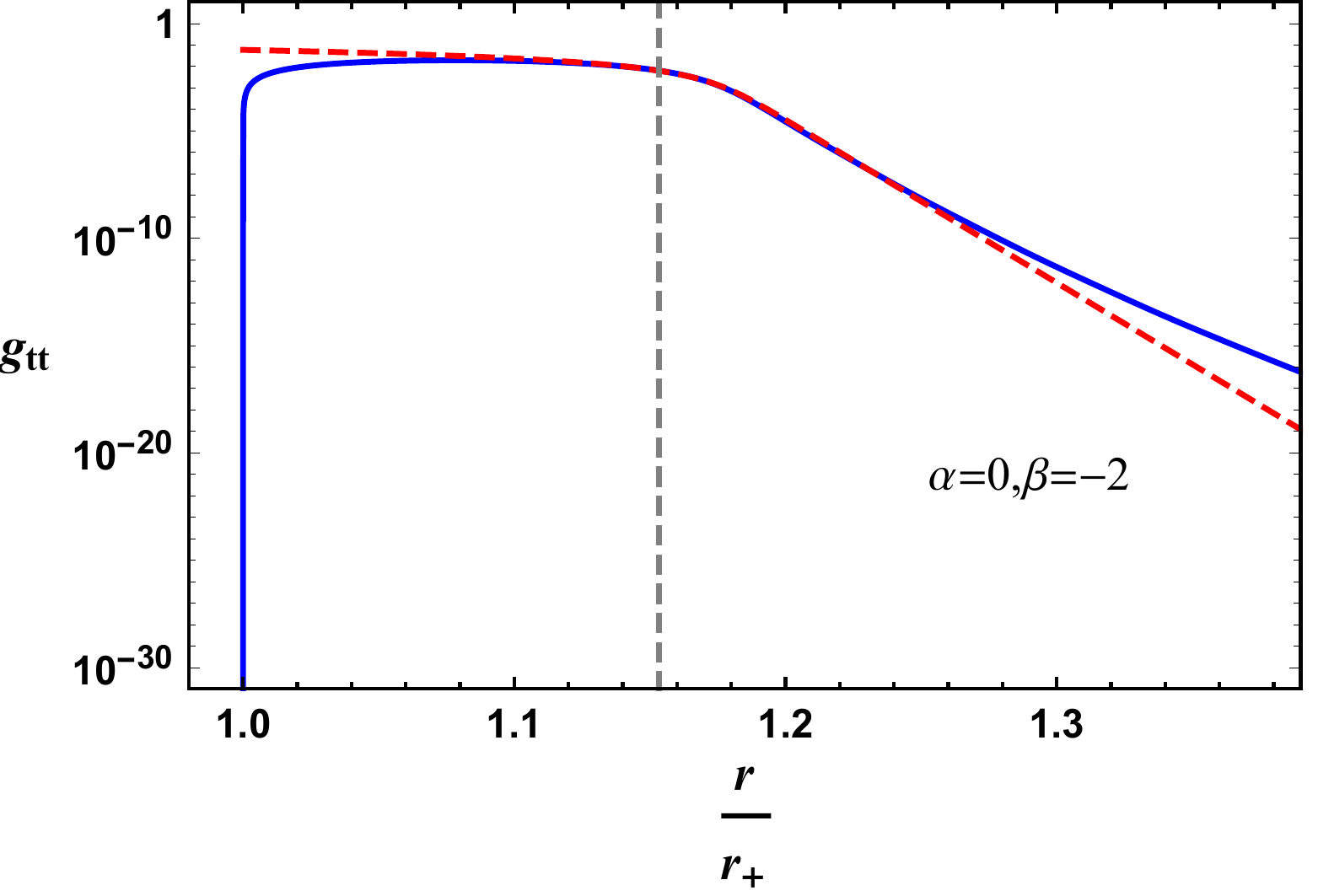}\hspace{.1cm}
 \includegraphics[width=0.45\linewidth]{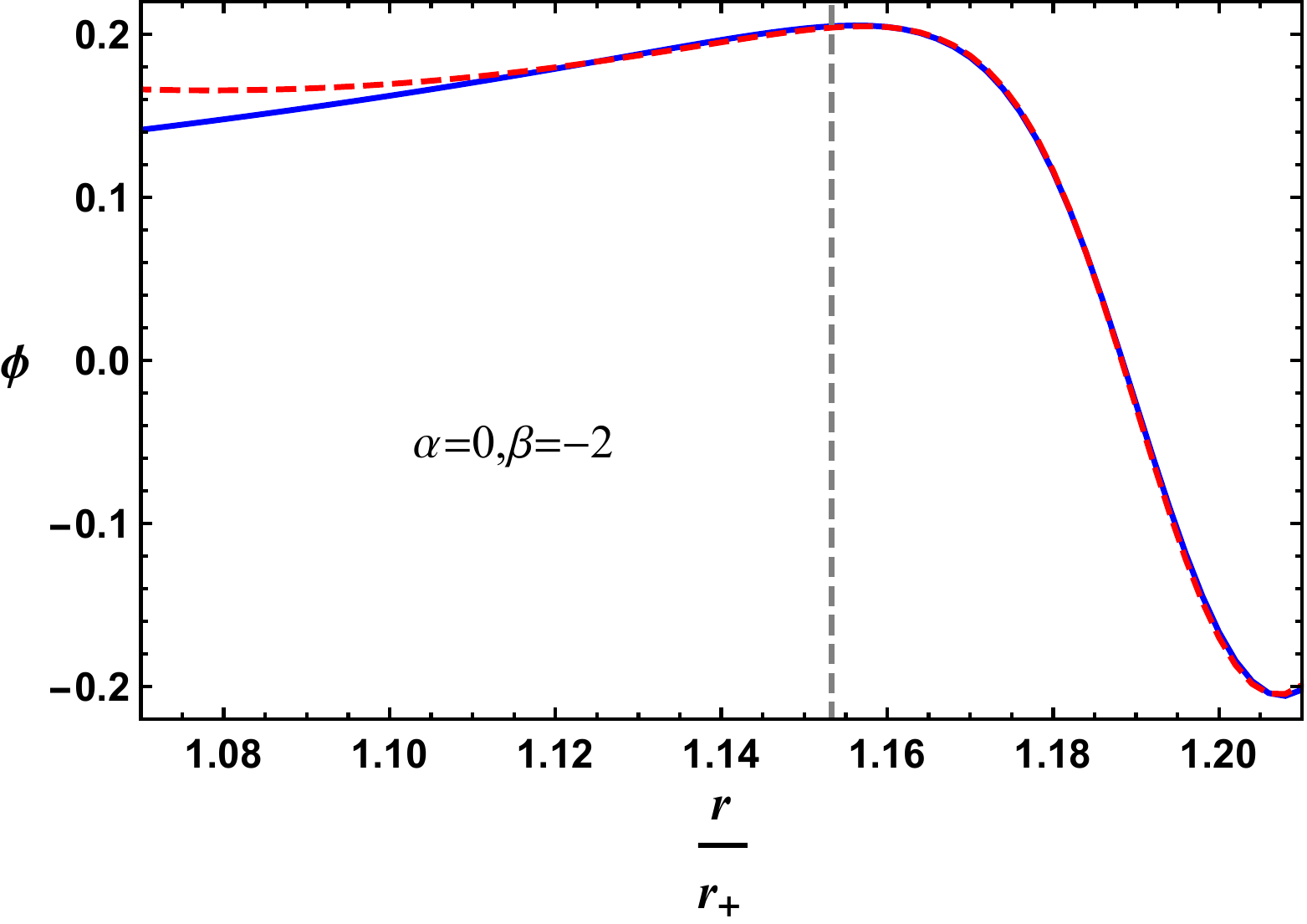}
 
 \caption{ Metric component $g_{tt}$ and scalar $\phi$ as a function of $r$ near the collapse of the ER bridge. The solid blue curves are the numerical data and the red dotted curves are the fits to the relations (\ref{phiER}) and \eqref{eq:gtt}. Here we take $T/T_{c}=0.995$ and
$\kappa/\mu=1$. The dashed vertical lines show the location of the would-be inner horizon $r_{\mathcal{I}}$.
 \label{Fig3}}
\end{figure}

Following a methodology similar to \cite{Hartnoll:2020fhc, Hartnoll:2020rwq}, just below the critical temperature $T_{c}$ when the scalar field is very small, the instability is so fast that we can essentially keep the $r$ coordinate fixed. Let us set $r= r_{\mathcal{I}}+\delta r$, so that $f$, $\chi$, $\Phi$ and $\phi$ are now functions of $\delta r$, while any explicit factors of $r$ in Eqs.\,(\ref{eq2})--(\ref{eq5}) are set to $r_{\mathcal{I}}$. 
 In this limit that $r_{I}$ is close to the inner horizon of the black hole solution \eqref{normalsol}, the potential function $\Phi$ is large compared to its derivative. 
Thus, in Eqs.\,(\ref{eqER2}) and
(\ref{eqER3}) we can set $\Phi=\Phi_0$, in which $\Phi_0$ is a constant.
Using this approximation, Eq.\,(\ref{eqER2}) can be solved explicitly yielding
\begin{equation}\label{phiER}
\phi(r)=\phi_0 \cos\left(q \Phi_0 \int_{r_{\cal{I}}}^{r} d r \frac{e^{\chi(r)/2}}{f(r)}
+\varphi_0\right),
\end{equation}
in which $\phi_{0}$ and $\varphi_{0}$ are two integration constants.
Substituting this relation into Eq.\,(\ref{eqER3}), we arrive at
$q^2 \phi^2 \Phi_0^2+e^{-\chi}f^2 \left(\phi'\right)^2=q^2 \phi_0^2 \Phi_0^2$. After some algebraic manipulations, we find that the metric component $g_{tt} = -f e^{-\chi}/r_{\mathcal{I}}^2$ must obey \cite{Hartnoll:2020rwq}
\beq\label{eq:gtt}
c_1^2 \log (g_{tt}) + g_{tt} = - \, c_2^2(r - c_3) \,,\quad \text{with} \quad c_1^2=\frac{2 q^2 \phi_0^2 \Phi_0^2}{r_{\mathcal{I}}^4 E_0^2-4 \kappa r_{\mathcal{I}} \Big(\alpha+ \beta \kappa r_{\mathcal{I}}\Big) -12},
\eeq
and $c_{2} > 0$ and $c_{3}$ are further constants of integration. Close to the would-be inner horizon, for $ r<c_{3}$, $g_{tt} \propto  (c_{3}-r)$ vanishes linearly disappearing towards $r_{\mathcal{I}}$, while for $r>c_{3}$ one observes a rapid collapse of $g_{tt}$ to an exponentially small value, i.e., $g_{tt} \propto e^{-(c_2/c_1)^2 (r-c_{3})}$.
Given $g_{tt}$, one can now obtain the solution for $\chi$ and $f$ as follows:
\begin{equation}\label{ER:fChi}
e^{-\chi}=\frac{c_2^2 g_{tt}^2}{\left(c_1^2+g_{tt}\right)^2} \frac{r_{\mathcal{I}}^3}{E_0^2 r_{\mathcal{I}}^4-4 \kappa r_{\mathcal{I}} \Big(\alpha+ \beta \kappa r_{\mathcal{I}}\Big) -12}\,, \qquad f=-r_\mathcal{I}^2 e^{\chi}  g_{tt}\,.
\end{equation}

\begin{figure}
\includegraphics[width=0.45\linewidth]{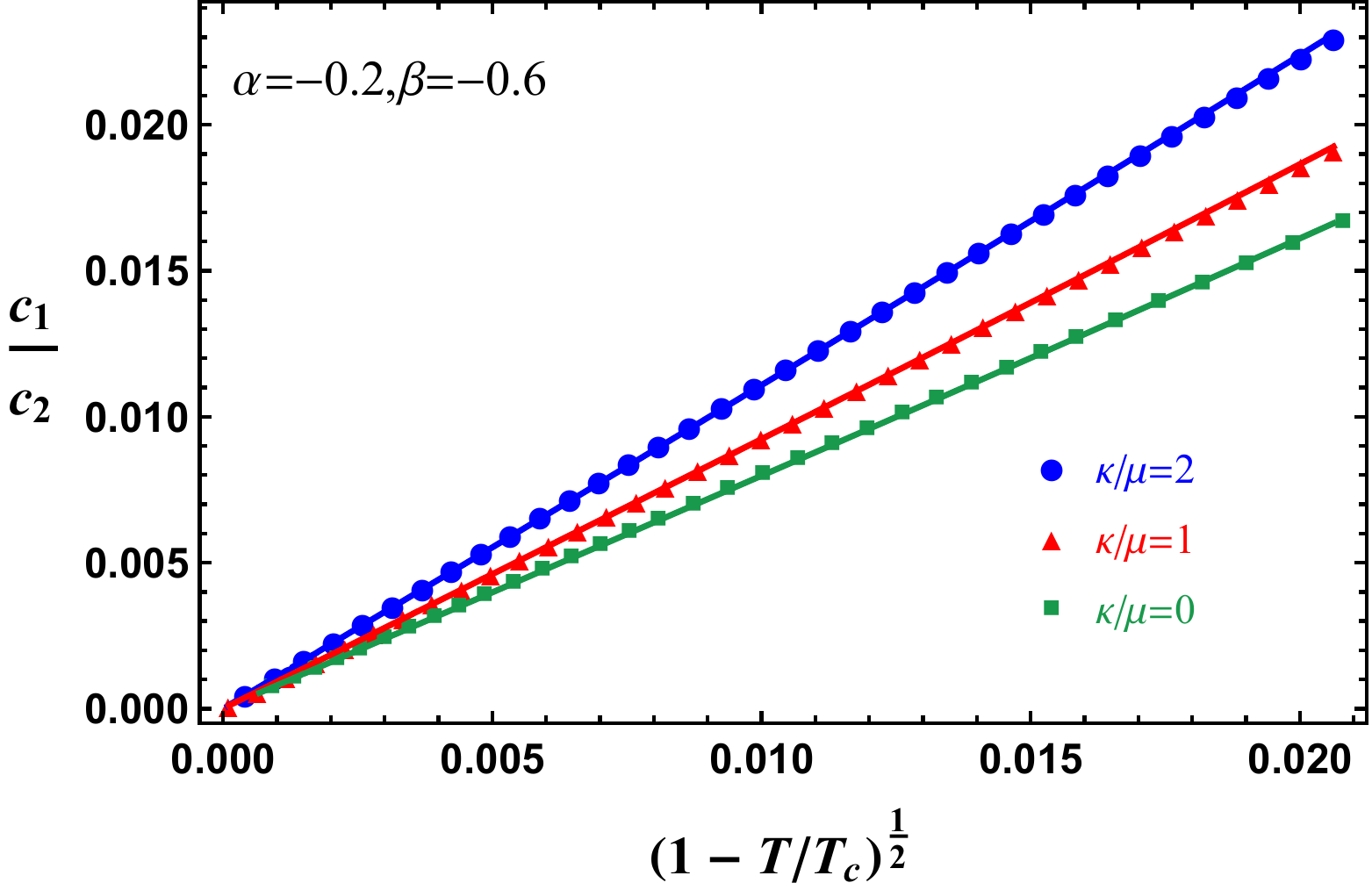}  \qquad
 \includegraphics[width=0.45\linewidth]{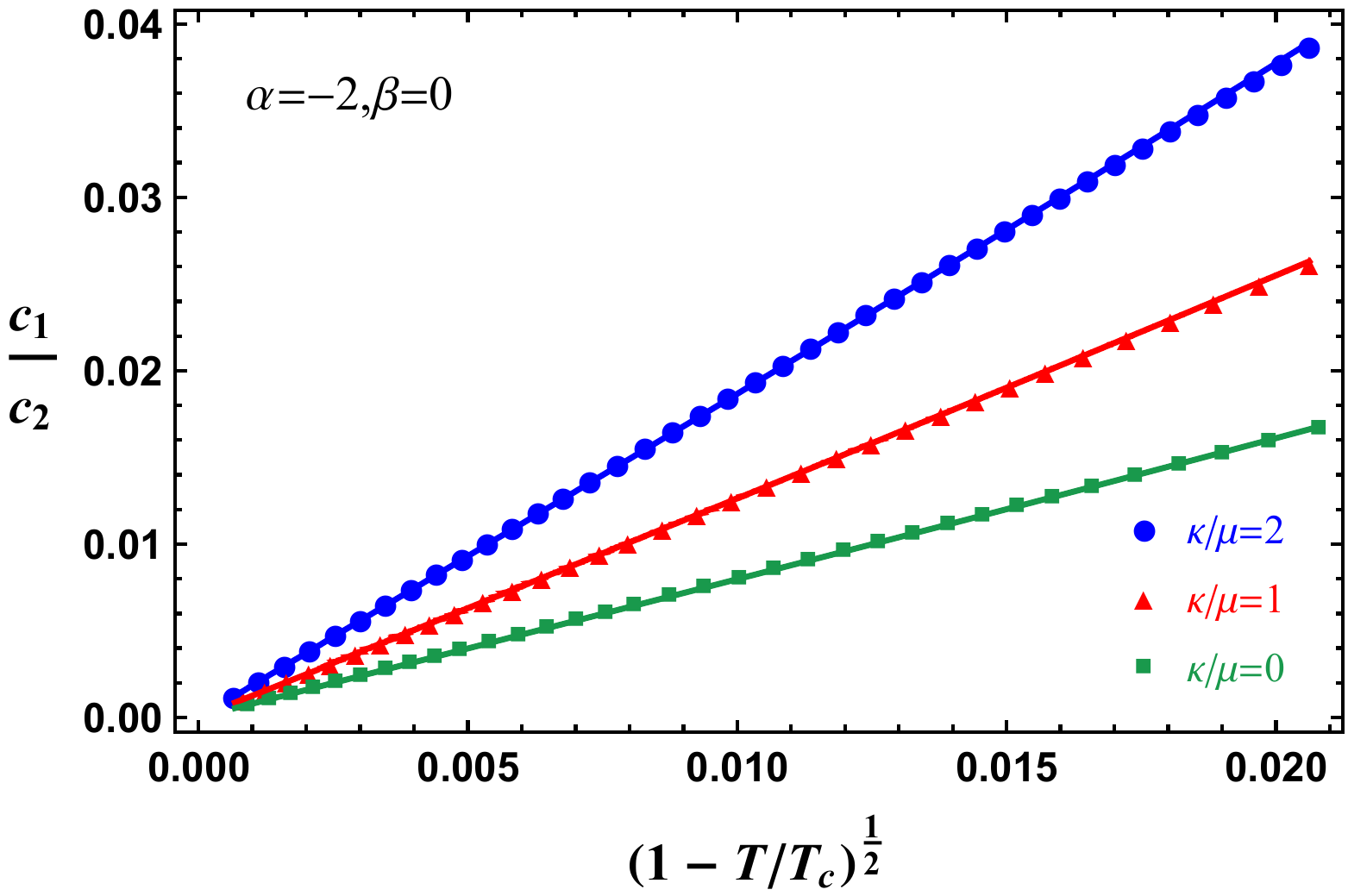}
 \begin{center}
 \includegraphics[width=0.45\linewidth]{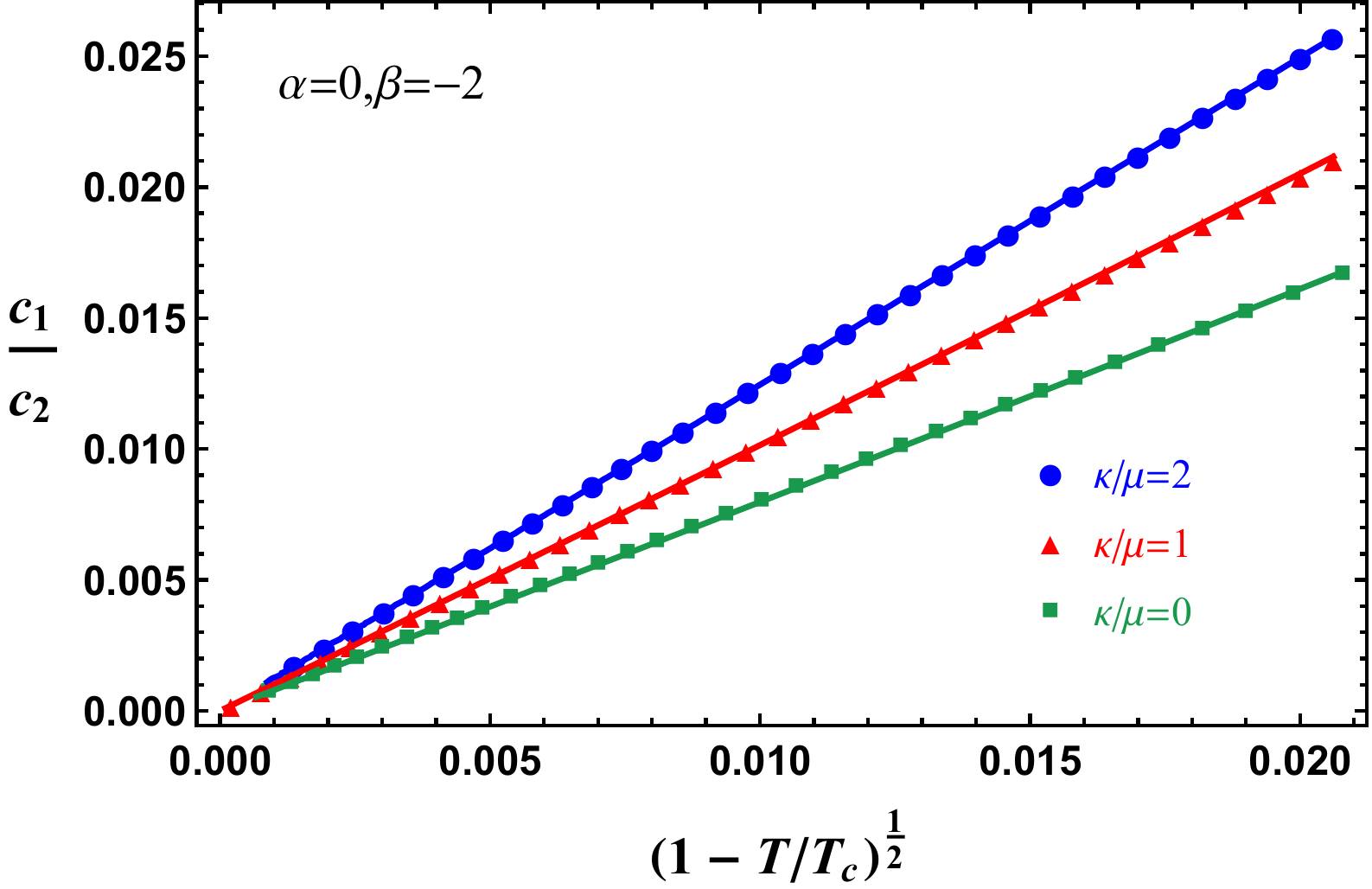}
 \end{center}
 \caption{ The variation of $c_{1}/c_{2}$ as a function of the temperature in various configurations of the massive gravity. The fitted straight lines for the data are determined by $c_{1}/c_{2}=a (1-T/T_{c})^{1/2}+b$. 
 \label{figrphi223}}
\end{figure}

\begin{figure}
 \includegraphics[width=0.45\linewidth]{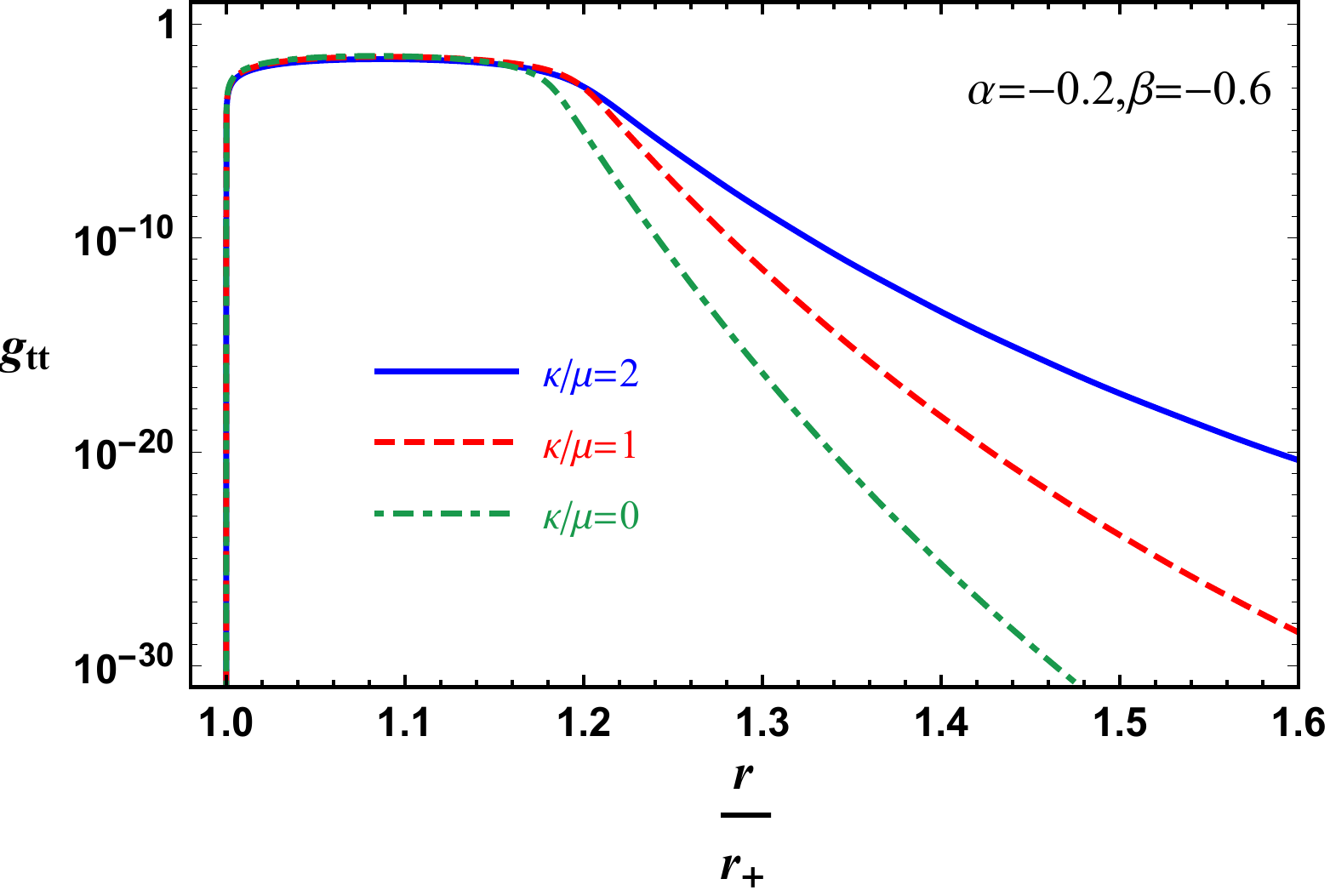} \quad \includegraphics[width=0.45\linewidth]{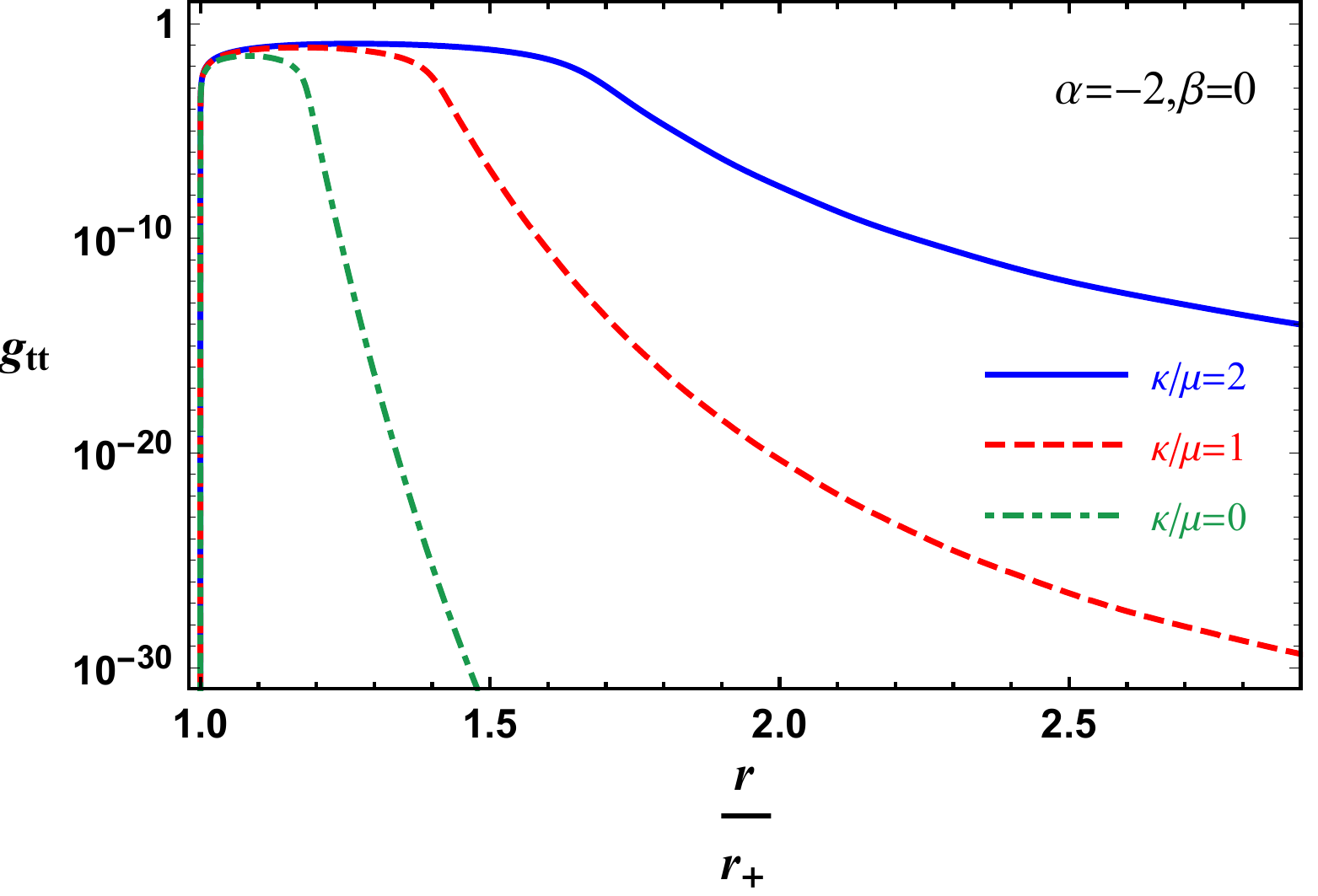}
 \begin{center}
 \includegraphics[width=0.45\linewidth]{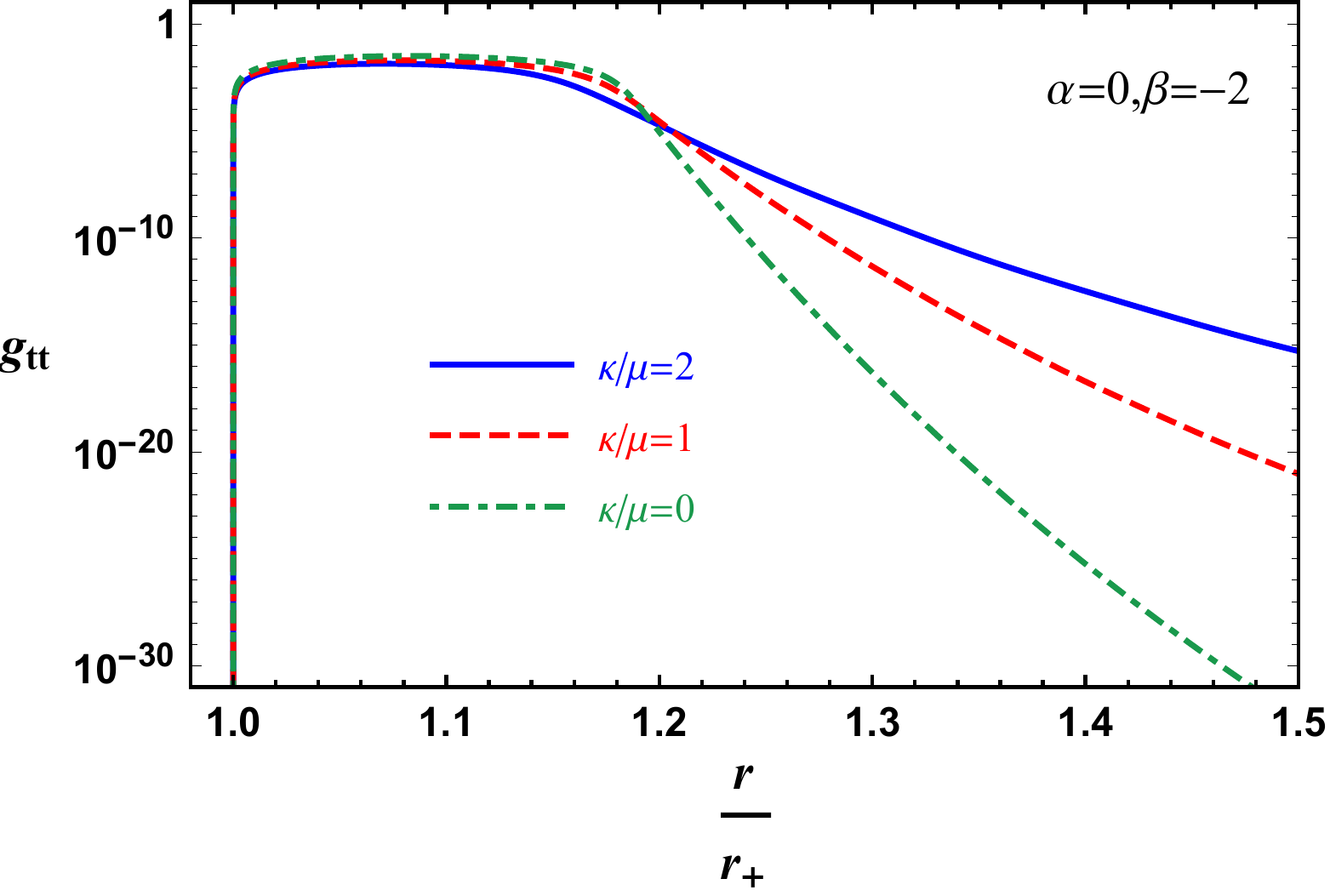}
 \end{center}
 \caption{ Metric component $g_{tt}$ as a function of $r$ near the collapse of the ER bridge for various values of $\kappa/\mu$. Here we take $T/T_{c}=0.995$.
 \label{Fig5}}
\end{figure}

For small values of the scalar field at the horizon, i.e., as $T \to T_{c}$, these solutions agree well with the full numerical evolution as shown in Fig.\,\ref{Fig3}.  
As pointed by~\cite{Hartnoll:2020rwq}, the value of $c_1/c_2$ will be large as $T\to T_{c}$. We check numerically in Fig.\,\ref{figrphi223} that indeed the ratio $c_1/c_2 \approx a (1-T/T_{c})^{1/2}+b$ near the critical temperature. Evidently, the slope of straight lines increases by growing the massive parameter $\kappa/\mu$.   In the limit $T\to T_{c}$, $(c_2/c_1)^2 \delta r$ in (\ref{eq:gtt}) can be very large, which in turn allows the metric component $g_{tt}$ to undergo a sudden change in the vicinity of $r_{\mathcal{I}}$.  From the straight lines depicted in Fig.\,\ref{figrphi223} it can be seen that the slope of these straight lines increases by enhancing the $\kappa/\mu$ values. It means that the collapse of the metric element $g_{tt}$ becomes faster at smaller values of $\kappa/\mu$ parameter. This is illustrated in Fig.\,\ref{Fig5}.

We present the collapse of the ER bridge for various values of $\kappa/\mu$ at $T/T_c=0.995$ for different configurations of the massive gravity in Fig.\,\ref{Fig5}.  
As can be seen, in the pure non-linear massive gravity configuration the collapse is shifted outward to a large radial coordinate $r$ by increasing $\kappa/\mu$ parameter, while for the other configurations, the ER collapse occurs approximately at the same would-be inner horizon with different slopes. Therefore, it makes sense that at this temperature the ER collapse could be completely disappeared at the large massive parameter $\kappa/\mu$ in the pure non-linear massive configuration with $\alpha \neq 0$ and $\beta =0$. We deal with this point in Section \eqref{largelimi}.  


\section{Josephson oscillation epoch}\label{Sec4}

Since inside the horizon, $r$ is a timelike coordinate while $t$ is spacelike, the cosine form of the scalar function in \eqref{phiER} can be written as $q \int A_{\hat t} \, d\tau$, where  $d\tau = \sqrt{g_{rr}} dr$ is  the proper time and $A_{\hat t} = A_t/\sqrt{g_{tt}}$ is the vector potential in locally flat coordinates. Therefore, a nonzero $A_{\hat t}$ determines a phase winding in the $t$ direction. In this respect, the scalar condensate $\phi$ indicates the superfluid stiffness. As a result, Eq.\,\eqref{phiER} determines the oscillations in time of the superfluid stiffness sourced by a background phase winding. This phenomenon is dubbed as the Josephson effect.  By the end of the collapse of the ER bridge epoch, these Josephson oscillations can be dominated near the critical temperature \cite{Hartnoll:2020fhc}. 

Just after the end of the collapse of the ER bridge described in the previous section, the derivative of the Maxwell field $\Phi'\propto e^{-\chi/2} $ is still very small which leads us to ignore the $( E_0^2-\frac{4}{r^3} \kappa  (\alpha +\beta \kappa r)-\frac{12}{r^4}) e^{-\frac{\chi}{2}}$  contribution in \eqref{eqER4}. It follows that the Maxwell field $\Phi$ and the gravitational field $f$ are given by 
\be\label{eq:pos}
 \Phi \simeq \Phi_o  \,, \qquad \frac{f e^{-\chi/2}}{r^3} \simeq - \frac{1}{d_1} \,,
\ee
where $d_1$ is constant. To fix it, one needs to match the Josephson oscillation solution \eqref{eq:pos} with the $ r > c_{3}$ solution \eqref{ER:fChi} of the collapse of the ER bridge. We thus obtain
  \be\label{c3}
d_1\simeq \frac{c_2}{c_1^2}\sqrt{\frac{r_\mathcal{I}^5}{E_0^2 r_{\mathcal{I}}^4-4 \kappa r_{\mathcal{I}} \Big(\alpha+ \beta \kappa r_{\mathcal{I}}\Big) -12}} \,.
\ee
Because $g_{tt}$ is very small near the inner horizon, here we have used the appropriate  approximation $c_1^2+g_{tt}^2(r_\mathcal{I}) \simeq c_1^2$.  The above relation confirms that $d_1$ goes to large values as $T \to T_{c}$. 

\begin{figure}
\includegraphics[width=0.45\linewidth]{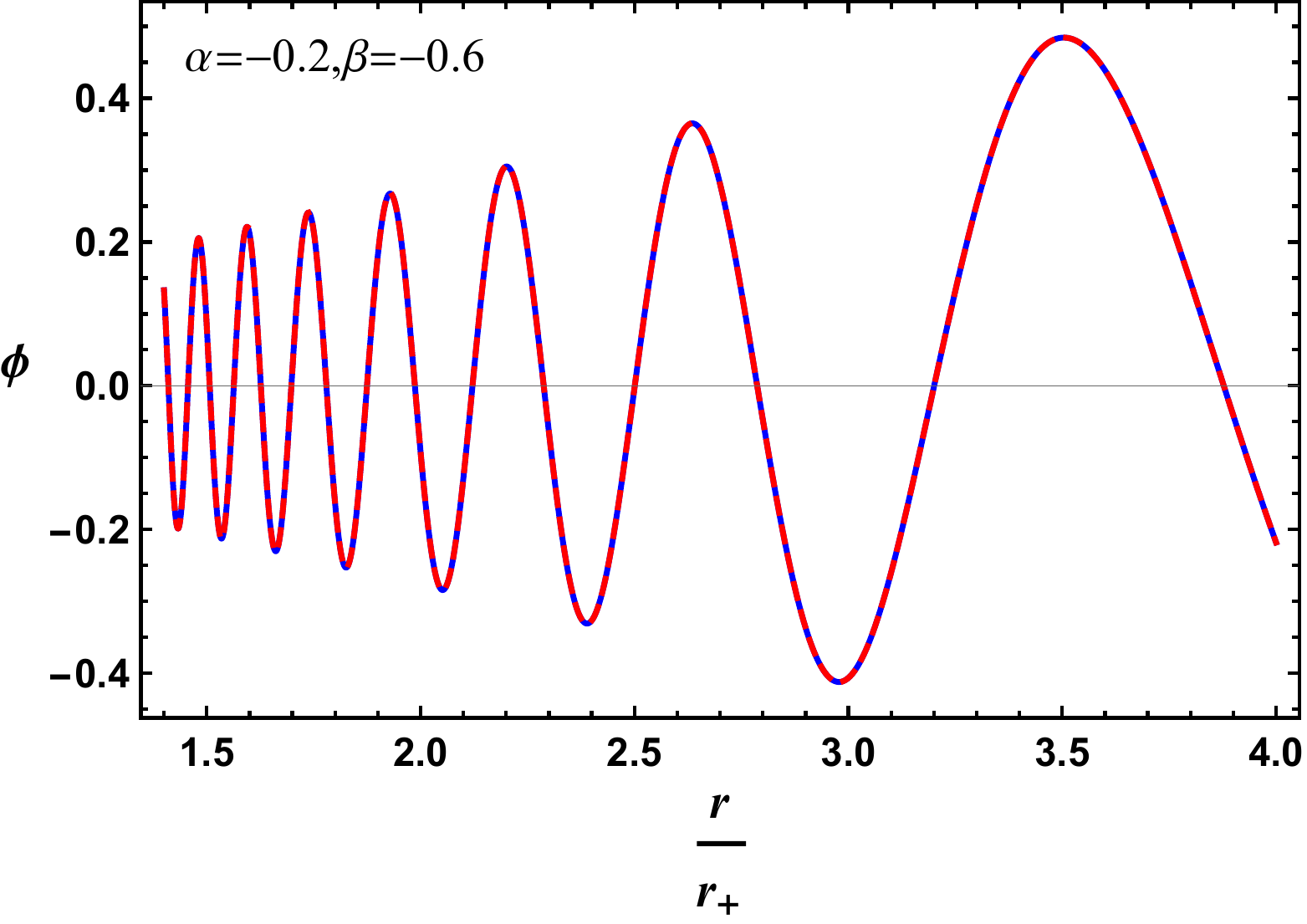} 
 \includegraphics[width=0.45\linewidth]{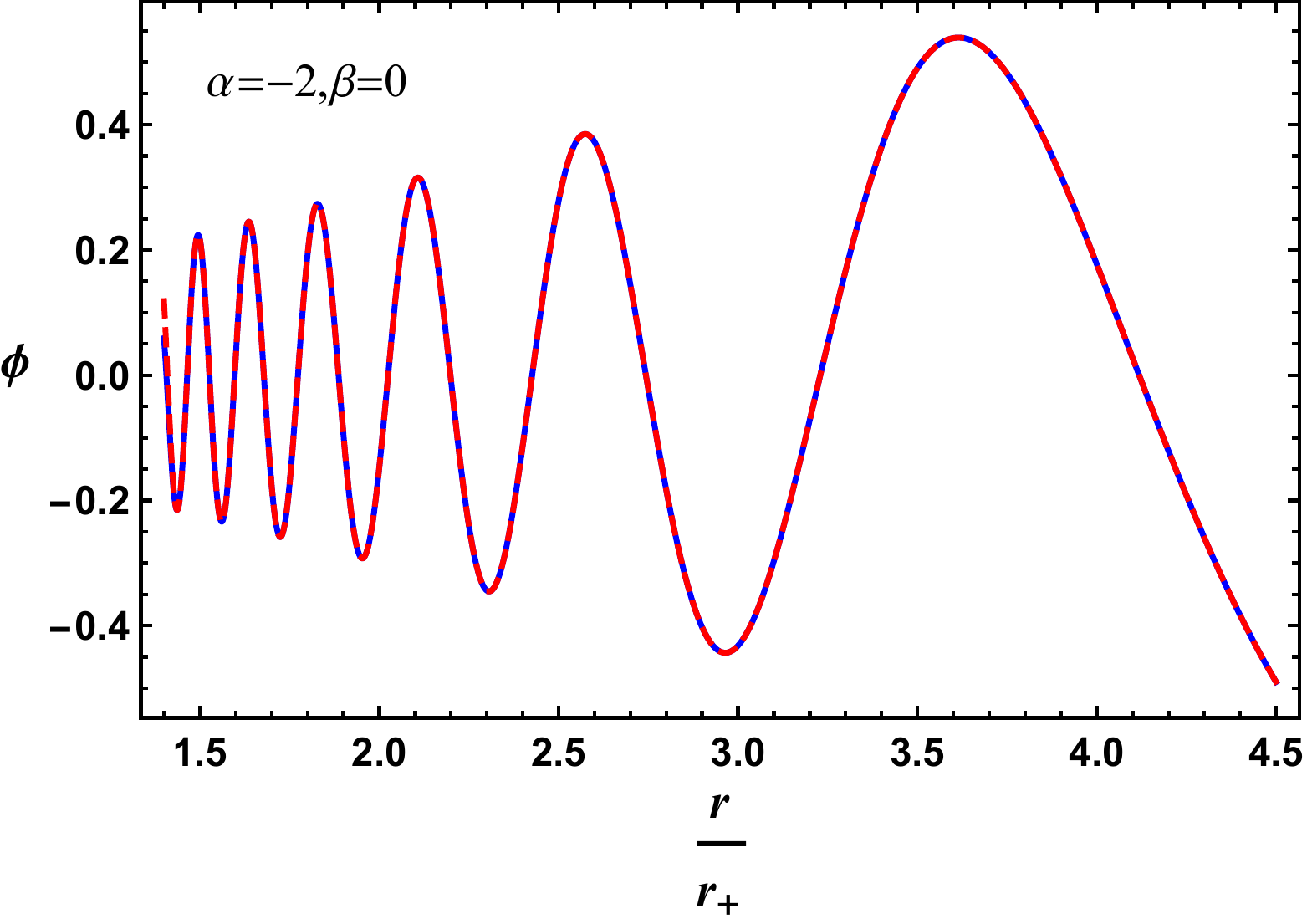}
 \begin{center}
 \includegraphics[width=0.45\linewidth]{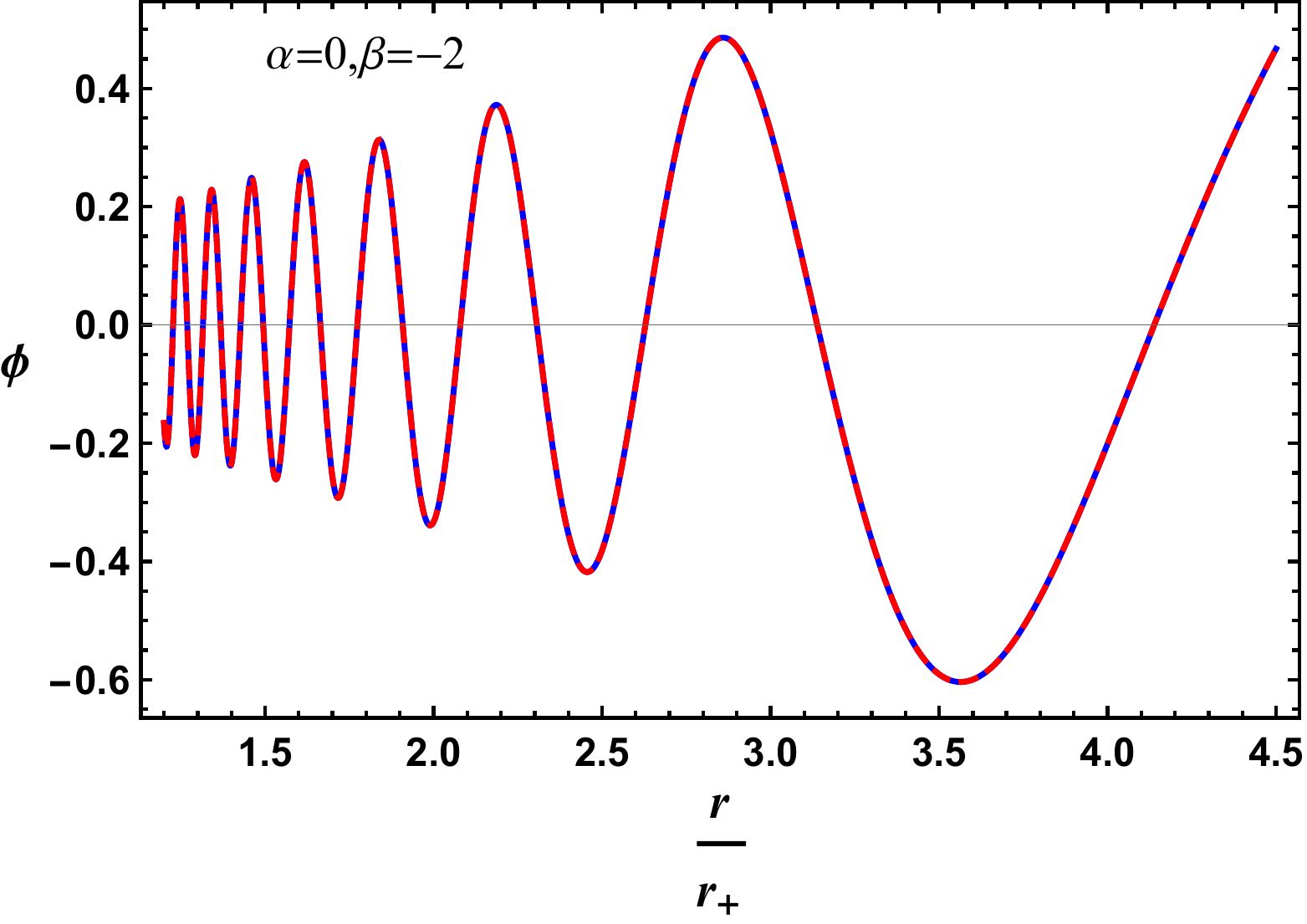}
 \end{center}
 \caption{ A comparison of the numerical solution (solid blue curves) of the scalar hair $\phi$ and fits to the analytic relation (\ref{Joseph:psi})(red dashed curves) describing the Josephson oscillations. Here  $T/T_{c}=0.995$ and
$\kappa/\mu=1$.
 \label{Fig8}}
\end{figure}

By the use of \eqref{eq:pos} Eqs.\,\eqref{eqER2} and \eqref{eqER4} can be solved in terms of Bessel functions. Thus, the scalar field solution is
\be\label{Joseph:psi}
\phi \simeq d_2 J_0\left(\frac{|q \Phi_o| d_1}{2 r^2}\right) + d_3  Y_0\left(\frac{|q \Phi_o| d_1}{2 r^2} \right),
\ee
in which $d_2$ and $d_3$ are integration constants.
In Fig.\,\ref{Fig8}, we compare the analytical approximation \eqref{Joseph:psi} (blue dashed lines) with the numerical data for the scalar field (solid lines) for $T/T_{c}=0.995$ and
$\kappa/\mu=1$ in different configurations of massive gravity. Clearly, the analytical expression \eqref{Joseph:psi} of the Josephson oscillations in the limit $T \to T_{c}$ is in excellent agreement to the numerical solution of the full  Eqs.\,(\ref{eq1})--(\ref{eq4}).

\begin{figure}
 \includegraphics[width=0.45\linewidth]{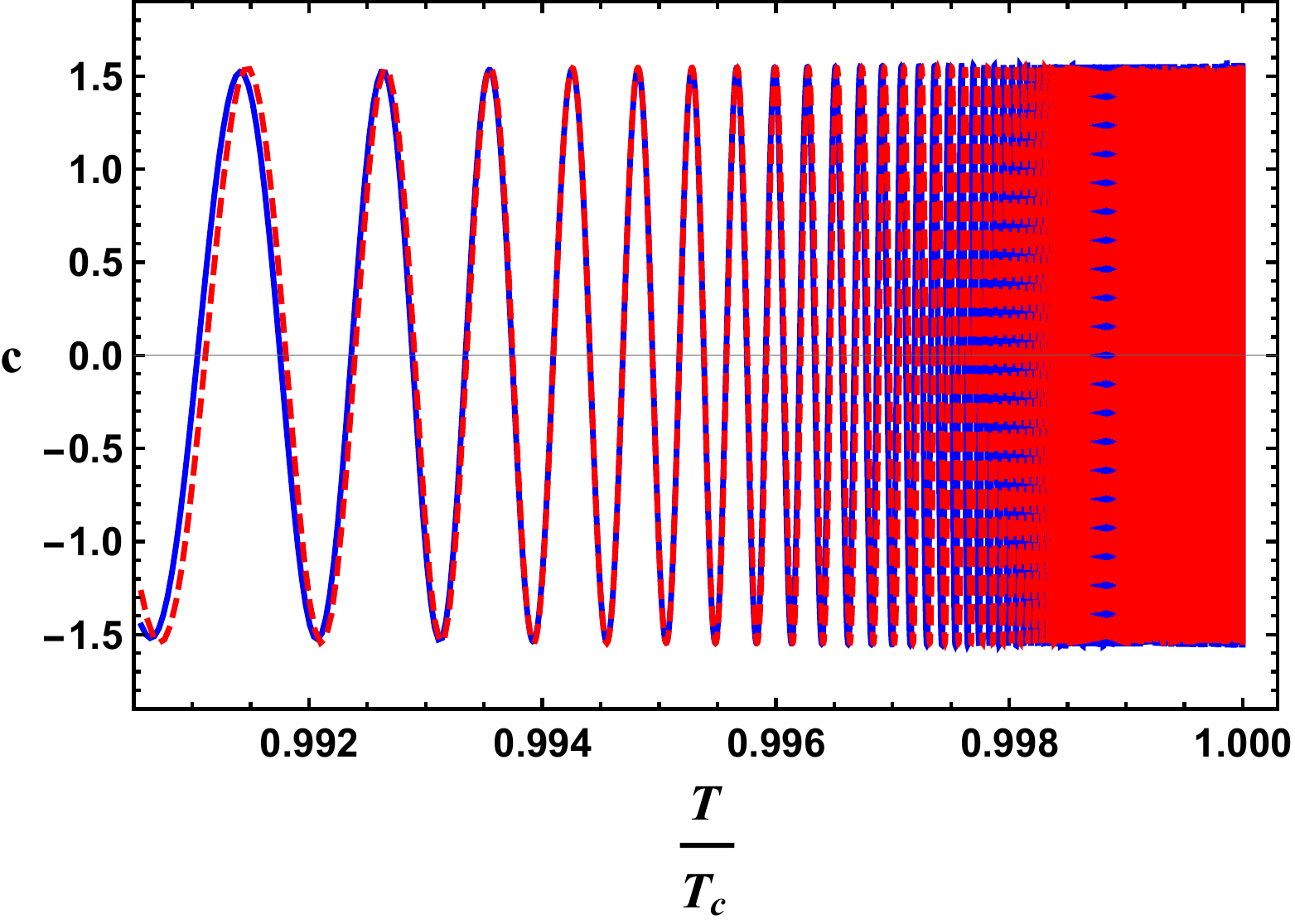}
 \includegraphics[width=0.45\linewidth]{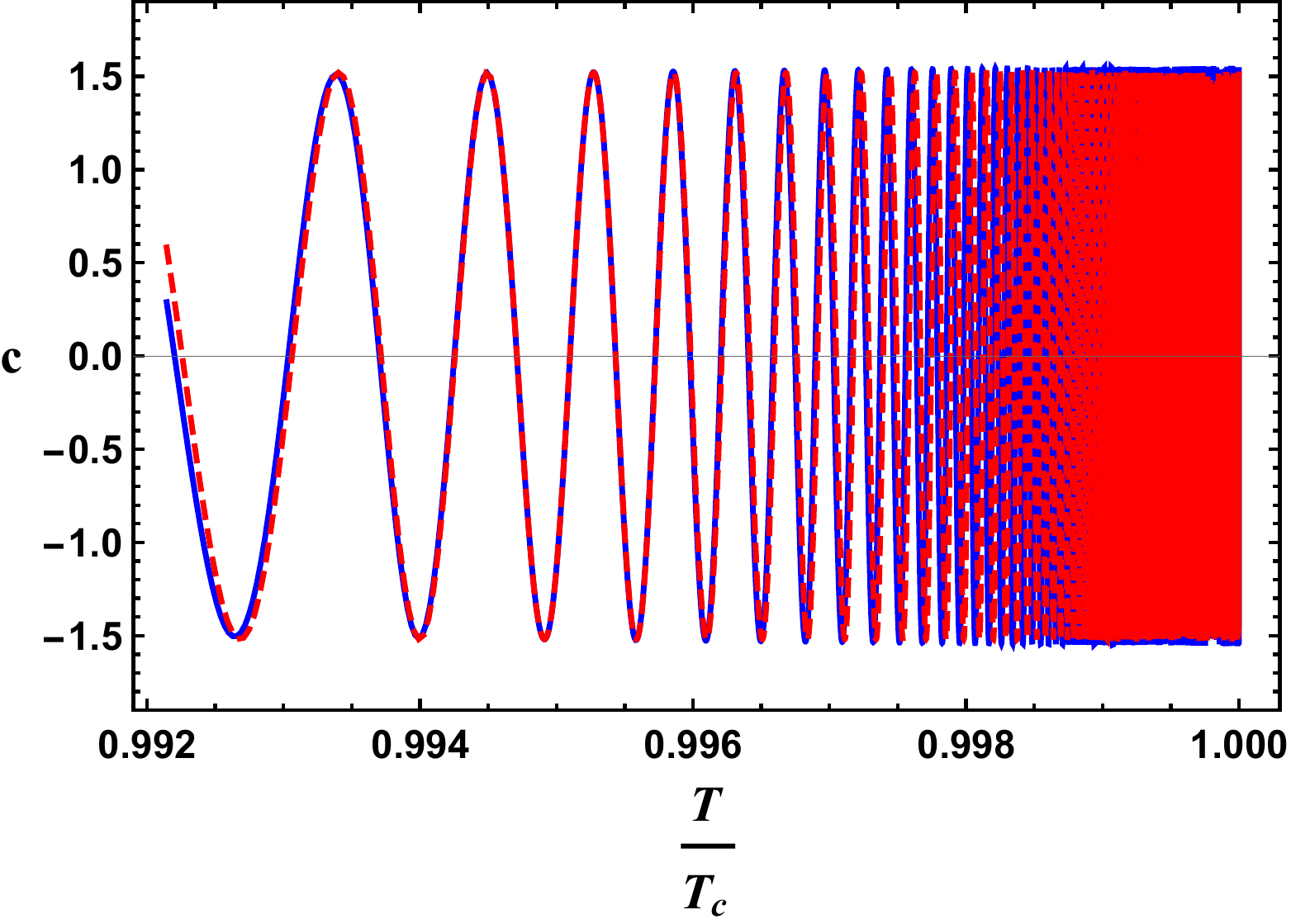}
 \begin{center}
 \includegraphics[width=0.45\linewidth]{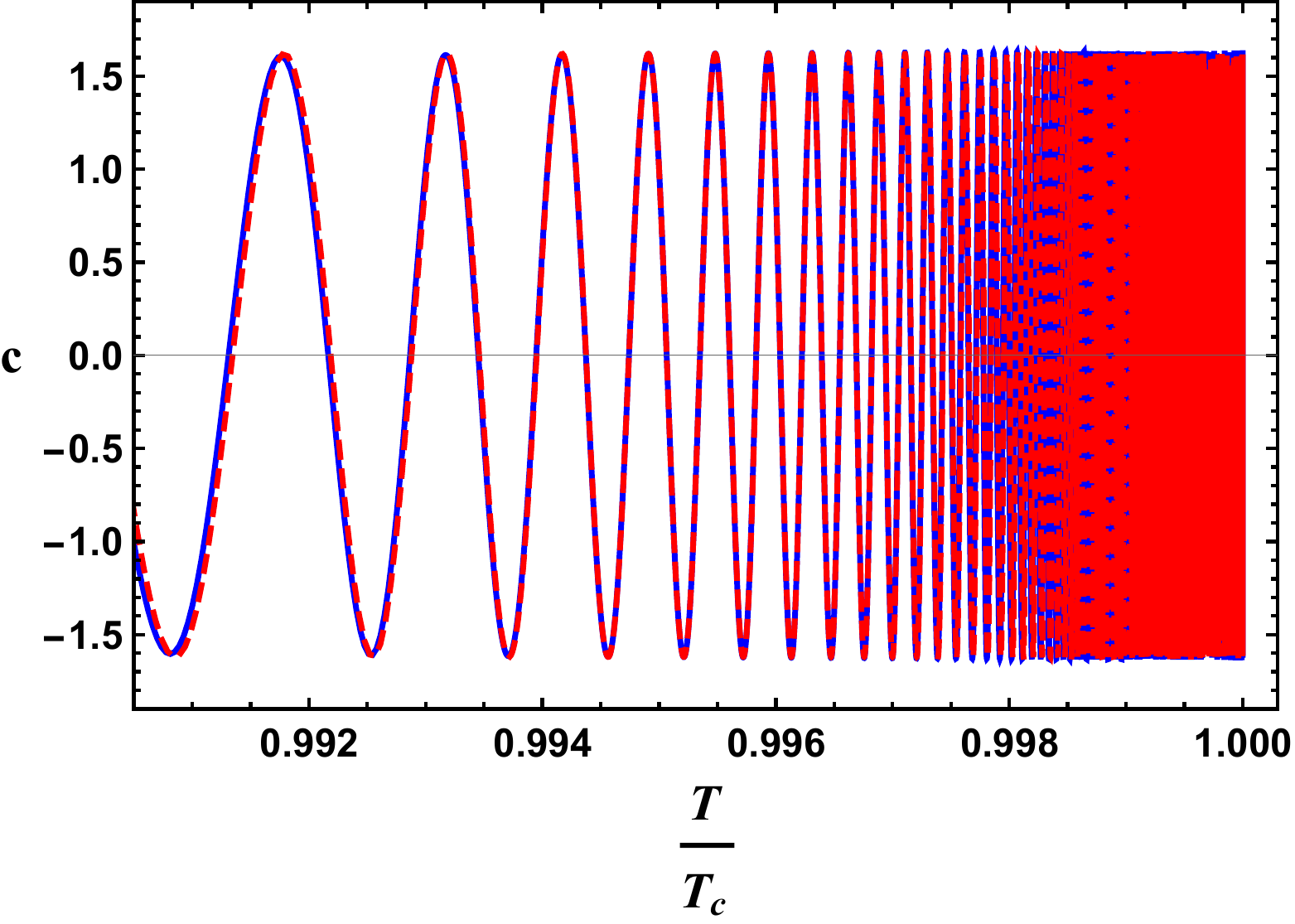}
 \end{center}
 \caption{ The oscillatory behavior of $c=-\sqrt{8} d_{3}/\pi$ determining the Kasner exponents as $T \to T_{c}$. The solid blue lines are the numerical data and the red dashed curves are fits to
the relation \eqref{C5relation}. From top to bottom, panels are associated with the massive configuration $\beta=3\alpha=-0.2$, $\beta=0,\alpha=-2$ and $\beta=-2,\alpha=0$, respectively. Here we have taken  $T/T_{c}=0.995$ and
$\kappa/\mu=1$.
 \label{Fig7}}
\end{figure}

 It is worth mentioning that the oscillations of the scalar field start in the collapse of the ER bridge regime and propagate continuously onto the Josephson oscillation epoch.
 Therefore, these constants can be found through matching the Josephson solution \eqref{Joseph:psi} and its derivative for the small $r$ with the ER bridge collapse solution \eqref{phiER} and its derivative at the large $r$ behavior in the overlapping region at $r_\mathcal{I}$. In this way, one finds
\begin{subequations}\label{c4c5} 
\begin{align}
& d_2 \simeq \left(\frac{r_\mathcal{I} \pi^2 c_2^2}{8} \, \frac{\phi_o^2}{c_1^2} \right)^{1/4} \sin \left(\frac{c_2 \sqrt{r_\mathcal{I}}}{\sqrt{2}}\, \frac{1}{\phi_o c_1} - \varphi_o + \frac{\pi}{4} \right),
\\
& d_3 \simeq \left(\frac{r_\mathcal{I}\pi^2 c_2^2}{8} \, \frac{\phi_o^2}{c_1^2} \right)^{1/4} \sin \left(\frac{c_2 \sqrt{r_\mathcal{I}}}{\sqrt{2}}\, \frac{1}{\phi_o c_1} - \varphi_o - \frac{\pi}{4} \right). 
\end{align}
\end{subequations}
Interestingly, the relation (\ref{c4c5}) confirms that $d_{3}$ is strongly oscillating with a constant amplitude as $T \to T_{c}$, i.e.
\begin{equation}\label{C5relation}
d_{3}=A \sin \Big(\frac{B}{1-\frac{T}{T_{c}}}+C\Big).
\end{equation}
As illustrated in Fig.\,\ref{Fig7}, this behavior agrees well with numerics over many oscillations. As an example, for the configuration of the massive gravity with $\beta=3\alpha=-0.6$, we find that the red dashed curve describing \eqref{C5relation} with $A=1.54553$, $B=0.331731$ and $C=1.3231$ which is in excellent agreement with the numerical data (blue solid line) over many oscillations as $T \to T_{c}$. 
 
 Near the singularity, the large $r$ behavior of the scalar field \eqref{Joseph:psi} is important. This yields 
\be\label{eq:tokasner}
\phi|_{\hbox{\tiny large} \, r} \simeq \frac{2 d_3}{\pi} \log \left(d_1 \frac{q e^{\gamma_E} \Phi_o}{4 r^2} \right) + d_2 + \cdots \,,
\ee
with $\gamma_E$ being the Euler-Mascheroni constant. This logarithmic behavior determines the onset of a Kasner regime which is discussed in the next section. 

 By substituting \eqref{eq:pos} into \eqref{eqER3}, one arrives at $\chi^\prime = 4\left(r{\phi^\prime}^2+ r^{-5}c_3^2 q^2 \Phi_o^2 \phi^2 \right)$. This allows us to obtain $\chi$. By plugging this solution into \eqref{eq:pos}, we finally obtain the following expression for the metric function $f$:
\begin{align}\label{Joseph:chif}
 f \simeq - f_o r^3 \exp \left[\frac{1}{2}\int_{r_\mathcal{I}}^r \left(\tilde{r} \phi^{\prime\,2} + \frac{q^2 \Phi_o^2 c_3^2 \phi^2}{\tilde{r}^5} \right) d\tilde{r} \right] \,, 
\end{align}
with $f_{o}$ being a constant. If we insert the Bessel functions solution \eqref{Joseph:psi} into the above integral,
it can be performed analytically in terms of Bessel functions.
These describe the small oscillations seen in $r f'/ f$ as shown in Fig.\,\ref{Fig10}.

\begin{figure}
 \includegraphics[width=0.45\linewidth]{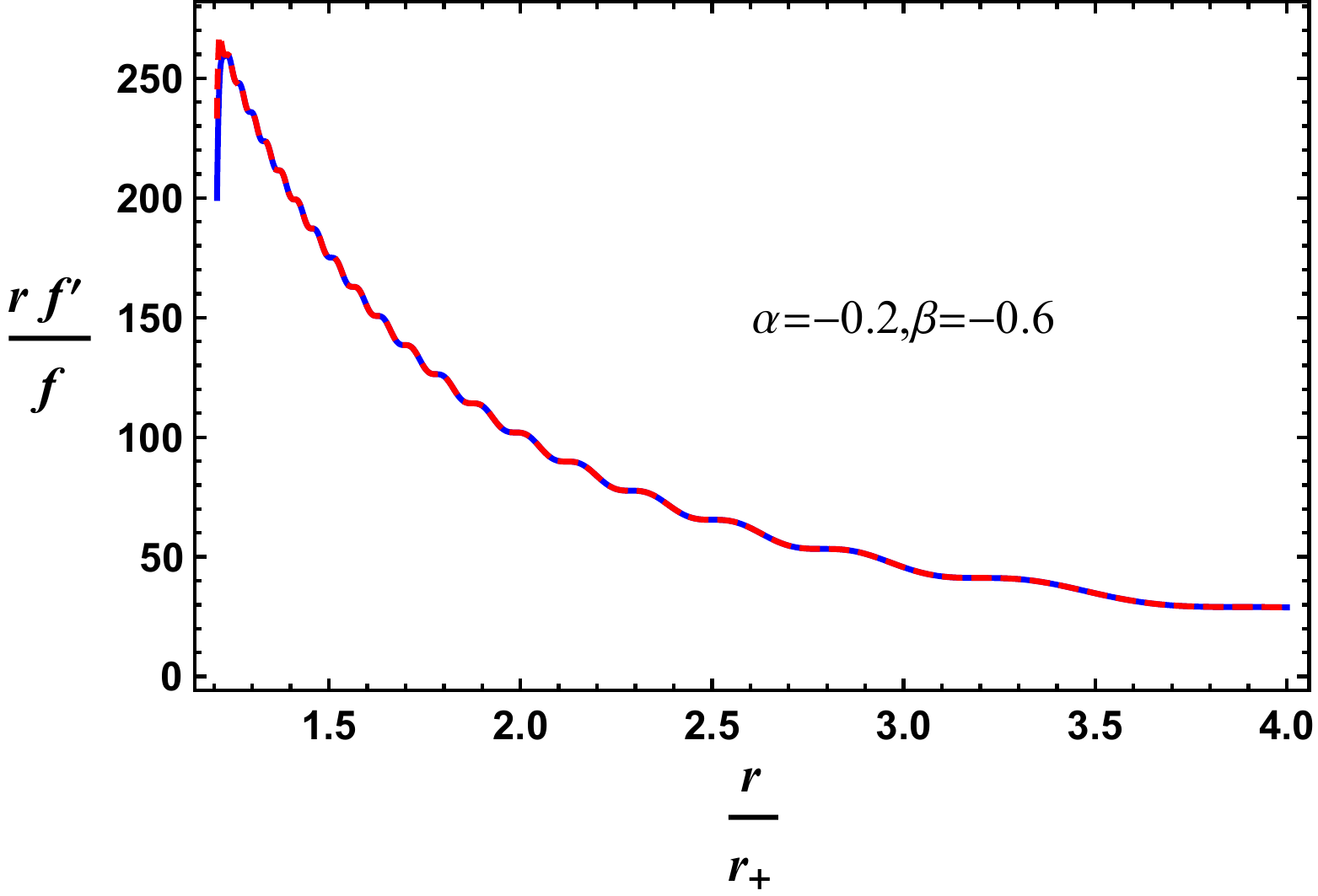}
 \includegraphics[width=0.45\linewidth]{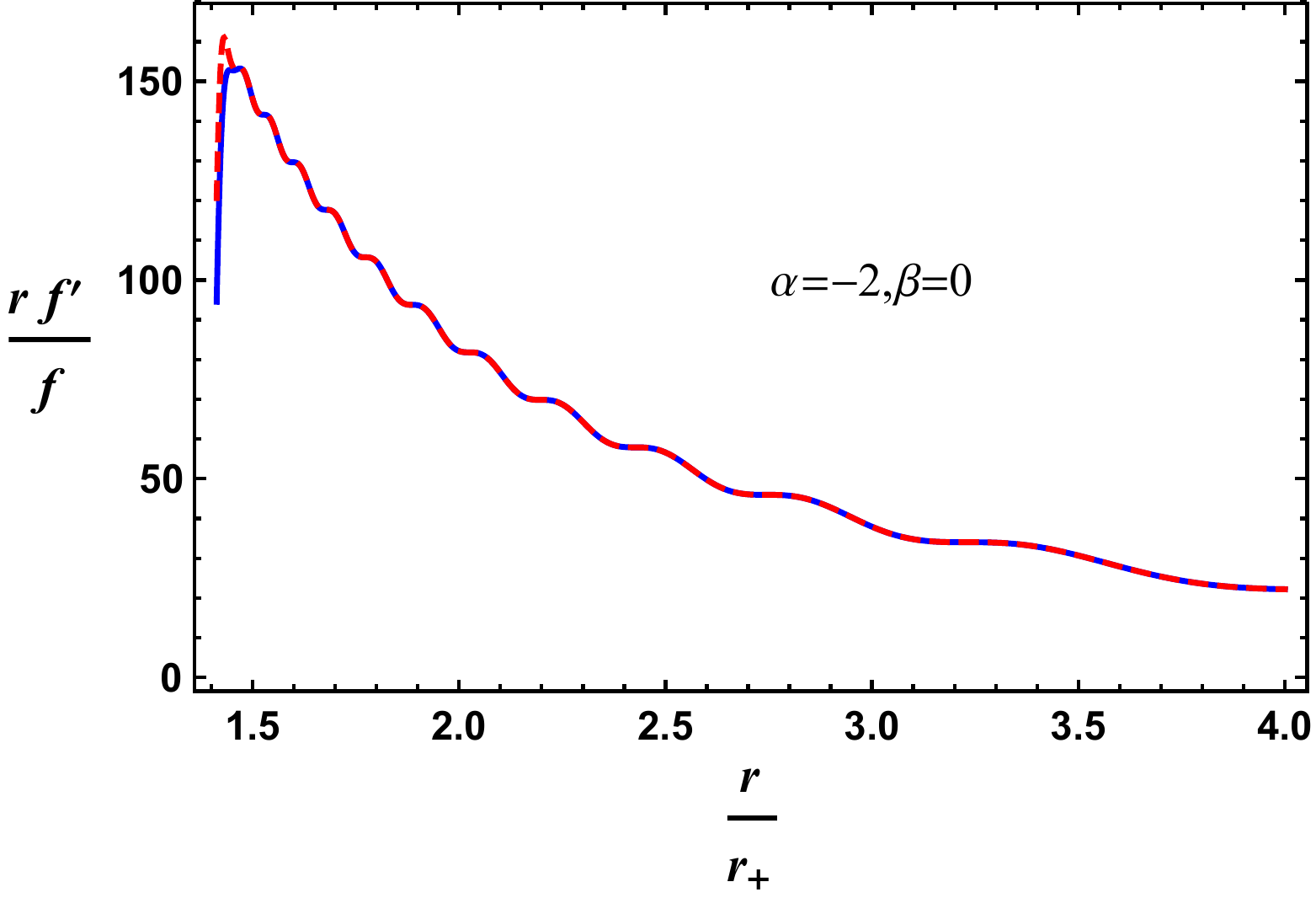}
 \begin{center}
 \includegraphics[width=0.45\linewidth]{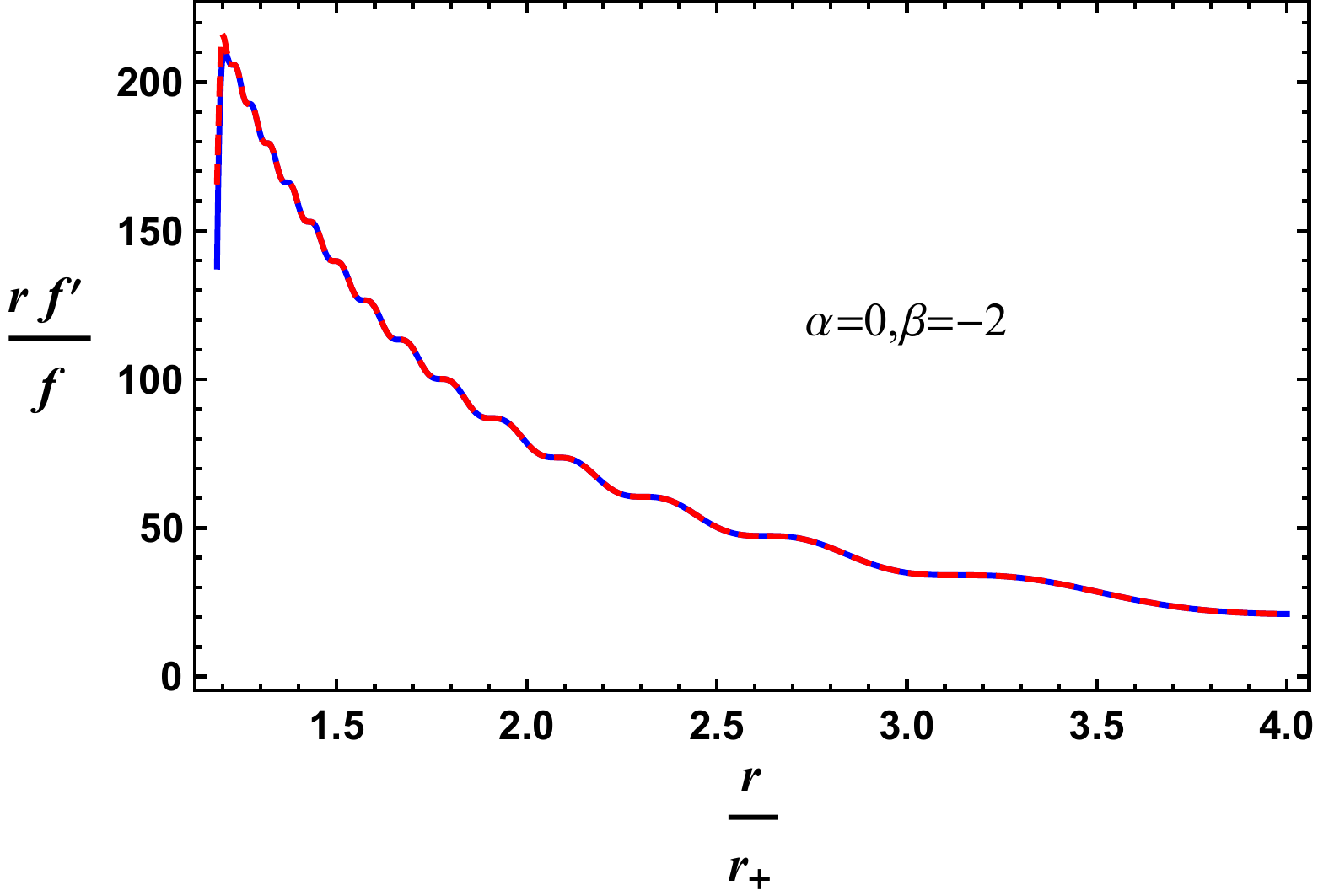}
 \end{center}
 \caption{ A comparison of the numerical solution (solid blue curves) of $r f'/f$ and fits to the analytic relation (\ref{Joseph:chif})(red dashed curves) describing Josephson oscillations. Here we have chosen  $T/T_{c}=0.995$ and
$\kappa/\mu=1$.
 \label{Fig10}}
\end{figure}
Before leaving this section, let us examine the influence of the massive parameter $\kappa/\mu$ on the Josephson oscillation amplitude. In Fig.\,\ref{Fig10}, we plot numerically the evolution of $\phi$ and $r f'/f$ with respect to $r$ for different configurations of the massive gravity and $T/T_{c}=0.995$. As can be seen, varying $\kappa/\mu$ can significantly divert the system behavior from the standard holographic superconductor case \citep{Hartnoll:2020fhc}. In fact, one can find that the Josephson oscillations can be even disappeared by the growth of the massive parameter $\kappa/\mu$ (see section \eqref{largelimi}).

\begin{figure}
\includegraphics[width=0.42\linewidth]{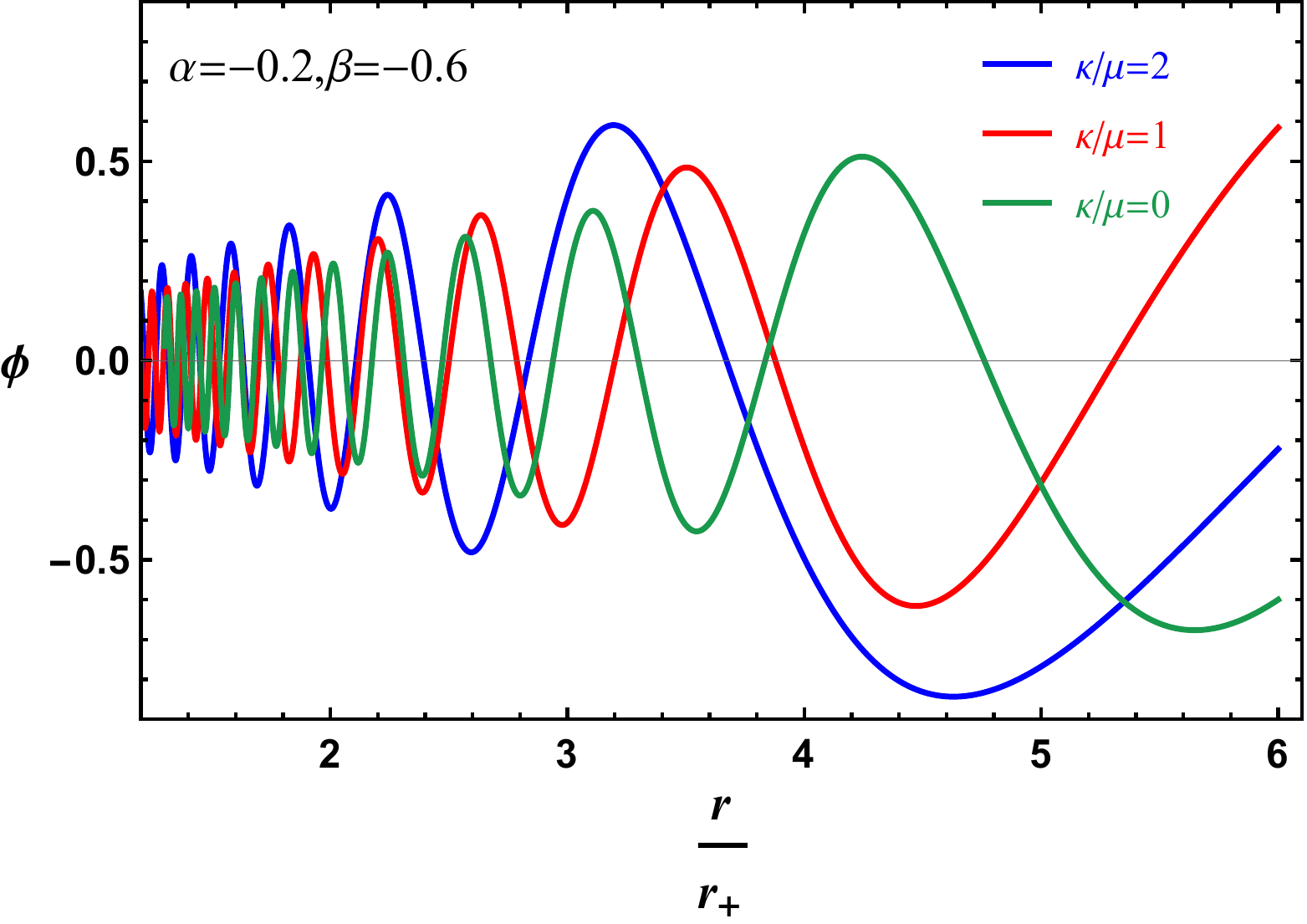}\hspace{.1cm}
 \includegraphics[width=0.45\linewidth]{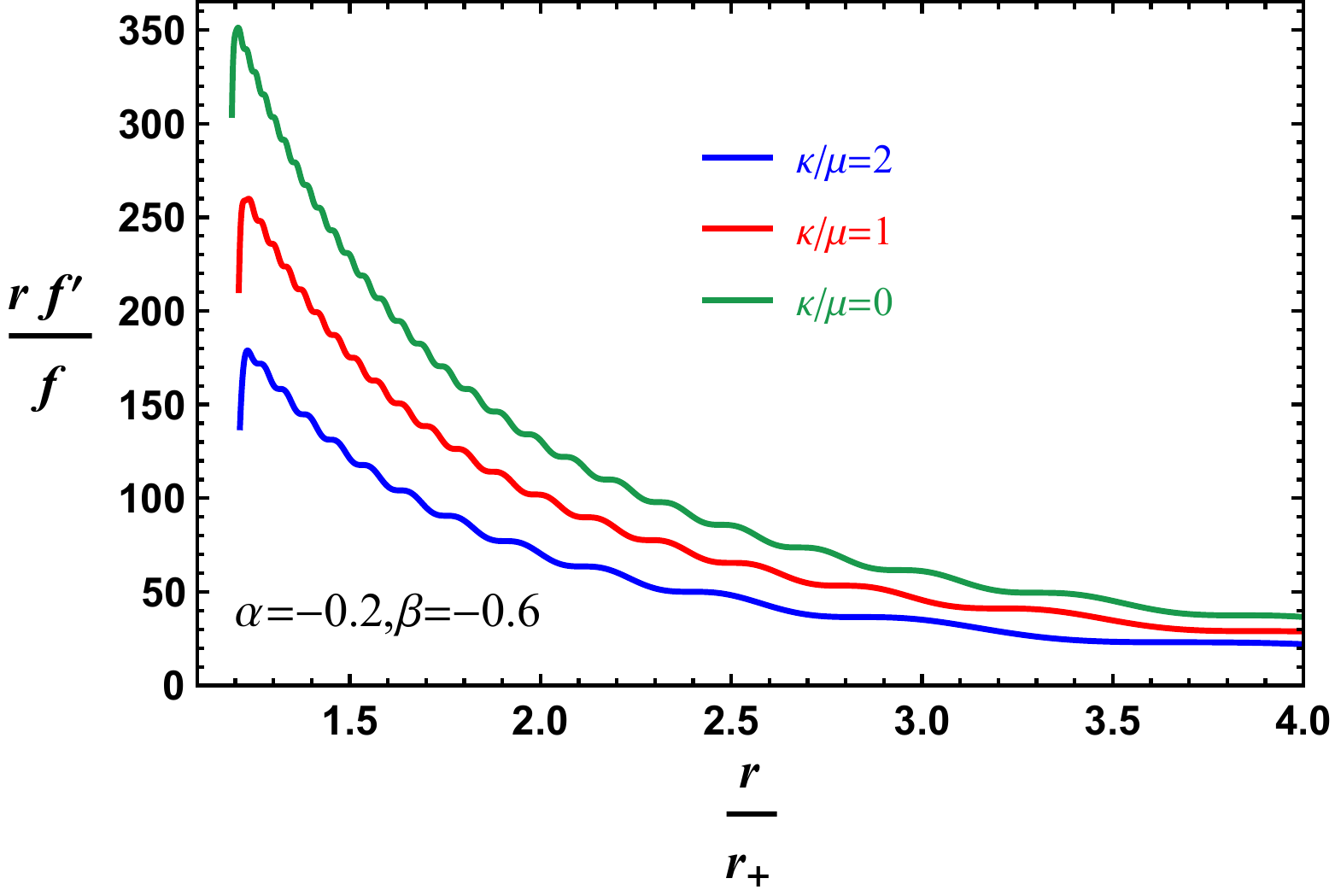}\\
 \includegraphics[width=0.42\linewidth]{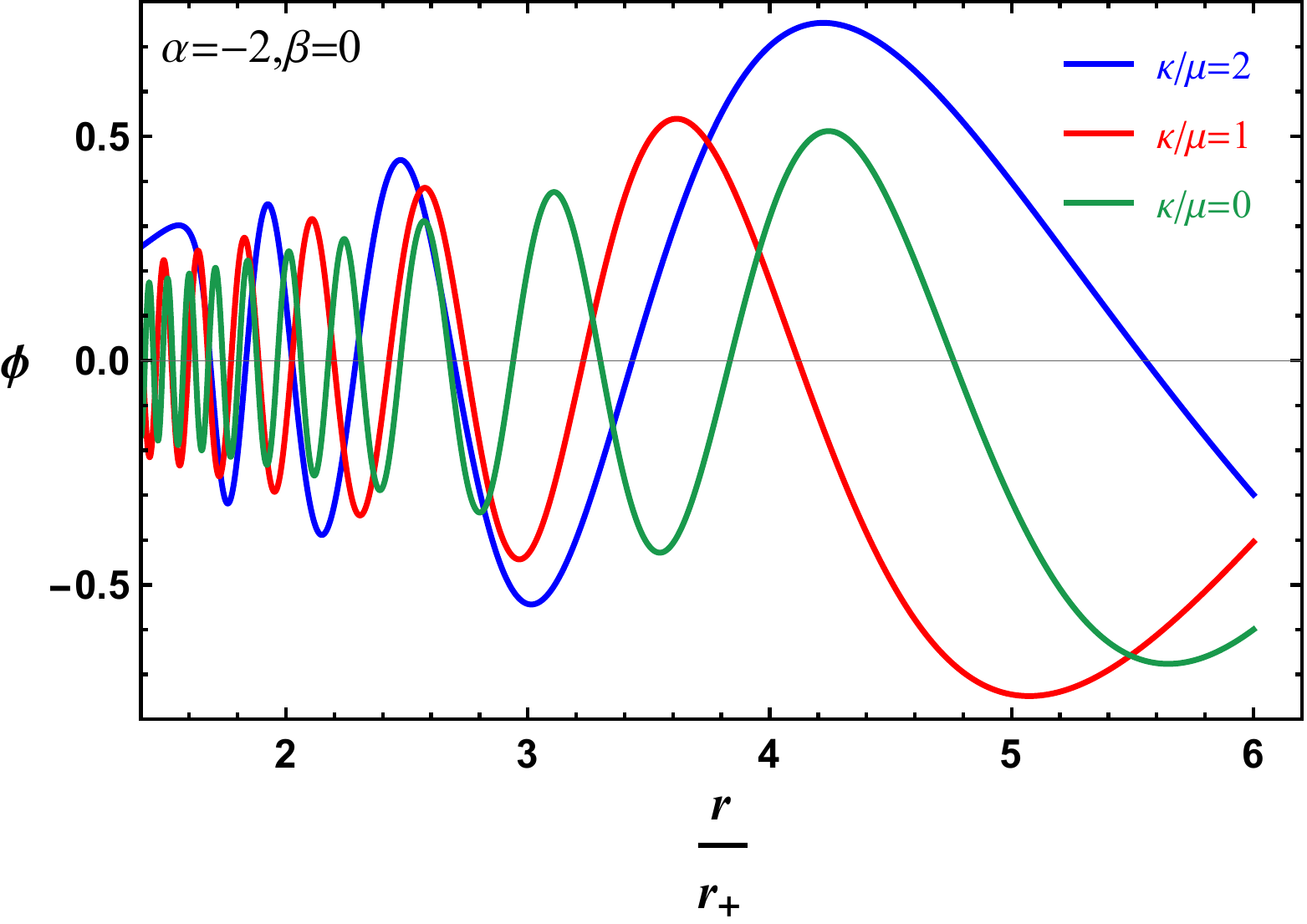}\hspace{.1cm}
 \includegraphics[width=0.45\linewidth]{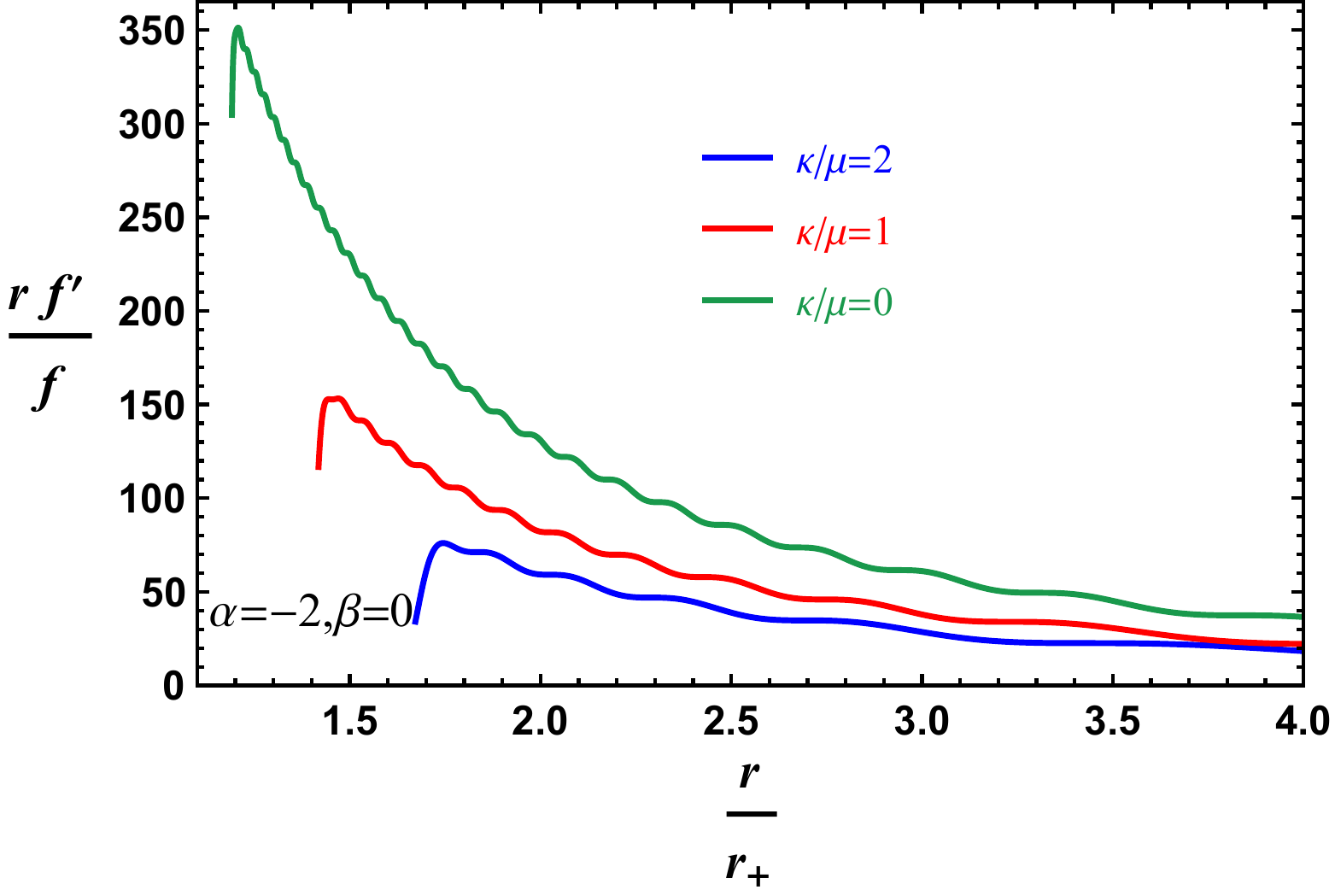}\\
 \includegraphics[width=0.42\linewidth]{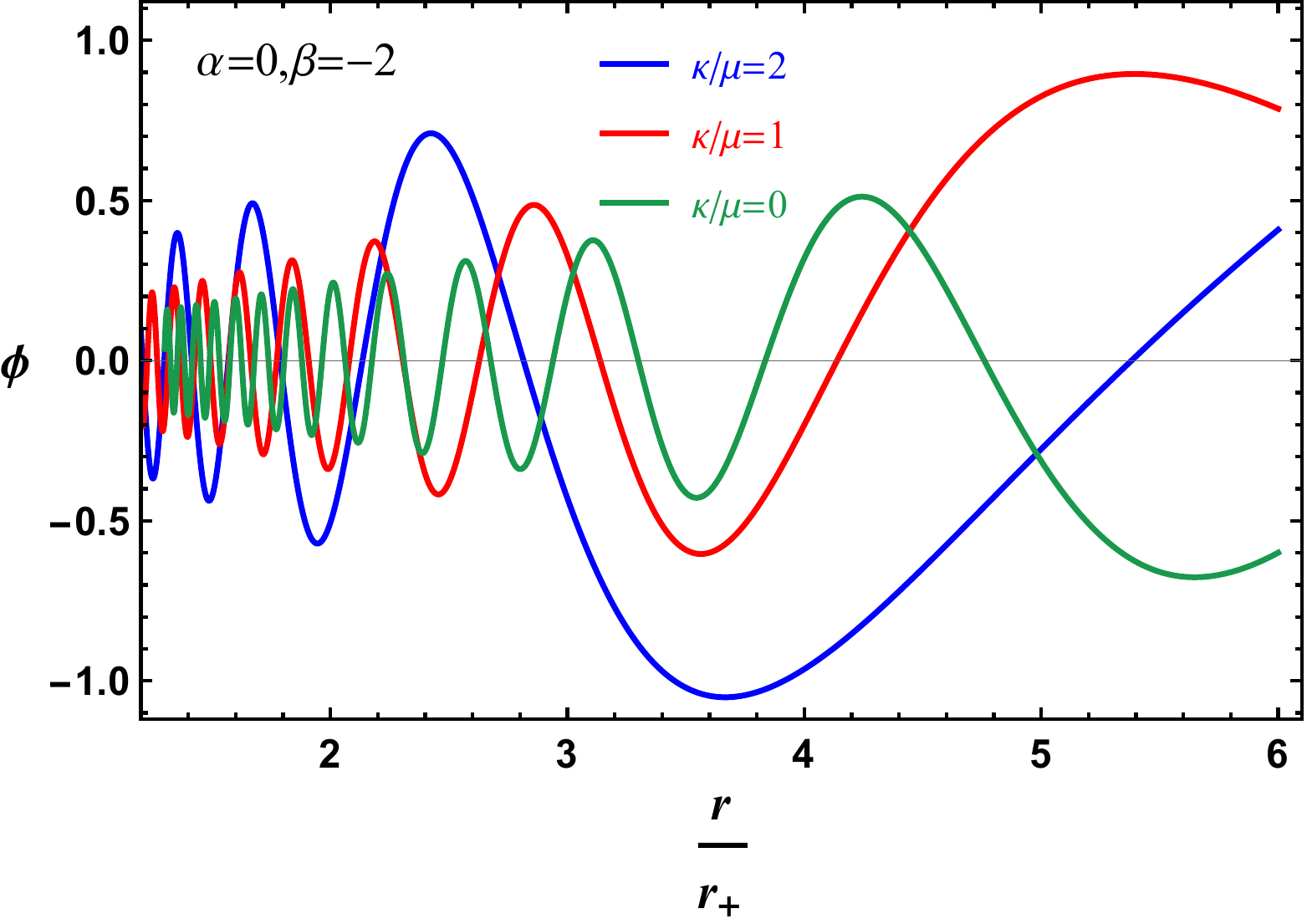}\hspace{.1cm}
 \includegraphics[width=0.45\linewidth]{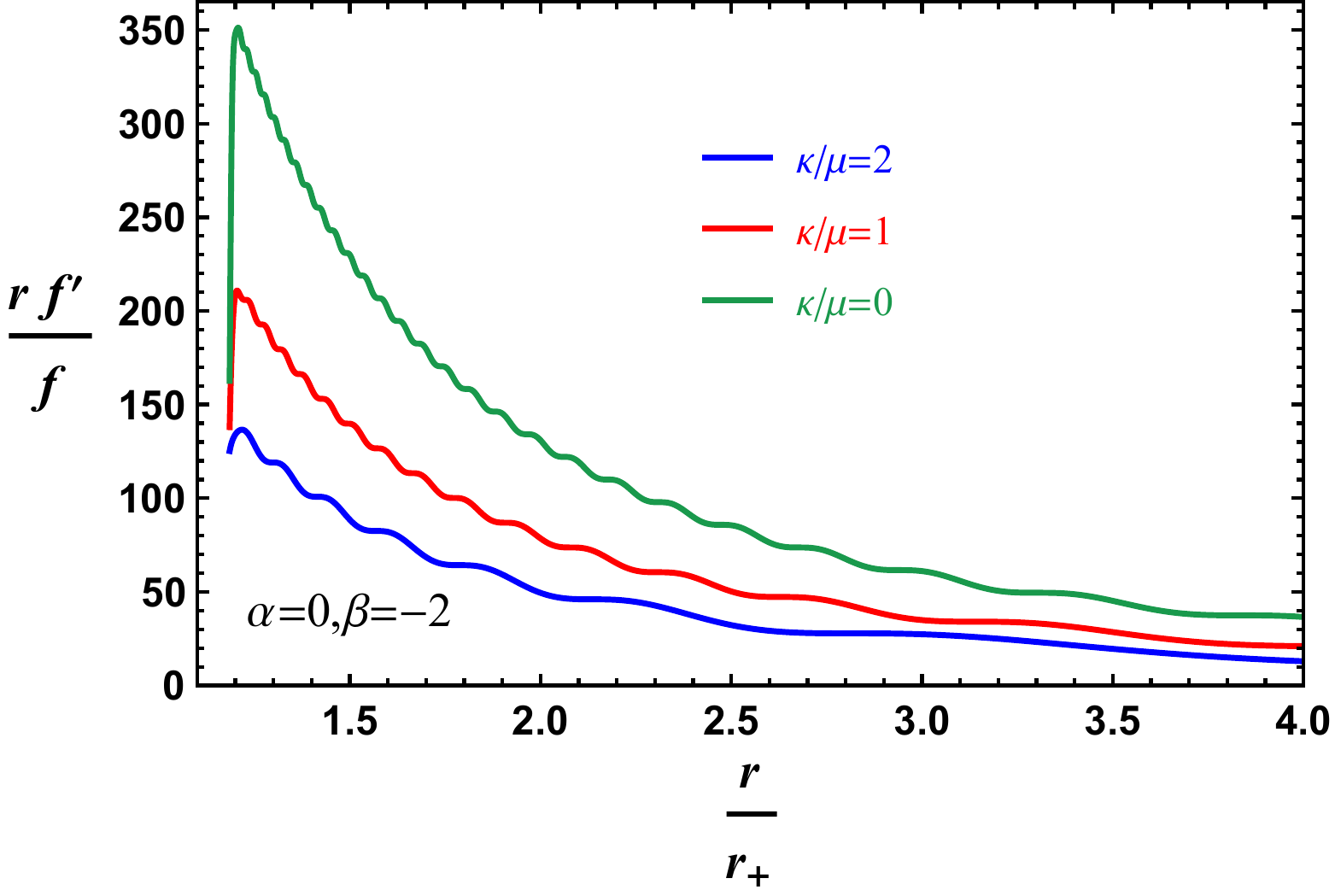}
 \caption{ The dynamical evolution of $\phi$ and $r f'/f$ as one changes the amount of the massive parameter $\kappa/\mu$. These graphs are at $T/T_{c}=0.995$.
 \label{Fig10}}
\end{figure}

\begin{figure}
\begin{center}
  \includegraphics[width=0.45\linewidth]{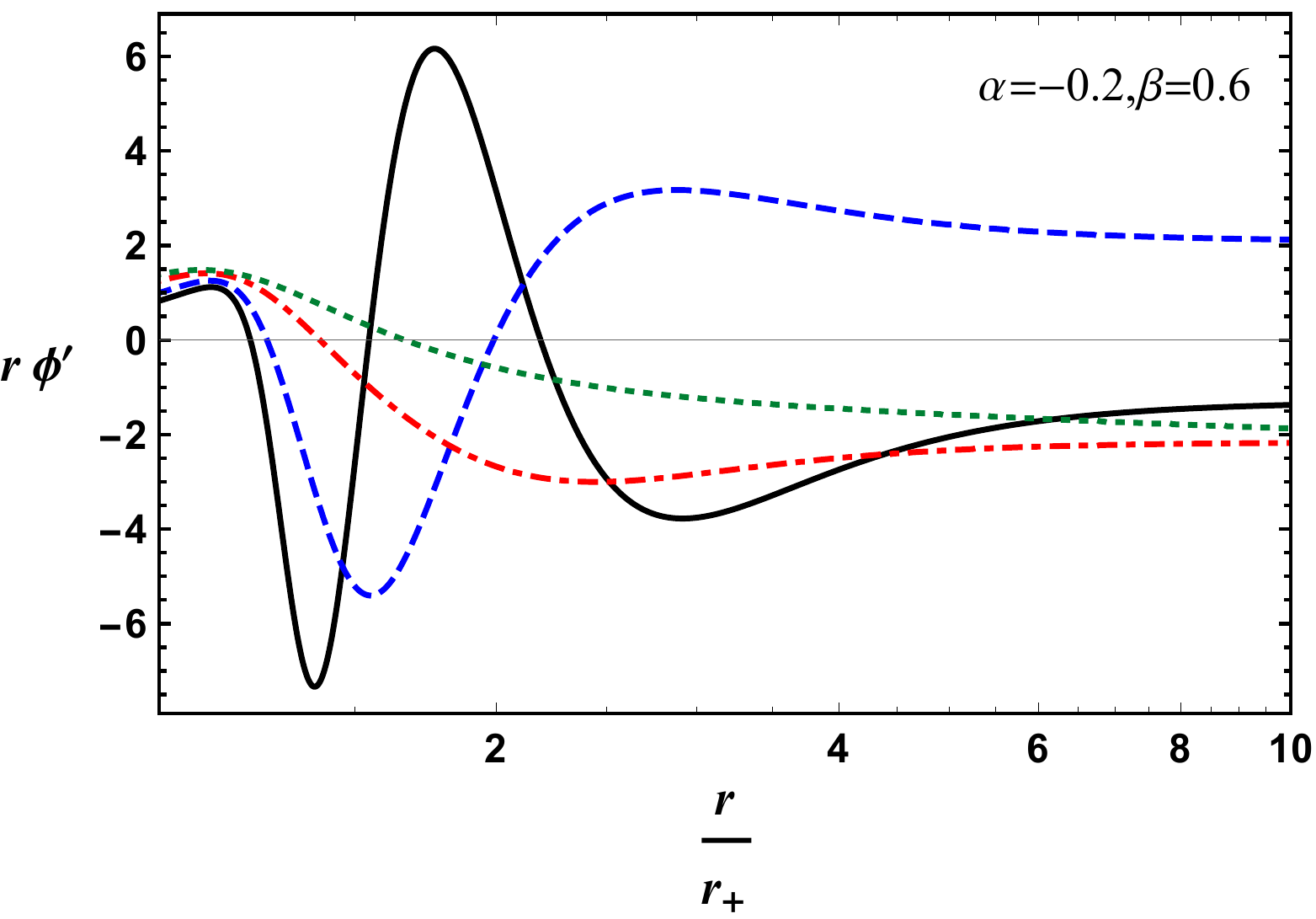}\hspace{.1cm}
 \includegraphics[width=0.45\linewidth]{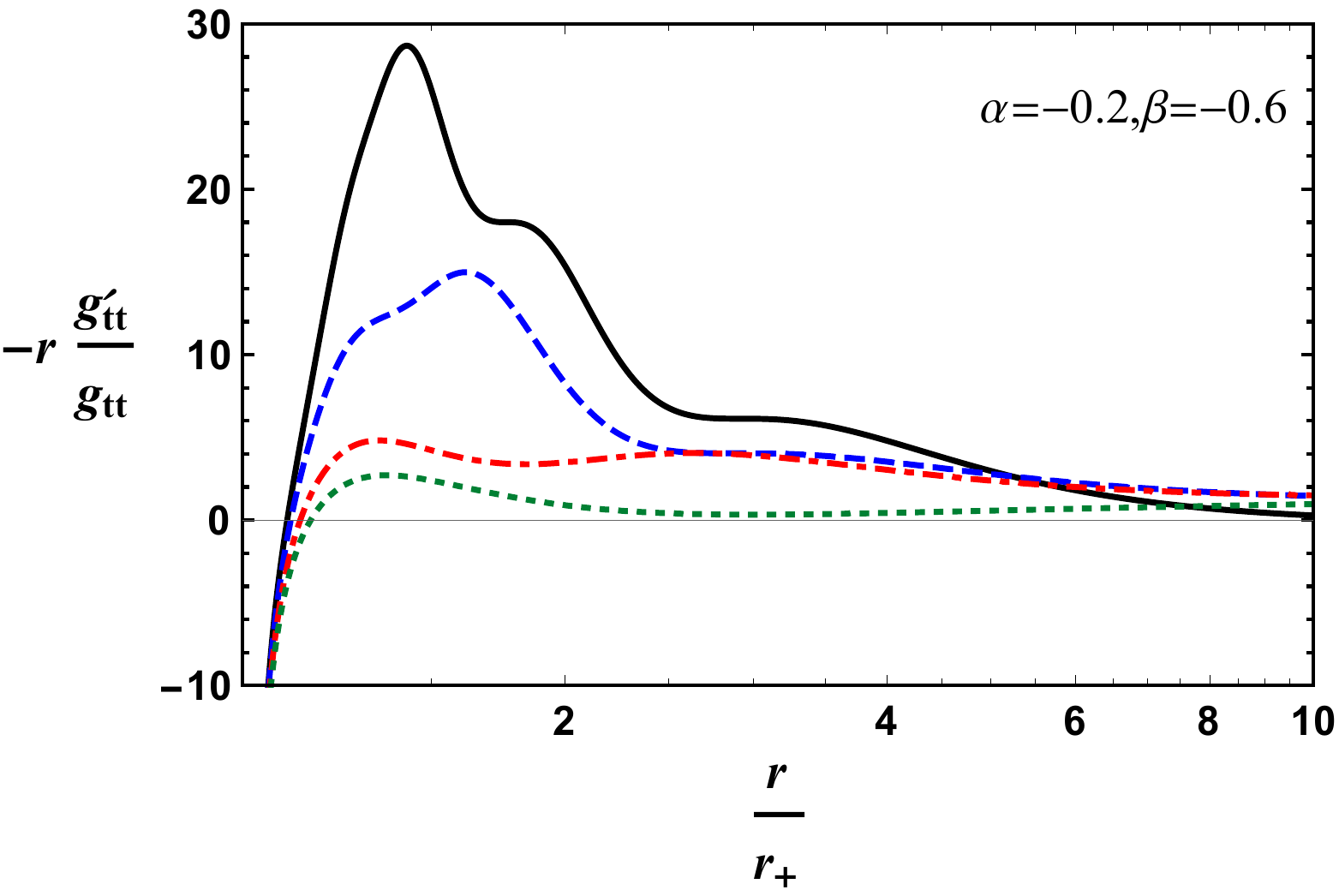}\\
 \includegraphics[width=0.45\linewidth]{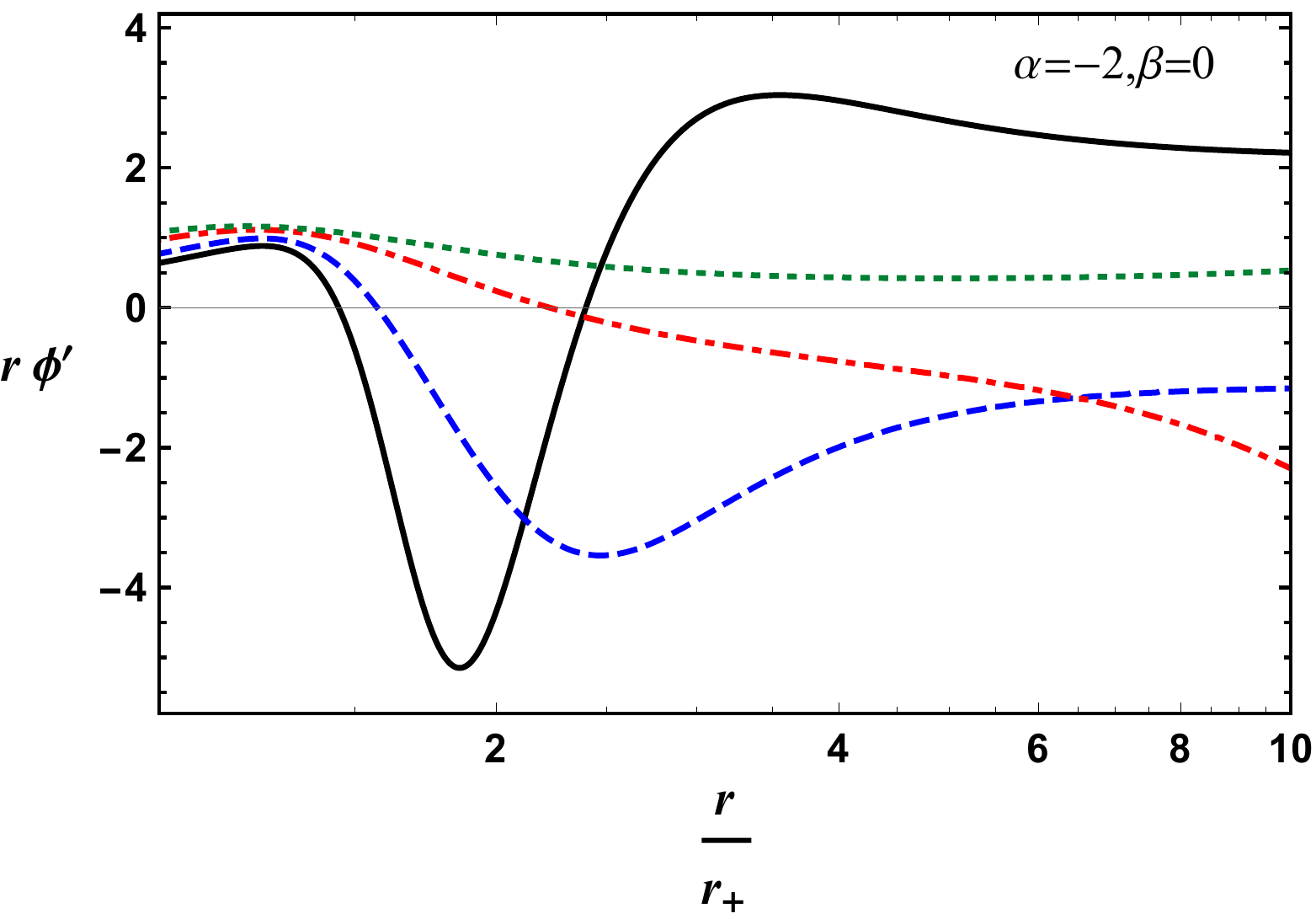}\hspace{.1cm}
 \includegraphics[width=0.45\linewidth]{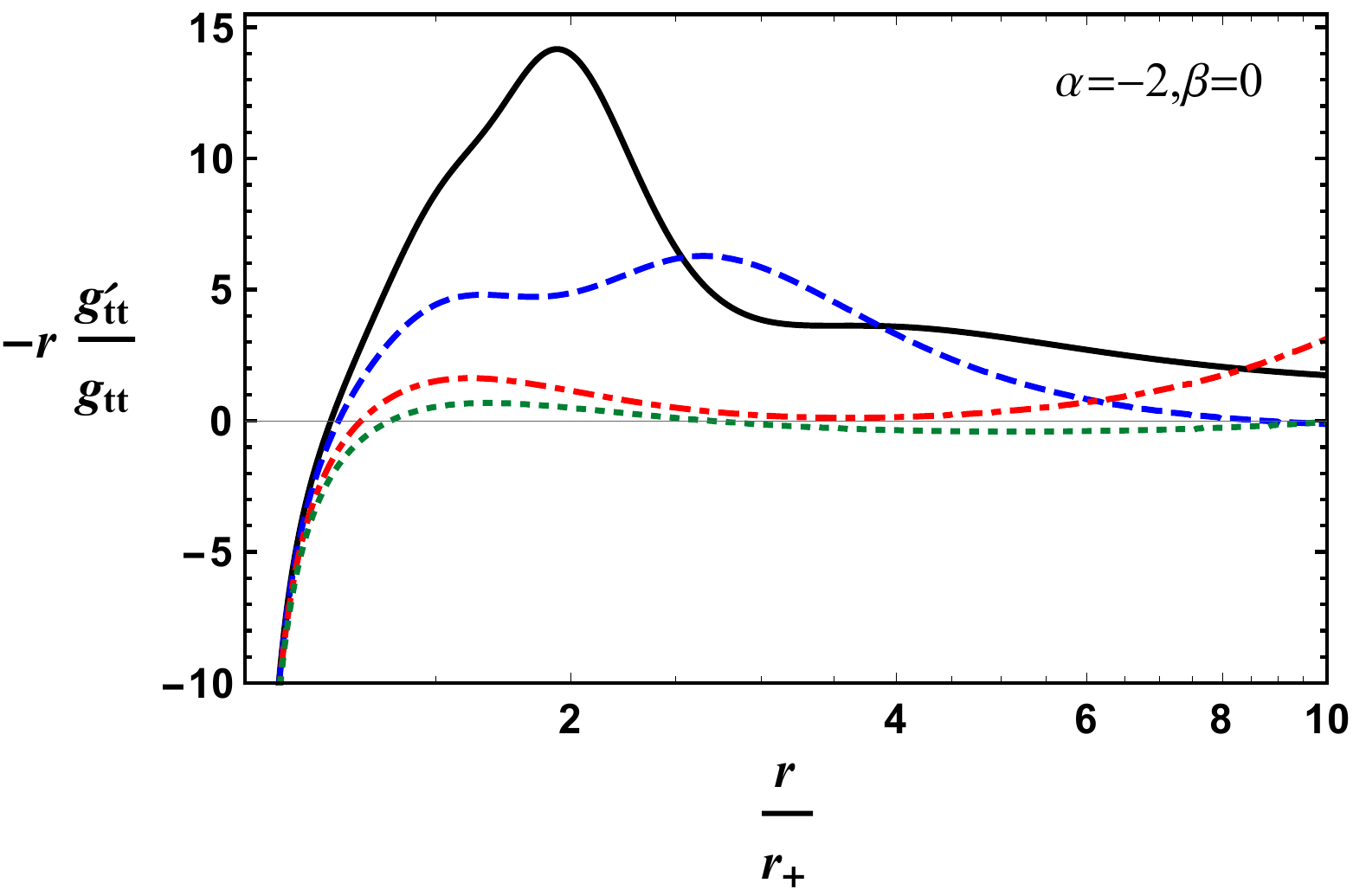}\\
 \includegraphics[width=0.45\linewidth]{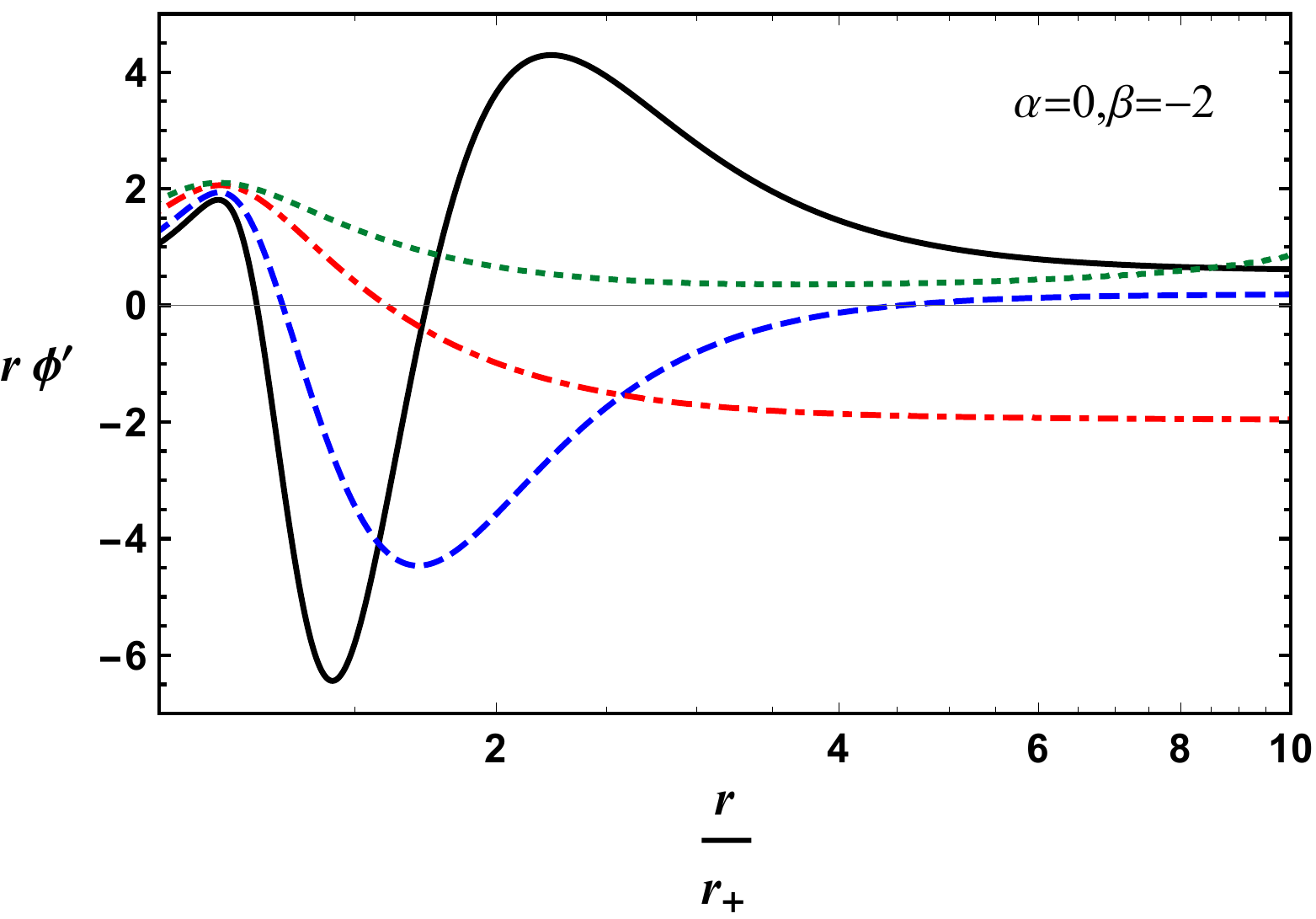}\hspace{.1cm}
 \includegraphics[width=0.45\linewidth]{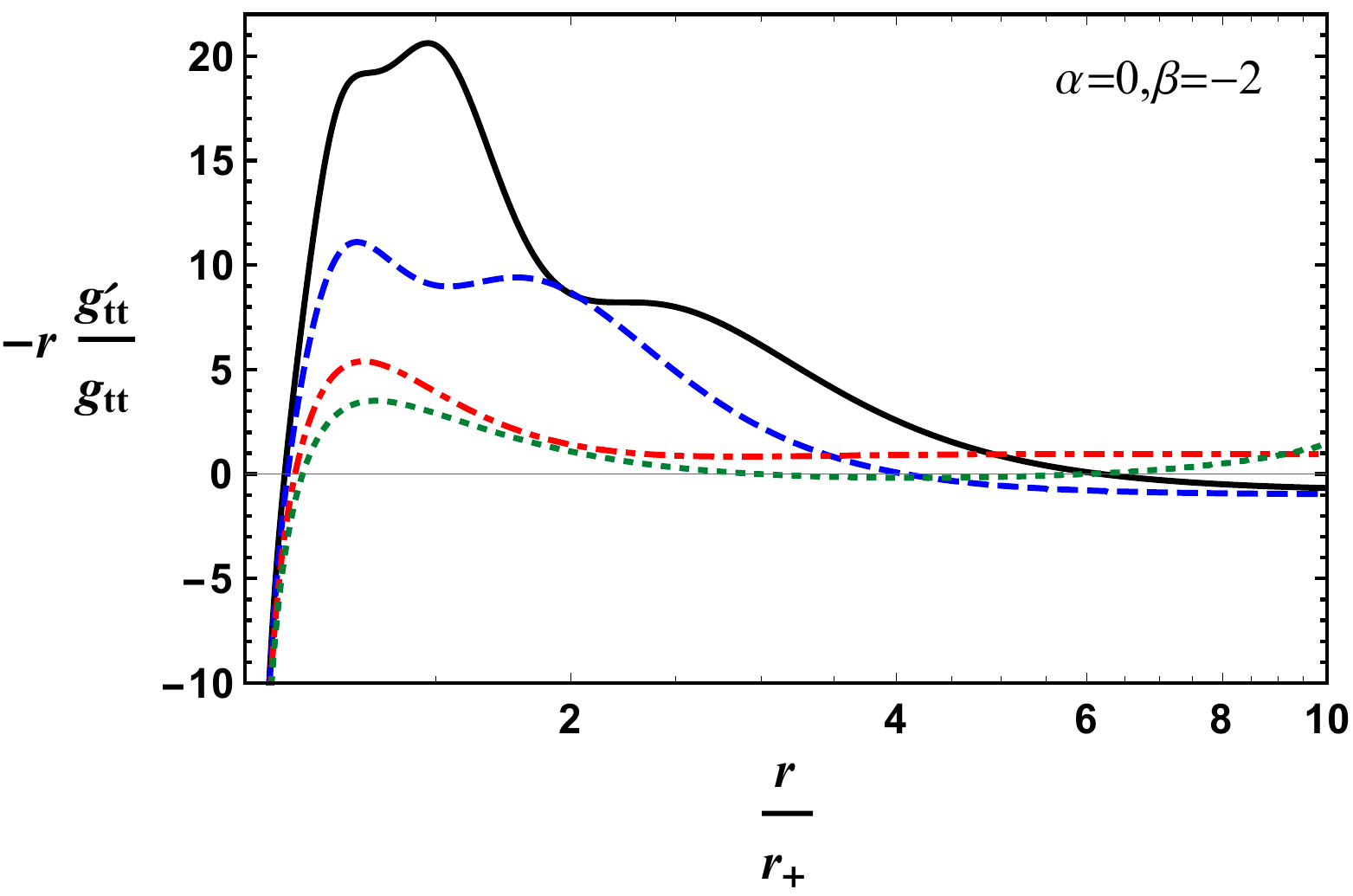}
\end{center}
 \caption{ The behavior of $r \phi'$ and $-r g'_{tt}/g_{tt}$ versus $r$ for the temperatures $T/T_{c} = 0.965$ (solid), $0.945$ (dashed), $0.895$ (dot-dashed) and $0.845$ (dotted). One can see fewer Josephson oscillations and the jump in the derivative of the metric element as the temperature is lowered. Here we take $\kappa/\mu=1$.
 \label{Fig11}}
\end{figure}

Summing up the results, it can be concluded that the growth of the massive parameter $\kappa/\mu$ assists to decrease the number of the Josephson oscillations epoch. Note that at large $\kappa/\mu$ limit, both ER collapse and Josephson oscillations epochs can be vanished (see Section \eqref{largelimi}). Interestingly, on the other hands, similar to the standard holographic superconductor case \cite{Hartnoll:2020fhc}, at temperatures which are far from $T_{c}$, the collapse of the Einstein-Rosen bridge and subsequent Josephson oscillations become less dramatic. For an example in Fig.\,\ref{Fig11}, the number of Josephson oscillations tends to decrease at smaller temperatures.  

\section{Kasner cosmology epoch}\label{Sec5}
At the end of the Josephson oscillations epoch, the scalar field $\phi$ grows logarithmically  (\ref{eq:tokasner}). It means that the system moves to the Kasner epoch. In this section, we are interested in finding what happens as the system evolves further inside the Kasner cosmology epoch. Nevertheless, it might be possible that
after the Josephson oscillations the solution enters a new Kasner epoch as
``\textit{Kasner inversion}'' observed in \cite{Hartnoll:2020fhc}.

Numerically, it is found that once the solution enters the Kasner epoch, the massive configuration coefficients  $\alpha$ and $\beta$, mass and charge terms in the equations of motion can be dropped. In this respect, the solutions at large $r$ take the following form:
\begin{equation}\label{kasner1}
\phi=c \sqrt{2}\log r+\dots, \hspace{0.5cm} \chi= 2c^2 \log r+\dots, \hspace{0.5cm} f=-f_{1} r^{3+2c^2}+\dots,
\end{equation}
where $f_{1}$ is constant and $c=-\sqrt{8} d_{3}/\pi$ according to \eqref{eq:tokasner}. In addition, using Eq.\,\eqref{eq1}, the Maxwell potential is

\begin{equation}\label{kasnerin}
\Phi=\Phi_{K}+E_{K} \int dr e^{\frac{-\chi}{2}}\approx \Phi_{K}+ E_{K} r^{1-c^2}+\dots .
\end{equation}
It is obvious that for $|c|>1$, the Maxwell field remains unimportant at large $r$, similar to that seen previously for a neutral scalar hair \cite{Hartnoll:2020rwq}. In this limit, the geometry near the singularity takes a Kasner form
\begin{equation}\label{kasnerm}
ds^{2} \sim -d \tau^2 + \tau^{2 p_{t}} dt^2+ \tau^{2 p_{x}} \Big(dx^2+dy^2\Big)\,, \hspace{0.5cm} \phi\sim - p_{\phi} \log \tau\,,
\end{equation}
where we have changed the $r$ coordinate to the proper time $\tau$ via $d\tau=\frac{dr}{r \sqrt{f}}$, and the Kasner exponents are obtained to be
\begin{equation}\label{exponent1}
p_{x}=\frac{2}{3+c^2}, \hspace{0.5cm} p_{t}=\frac{c^2-1}{3+c^2}, \hspace{0.5cm}p_{\phi}= \frac{2 \sqrt{2} c}{3+c^2}\,,
\end{equation} 
which obey $p_{t}+2p_{x}=1$ and $p_{\phi}^2+p_{t}^2+2 p_{x}^2=1$. 

\begin{figure}
 \includegraphics[width=1\linewidth]{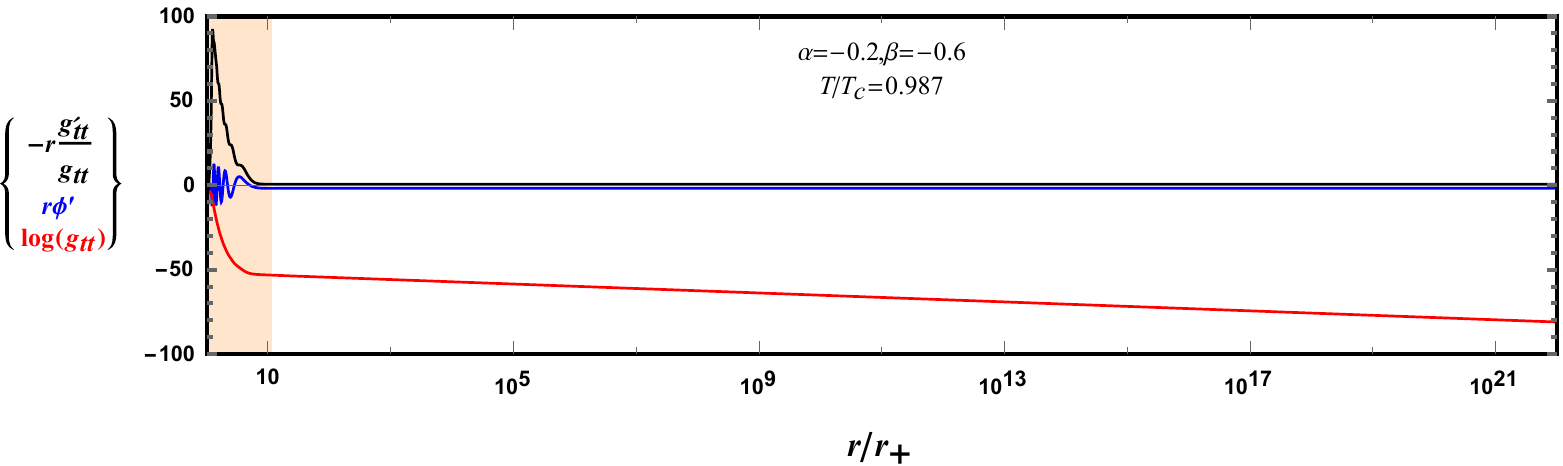}
 \includegraphics[width=1\linewidth]{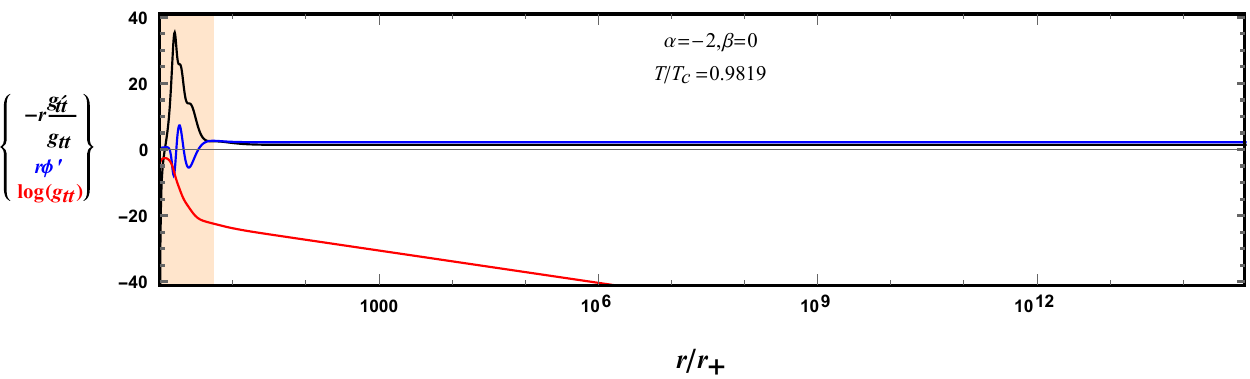}
 \includegraphics[width=1\linewidth]{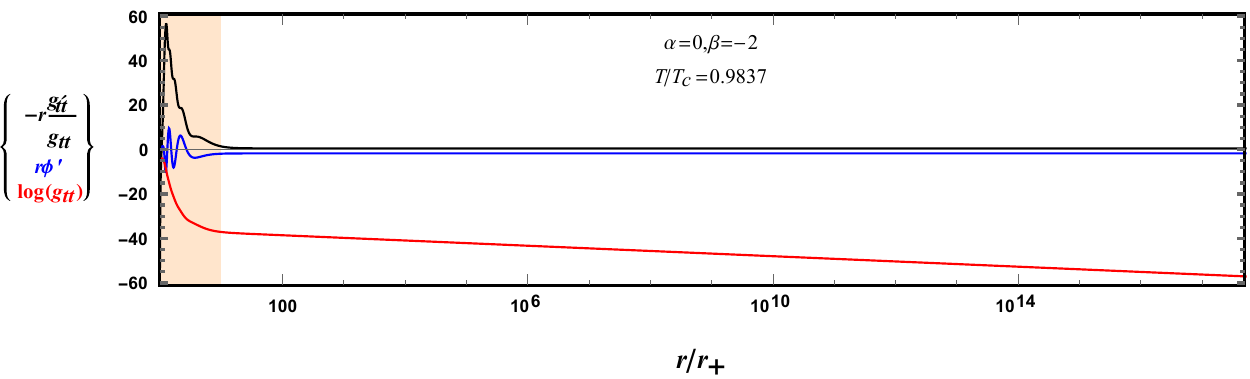}
 \caption{ Typical examples of  the interior evolution with no Kasner inversion in different configurations of the massive gravity. In $1<r<10$ (orange shaded area) the system experiences the ER bridge collapse and Josephson oscillations epoch, then it stays in Kasner regime for ever. Clearly, in all panels, the function $\log (g_{tt} )$ begins to collapse, while the other functions $r \phi'$ and $-rg'_{rr}/g_{tt}$ become constants in the Kasner epoch. In all these graphs  
$\kappa/\mu=1$.
 \label{FignoKas}}
\end{figure}

In Fig.\,\ref{FignoKas}, an interior evolution is illustrated that does not deviate from the Kasner cosmology \eqref{kasner1}, because the intermediate Kasner exponent is already positive, i.e. $p_{t}>0$. In other words, we observe that when $|c|>1$ or ($p_{t}>0$) at the beginning of the Kasner epoch, the system remains described by the Kasner geometry \eqref{kasner1} with the exponents \eqref{exponent1} during  a journey to the actual singularity as $r \to \infty$.

\begin{figure}
\includegraphics[width=1\linewidth]{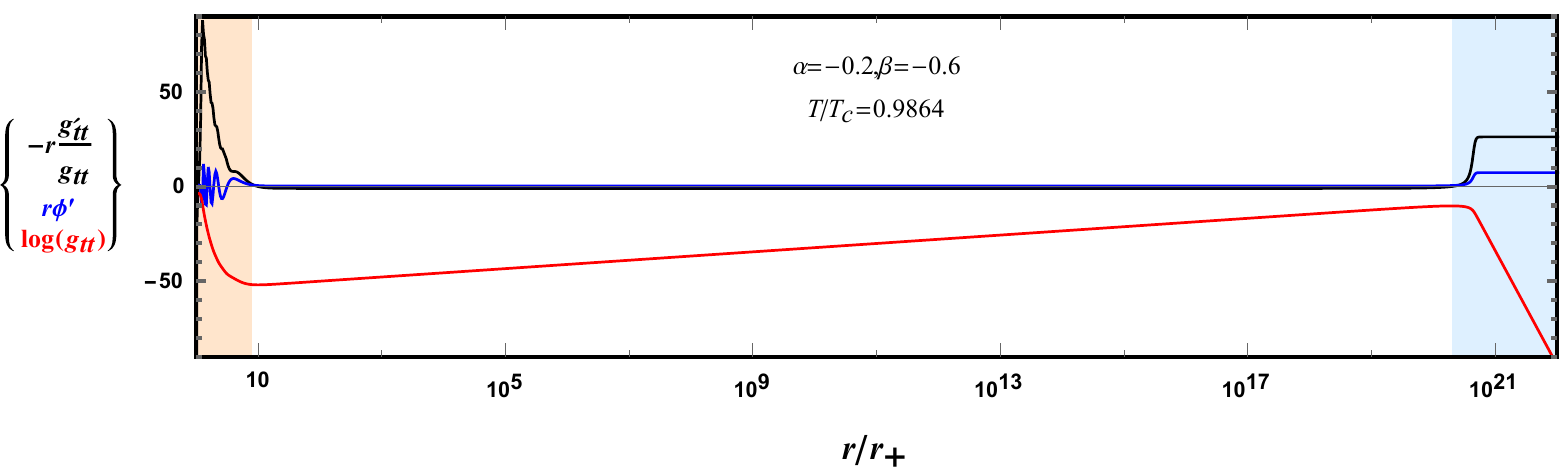}
 \includegraphics[width=1\linewidth]{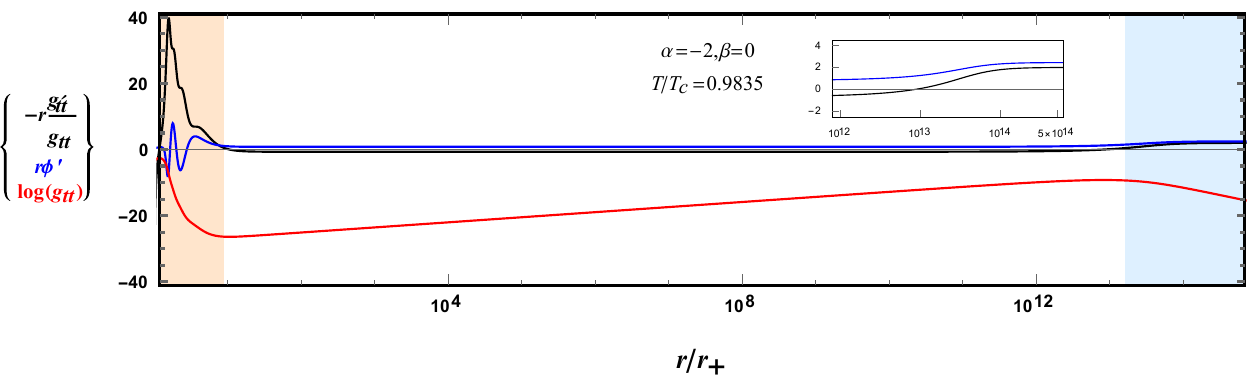}
 \includegraphics[width=1\linewidth]{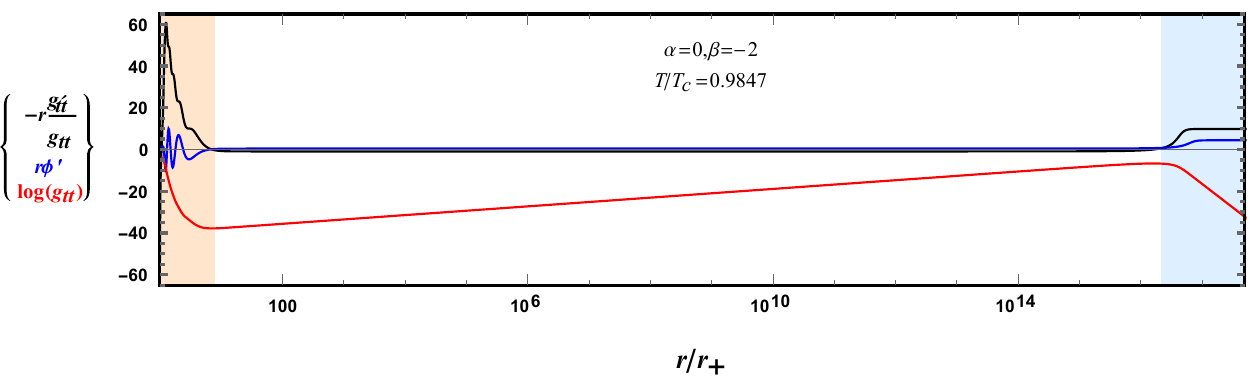}
 \caption{ An example of an interior evolution with Kasner inversion in different configurations of the massive gravity. The system evolves from the outer event horizon at $r_{+}=1$ towards the final singularity at $r \to \infty$. In the range $1<r<10$ (orange shaded regime) the system goes quickly through the ER bridge collapse and Josephson oscillations epoch to the intermediate Kasner regime (white area) and stays relatively steady for an exponentially long range of $r/r_{+}$ with $|c|>1$ (or $p_{t}^{int}<0$), and finally there is a Kasner inversion (blue shaded area) with $|c_{new}|>1$ (or $p_{t}=-p^{int}_{t}/(2 p^{int}_{t}+1)>0$). Here 
$\kappa/\mu=1$ in all panels.
 \label{Figinvert2}}
\end{figure}

\begin{figure}
  \includegraphics[width=0.45\linewidth]{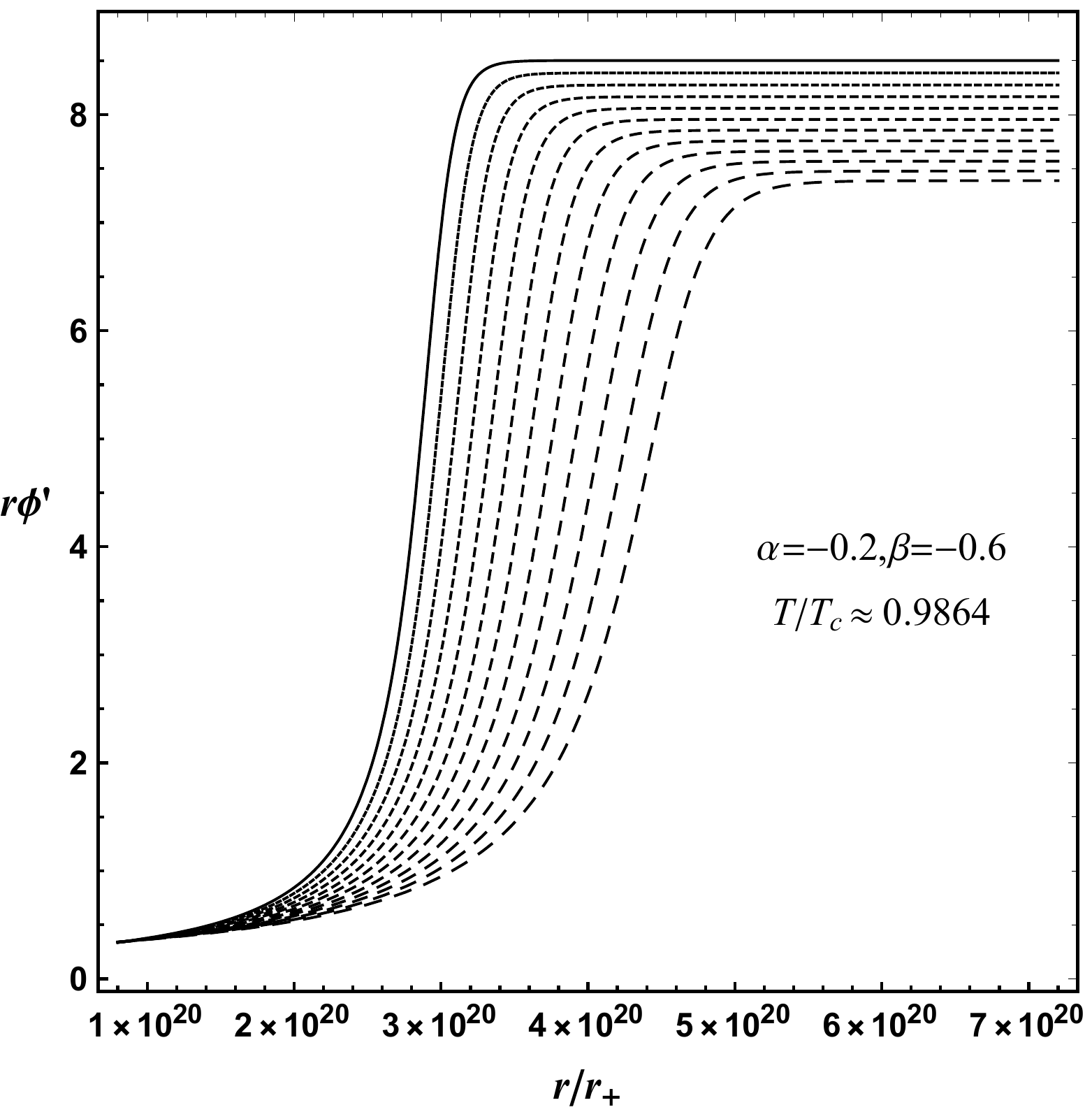}
 \includegraphics[width=0.45\linewidth]{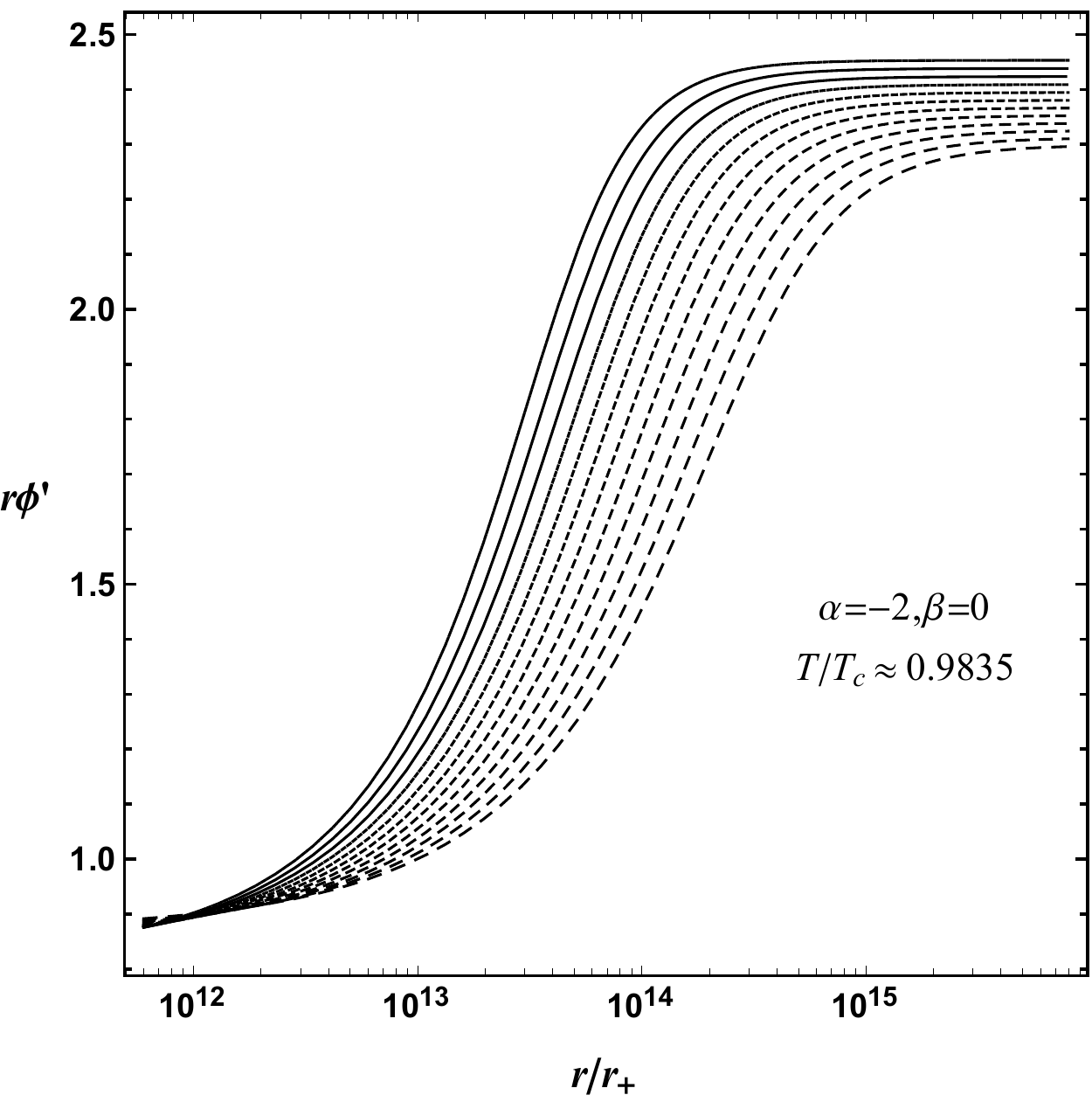} 
 \begin{center}
\includegraphics[width=0.45\linewidth]{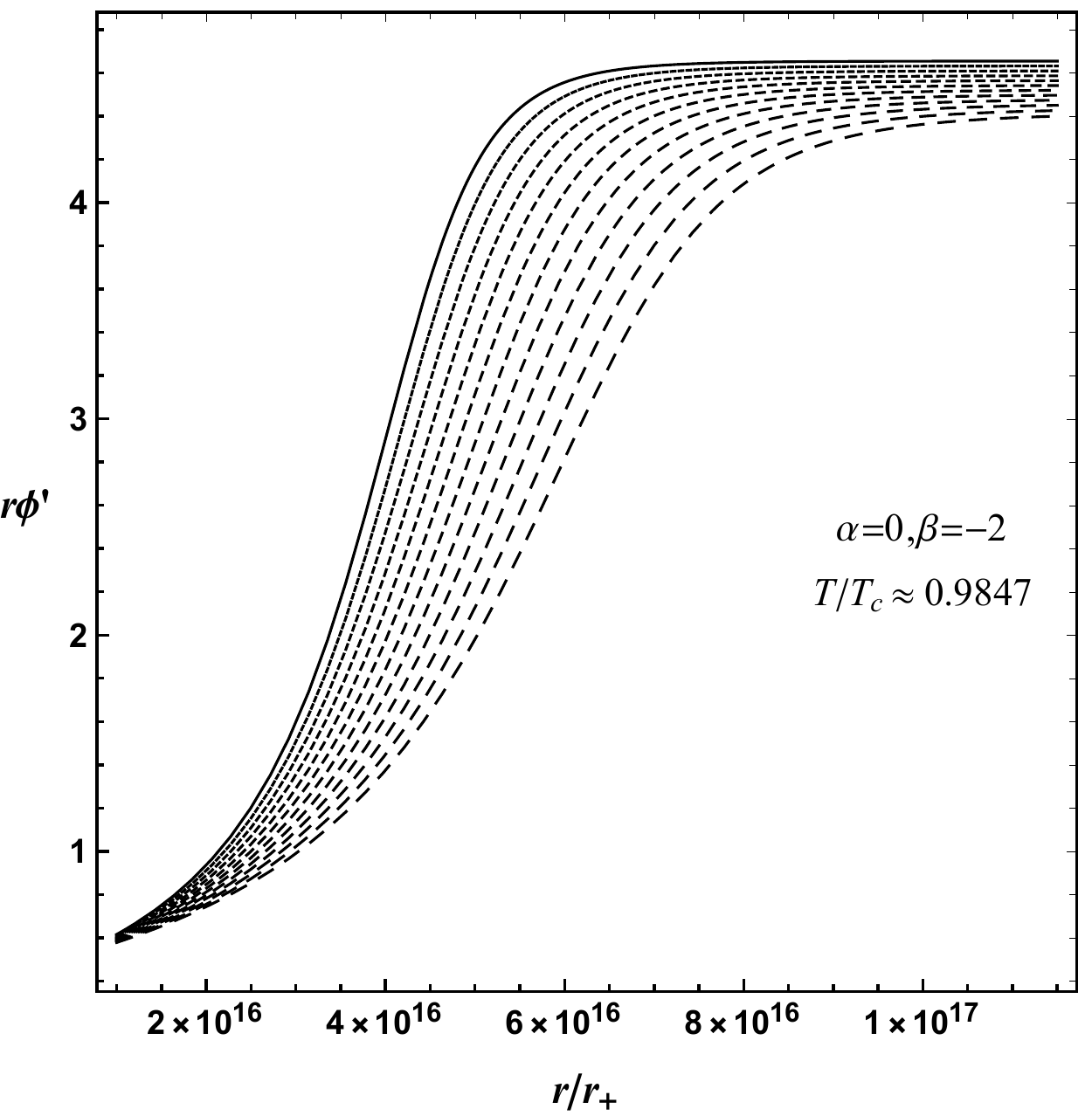}
 \end{center}
 \caption{ The value of $r \phi'$ near the kasner inversion from $c$ to $c_{new}=1/c$ at $r_{inv}$ which is the location of the transition. Form the left hand side and top to bottom, panels are drawn for the Kasner inversion in the range $c \approx 0.1664- 0.1914$, $c \approx 0.5792 - 0.6163$ and $c \approx 0.4336 - 0.4653$, respectively.  
 \label{Figinver1}}
\end{figure}

On the other hand, when $|c|<1$, the growing Maxwell field \eqref{kasnerin} results in a transition to a different Kasner solution at $r_{inv}$ with new exponents. This is called the Kasner inversion found previously in \cite{Hartnoll:2020fhc}. As proved analytically in \cite{Hartnoll:2020fhc}, near $r_{inv}$, the Maxwell field growth (\ref{kasnerin}) triggers a transition to a new Kasner regime
with exponent $c \to c_{new}=1/c$, or equivalently 
$p_{t} \to p_{t}^{new}=-p_{t}/(2p_{t}+1)$. It should be noted that the effect of the massive gravity contribution is negligible in the Kasner epoch (we refer to this point at the end of this section), and the above results are still satisfied numerically for our model. For the resulting plots, see Figs.\,\ref{Figinvert2} and \ref{Figinver1}. 

As follows from all panels shown in Fig.\,\ref{Figinver1}, at a value of $r$ near
 the inner horizon of the non-hair solution of the massive gravity \eqref{normalsol}, $r g'_{tt}/g_{tt}$ (black curve) experiences a large kick at which $g_{tt}$ (red curve)
becomes very small. This is the collapse of the Einstein-Rosen bridge. Subsequently, a series of Josephson oscillations in the scalar field $\phi$ and corresponding imprint in a series of short steps on the metric derivative $r g'_{tt}/g_{tt}$ (or equivalently $r f'/f$) are observed. These all occur at relatively small $r/r_{+}$ (orange
shaded area). Immediately
afterwards, the oscillations settle down to an intermediate Kasner
regime with the Kasner exponent $p_{int}$. For an exponentially long range of $r/r_{+}$, this Kasner epoch remains just before  another kick (whenever $p^{int}_{t}<0$) happens in $g_{tt}$. After that the 
system enters the final Kasner epoch with positive exponent $p_{t}=-p^{int}_{t}/(2 p^{int}_{t}+1)>0$ (blue shaded area). Because $g_{tt}$ is a power law function, $-r g'_{tt}/g_{tt}$ is constant in both sides of the Kasner inversion location.

To have concrete evidence of the Kasner inversion, we have also plotted the dynamical evolution of $r \phi'$ for a certain range of $c$ in Fig.\,\ref{Figinver1}. As an example, in the massive gravity configuration with $3\alpha= \beta=0.6$,  before inversion $c_{\text{before}}=0.1914$, after inversion $c_{\text{after}}=5.225$ which is in agreement with the inversion limit formula, with the error $|c_{\text{after}}- 1/c_{\text{before}}|=4.451 \times 10^{-9}$.

\begin{figure}
 \includegraphics[width=0.45\linewidth]{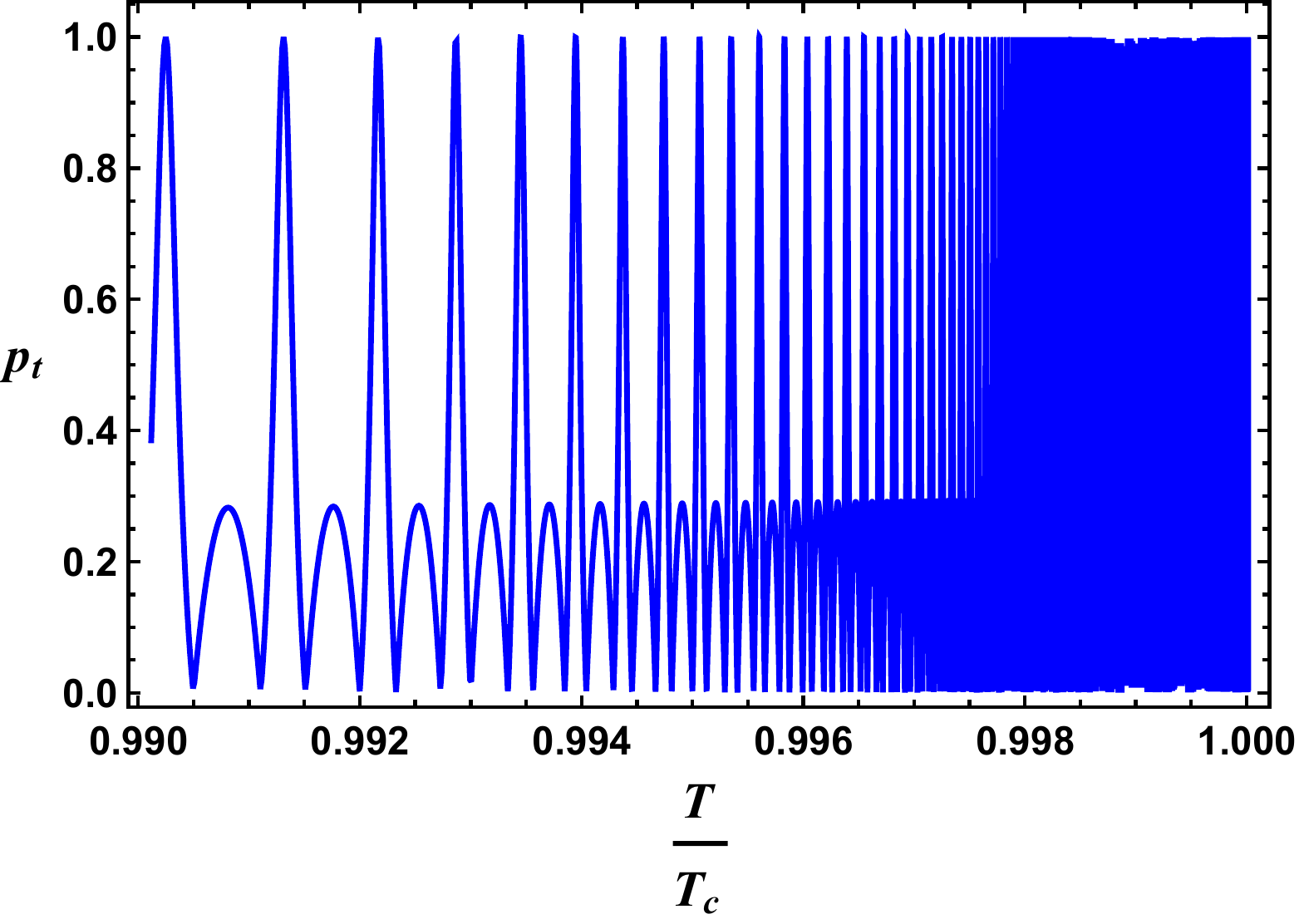}
 \includegraphics[width=0.45\linewidth]{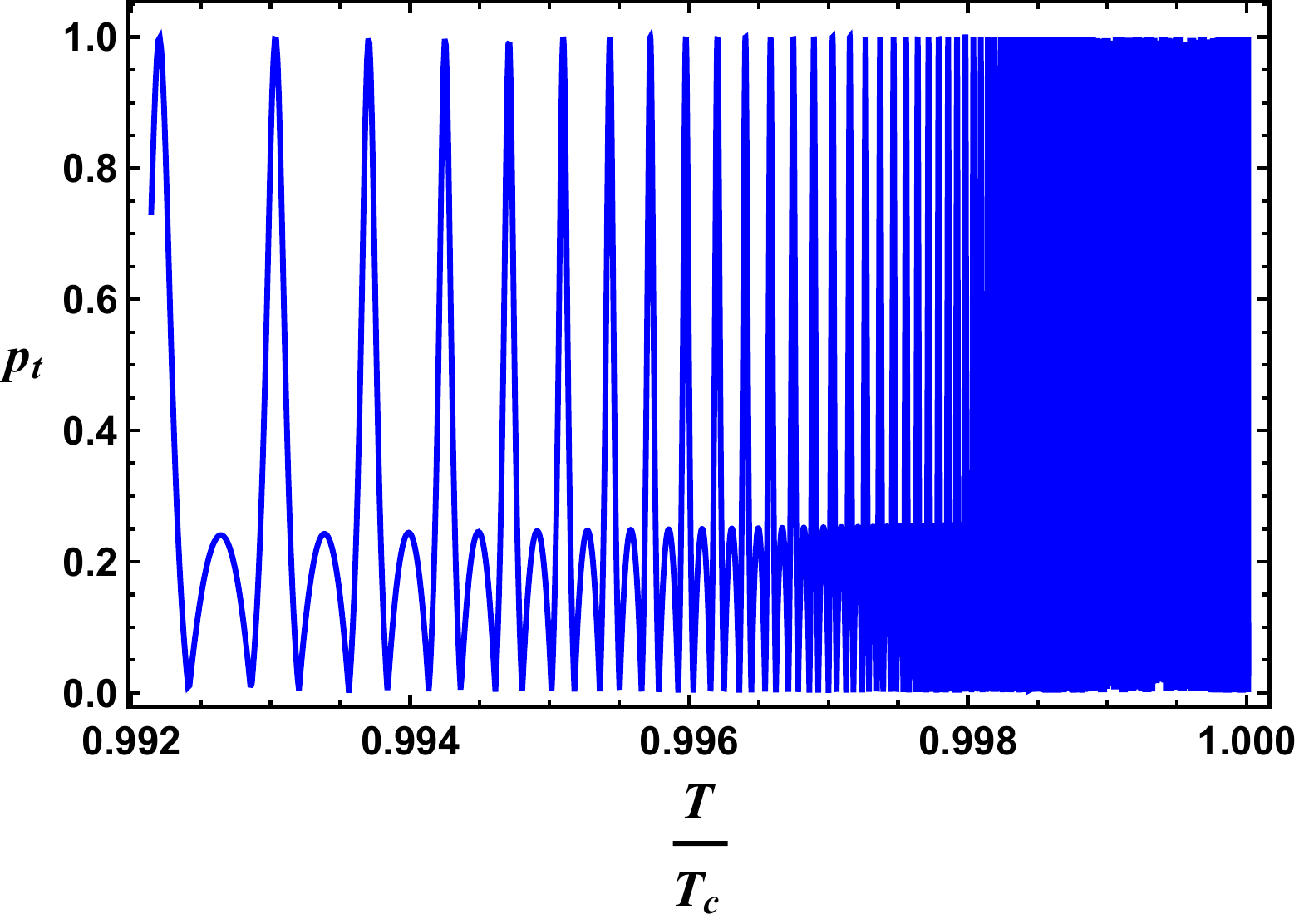}
 \begin{center}
 \includegraphics[width=0.45\linewidth]{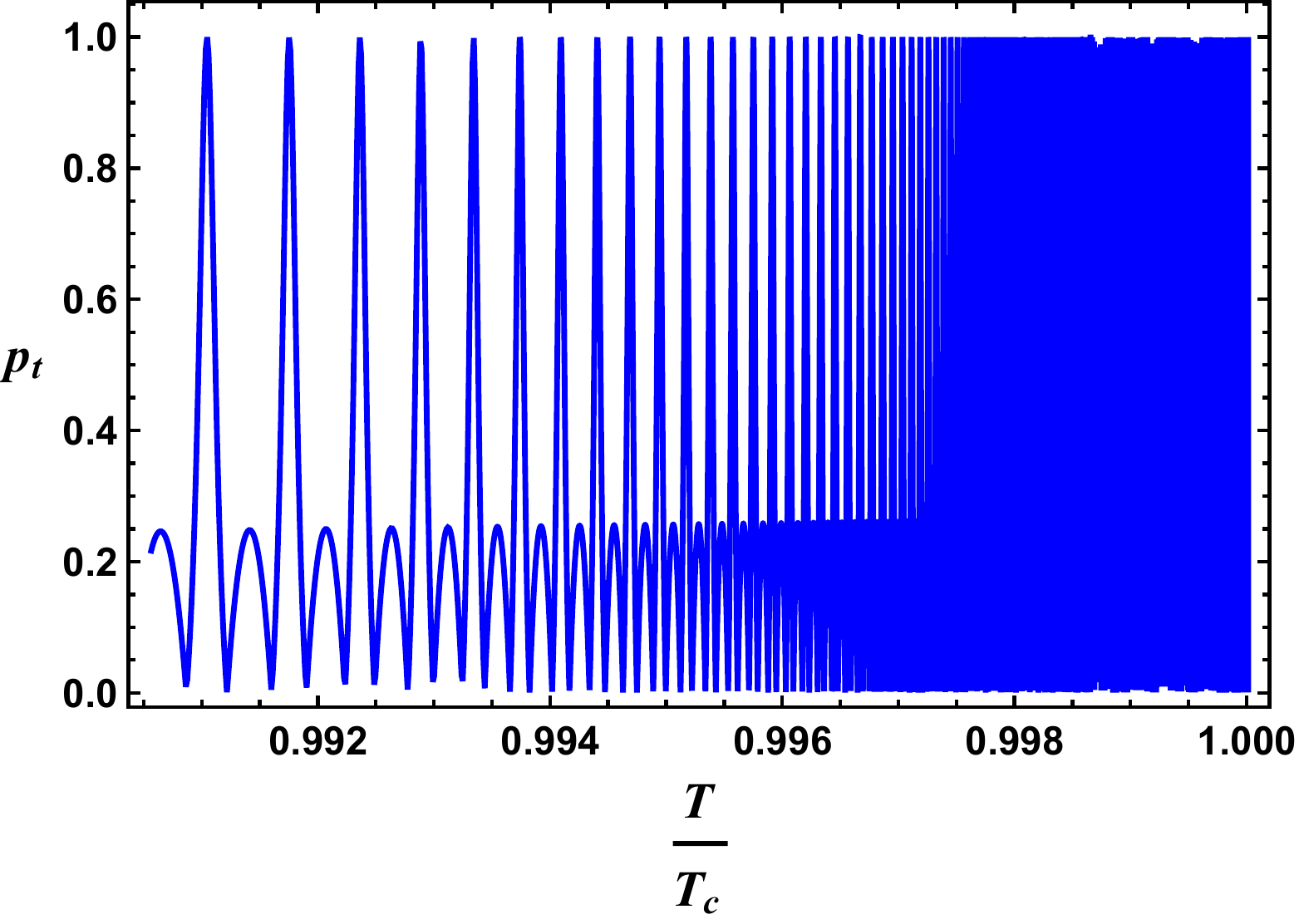}
 \end{center}
 \caption{ The Kasner exponent $p_{t}$ after inversion as a function of the temperature in various configuration of the massive gravity. Notice that the accumulation of the oscillations near the critical temperature $T_{c}$ is described by the analytic expression \eqref{C5relation} for $d_{3}$ which is plotted numerically in Fig.\,\ref{Fig7}. They are for 
$\kappa/\mu=1$.
 \label{Fig10pt}}
\end{figure}
 We also present the value of the Kasner exponent $p_{t}$ after the inversion for all temperatures up to $T_{c}$ in Fig.\,\ref{Fig10pt}. Obviously, close to the critical temperature $T_{c}$, there are strong oscillations. This is due to an accumulation of the oscillations in the scalar field just beyond the  inner horizon of the non-hair black hole solution. More precisely, the analytic result \eqref{C5relation} indicates the oscillatory Kasner exponent after a first collapse of the Einstein- Rosen bridge.

Before ending this section, let us discuss whether or not the massive gravity contributions can violate the Kasner form  \eqref{kasnerm} near the spacelike singularity at $r \to \infty$.
Generally, a $(d+2)$-dimensional massive gravity is given by the following action \cite{cai2015thermodynamics}.
\begin{equation}
S_{\text{massive gravity}}=\int d^{d+2} x \sqrt{-g} \sum_{i=1}^{4} n_{i} U_{i}(g,\gamma),
\end{equation}
where $n_{i}$ are dimensionless constants\footnote{For stability of the theory, all those coefficients might be required to be negative.} and $U_{i}$ are symmetric polynomials of the eigenvalues of the matrix $\mathcal{K}$ which satisfy the following relations.
\begin{eqnarray}
\nonumber U_{1}&=&\Tr \mathcal{K},\\
 U_{2}&=&\Big(\Tr \mathcal{K}\Big)^2-\Tr \mathcal{K}^2,\\
\nonumber U_{3}&=&\Big(\Tr \mathcal{K}\Big)^3-3 \Tr \mathcal{K} \Tr \mathcal{K}^2+2 \Tr \mathcal{K}^3,\\
\nonumber U_{4}&=&\Big(\Tr \mathcal{K}\Big)^4-6 \Tr \mathcal{K}^2 \Big(\Tr \mathcal{K}\Big)^2+8 \Tr \mathcal{K}^3 \Tr \mathcal{K}+3\Big(\Tr \mathcal{K}^2\Big)^2 -6 \Tr \mathcal{K}^4.
\end{eqnarray}
Now by taking the general reference metric $f_{\mu \nu}=\text{diag}\Big(0,0, \kappa^2 \delta_{ij}\Big)$ in $d+2$ dimensions, one obtains
  \begin{eqnarray}
 \nonumber  U_{1}&=&d \kappa r,\\
  U_{2}&=& d(d-1) \kappa^2 r^2,\\
\nonumber   U_{3}&=& d(d-1)(d-2) \kappa^3 r^3,\\
\nonumber   U_{4}&=& d(d-1)(d-2)(d-3) \kappa^4 r^4.
  \end{eqnarray}
  It is clear that in the 4-dimensional case with $d=2$, we have identically $U_{3}=U_{4}=0$ as seen in \eqref{action1}, whereas in the 5-dimensional case with $d=3$, one takes $U_{4} = 0$ and so on.
  
In order to get the Kasner solutions \eqref{kasner1}, we require the kinetic term of the scalar field to be dominant with respect to the massive gravity contributions near the singularity. It means that
\begin{equation}\label{condi}
\lim_{r \to \infty} \frac{|\sum_{i=1}^{4} U_{i}|}{r^{d+1+\alpha^2}}\ll 1.
\end{equation}
Clearly, this condition is never violated in any dimensions. Therefore, once the solution enters the Kasner epoch, the massive gravity contributions are not important. 

\section{Dynamical epochs at the large $\kappa/\mu$ limit}\label{largelimi}
 
\begin{figure}
\includegraphics[width=0.45\linewidth]{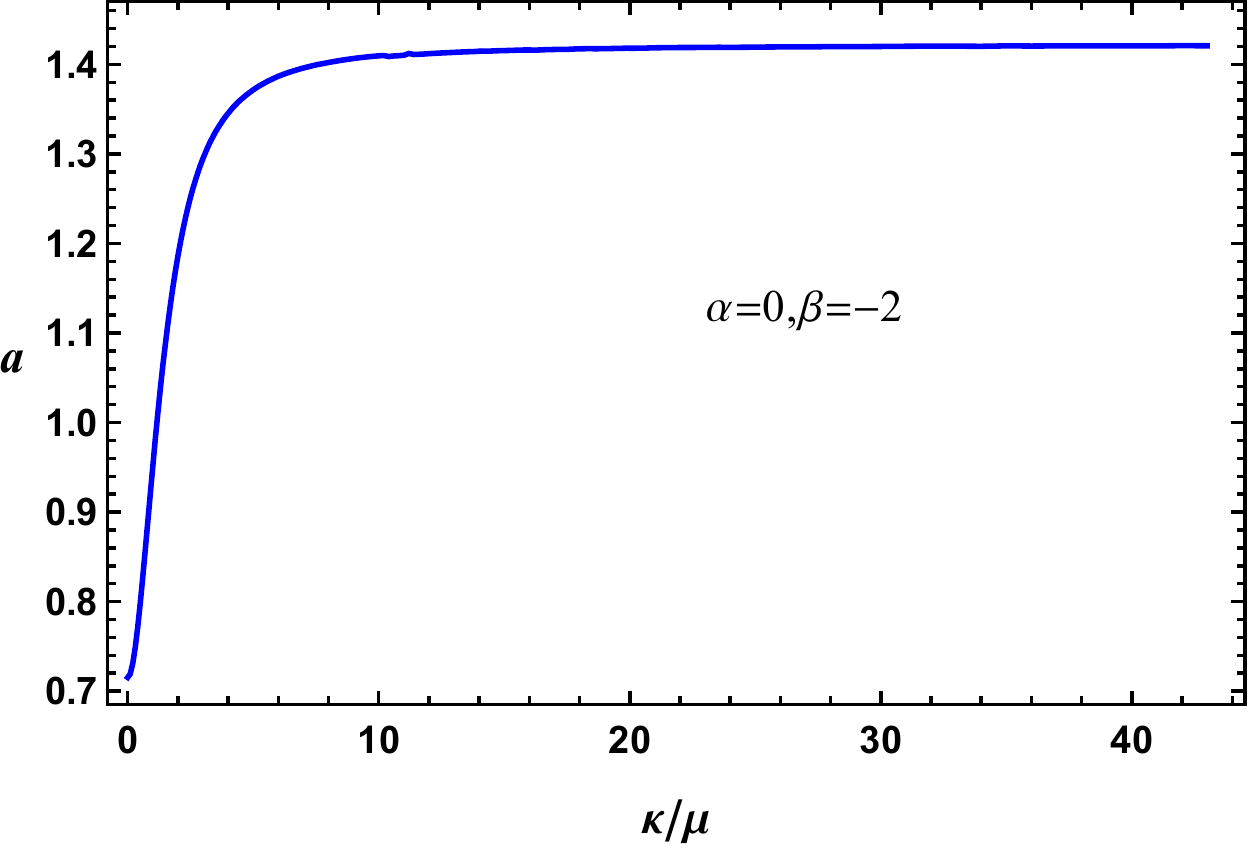}  \qquad
  \includegraphics[width=0.45\linewidth]{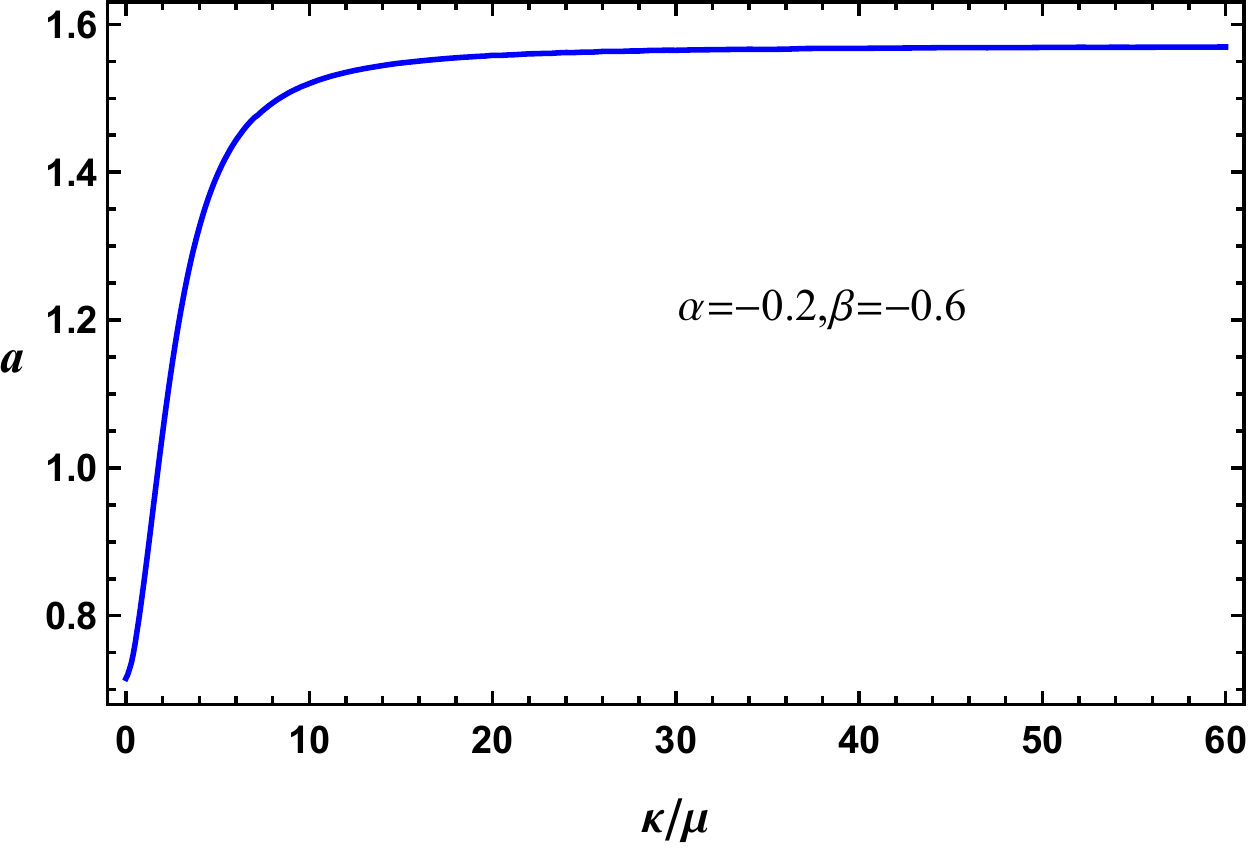}
 \begin{center}
\includegraphics[width=0.45\linewidth]{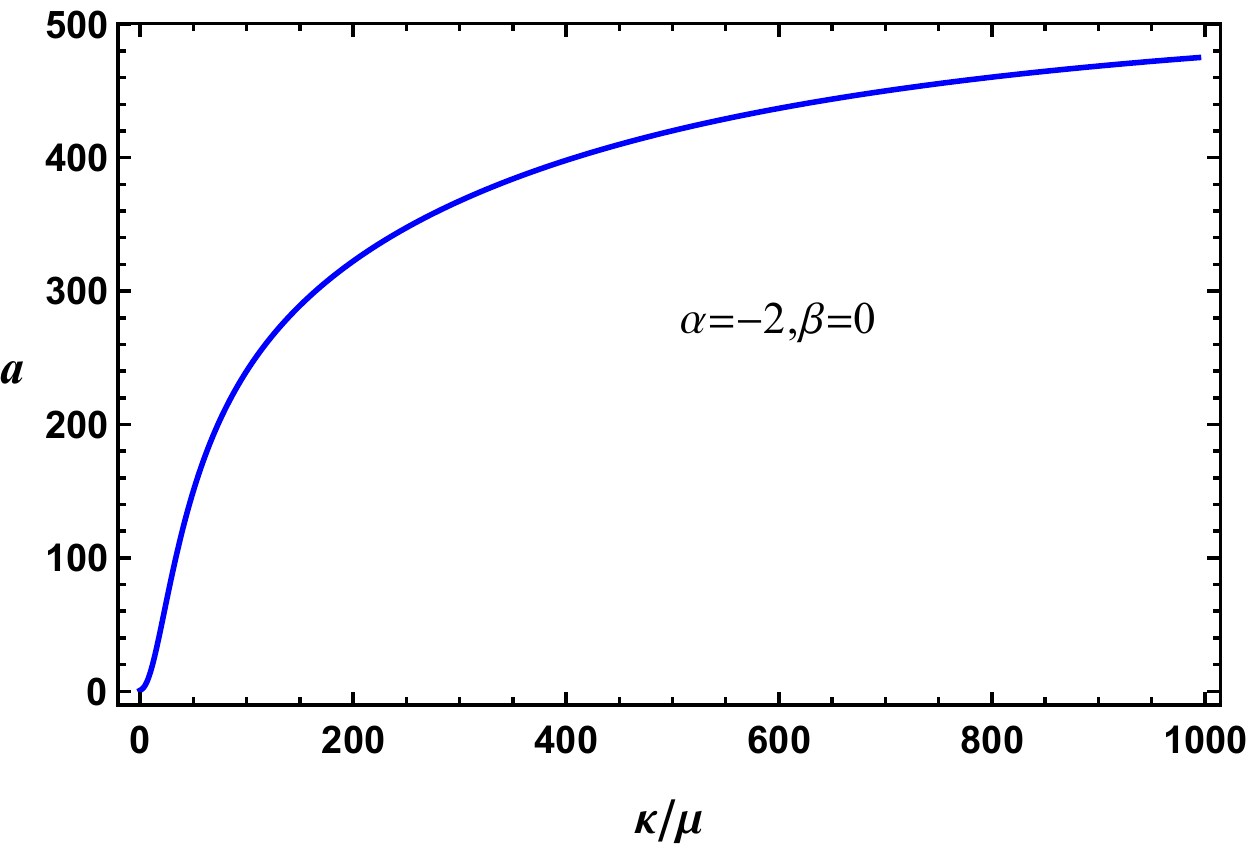}
 \end{center}
 \caption{ The slope of $c_{1}/c_{2}$ with respect to $\kappa/\mu$.  
 \label{afig}}
\end{figure}

In this section, let us focus our attention on the interior dynamics in the presence of large amounts of the massive gravity parameter. As mentioned, the quantity $c_{1}/c_{2}$ determines the intensity of the occurrence of the ER collapse at the would-be inner horizon.  As shown in Fig.\,\ref{figrphi223}, the ratio $c_1/c_2$ is described numerically by the straight line $a(1-T/T_{c})^{1/2}+b$ in which its slope $a$ is growing by increasing $\kappa/\mu$. Therefore, we have depicted the behavior of the slope $a$ for large amounts of the massive parameter $\kappa/\mu$ in Fig.\,\ref{afig}. As illustrated in Fig.\,\ref{afig}, for massive configurations $\alpha =0$, $\beta\neq 0$ and $\alpha, \beta\neq 0$ one observes the slope $a$ shifts to a constant at large $\kappa/\mu$ values, whereas it continues to grow even for so large amounts of $\kappa/\mu$ for the pure non-linear massive configuration $\alpha \neq 0$ and $\beta =0$. Notice that the slope values presented here have been read for all range of temperature as presented in Fig.\,\ref{figrphi223}. In other words, the slope $a$ depends only on the massive parameter, i.e. $a=a(\kappa/\mu)$.

  This finding confirms that the amount of $c_{1}/c_{2} \propto a(\kappa/\mu) (1-T/T_{c})^{1/2}$ is affected by varying both the massive parameter and the temperature. In comparison, in the standard holographic model \cite{Hartnoll:2020fhc}, the slope $a$ will be constant and $c_{1}/c_{2}$ varies only by temperature. 
  
  As a consequence of this finding, for a sufficiently large massive parameter, by fine tuning the temperature, one could get the largest amount for $c_{1}/c_{2}$ in all massive configurations. 
 Hence, in this sense, we expect that the time component of the metric $g_{tt}$, which in general cases is exponentially vanishing after $r_{\mathcal{I}}$ according to the discussion bellow Eq.\,(\ref{eq:gtt}), undergoes a linear function,
\begin{equation}\label{gttexpand}
g_{tt} \propto e^{-(c_2/c_1)^2 (r-c_{3})}\approx  \Big(\frac{c_{2}}{c_{1}}\Big)^2 (c_{3}-r)+\dots .
\end{equation} 
both sides of the would-be inner horizon. Namely, the ER collapse disappears from intrior dynamics epoches \footnote{A similar result has been
reported in \cite{Sword:2021pfm} for the holographic superconductor model including an axion field term and an Einstein-Maxwell-scalar (EMS) coupling term. At the large EMS coupling parameter, the ER collapse does not occur near the would-be inner horizon and is shifted outward to a large radial value. In fact, there is a delay in the collapse of the ER and it is not completely removed from the intrior dynamical epochs.}. In fact, unlike the standard holographic superconductor \cite{Hartnoll:2020rwq}, $g_{tt}$ does not exponentially collapse after the would-be inner horizon, while one sees a smooth crossing at this horizon. 

\begin{figure}
\includegraphics[width=0.45\linewidth]{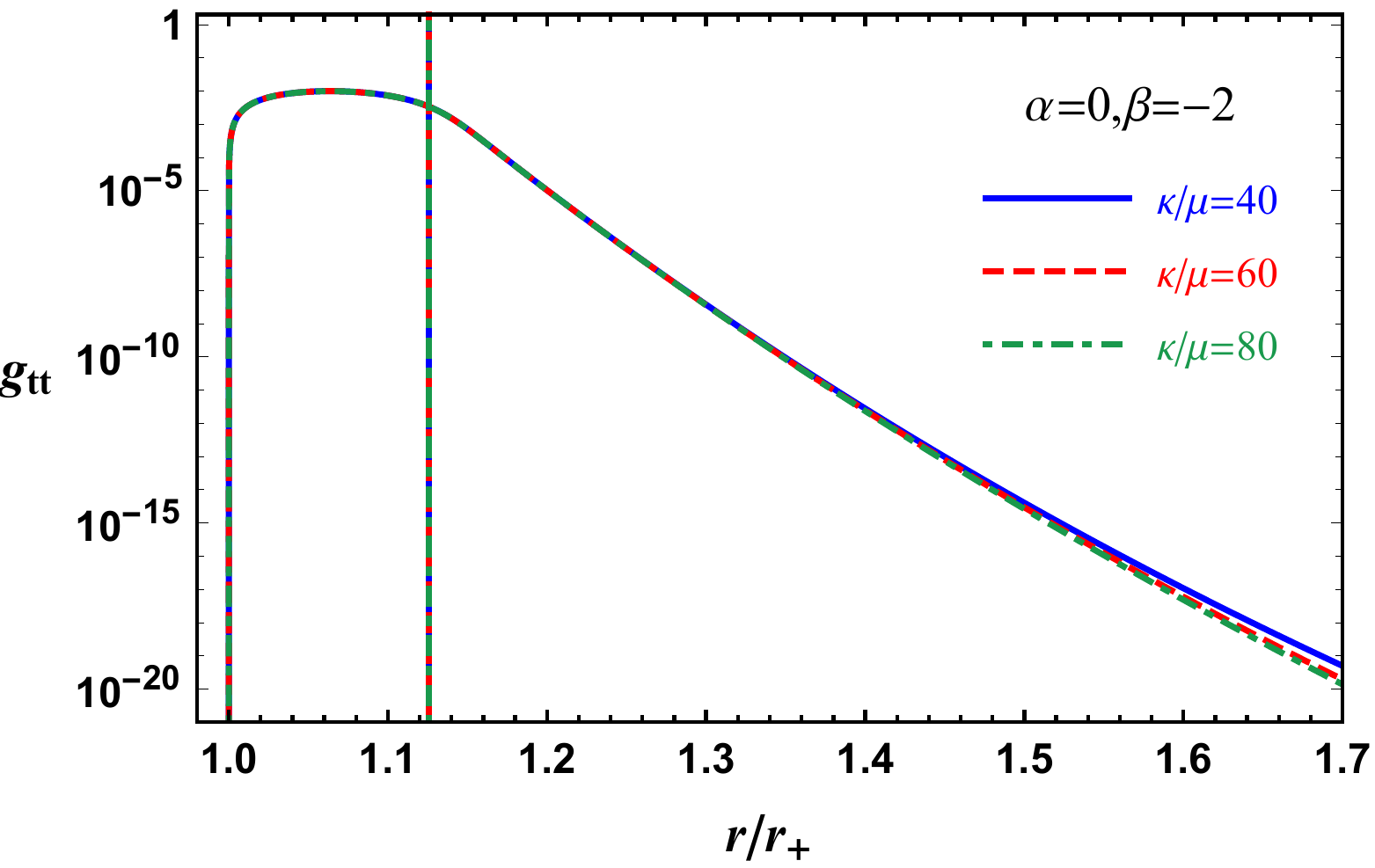}  \qquad
  \includegraphics[width=0.45\linewidth]{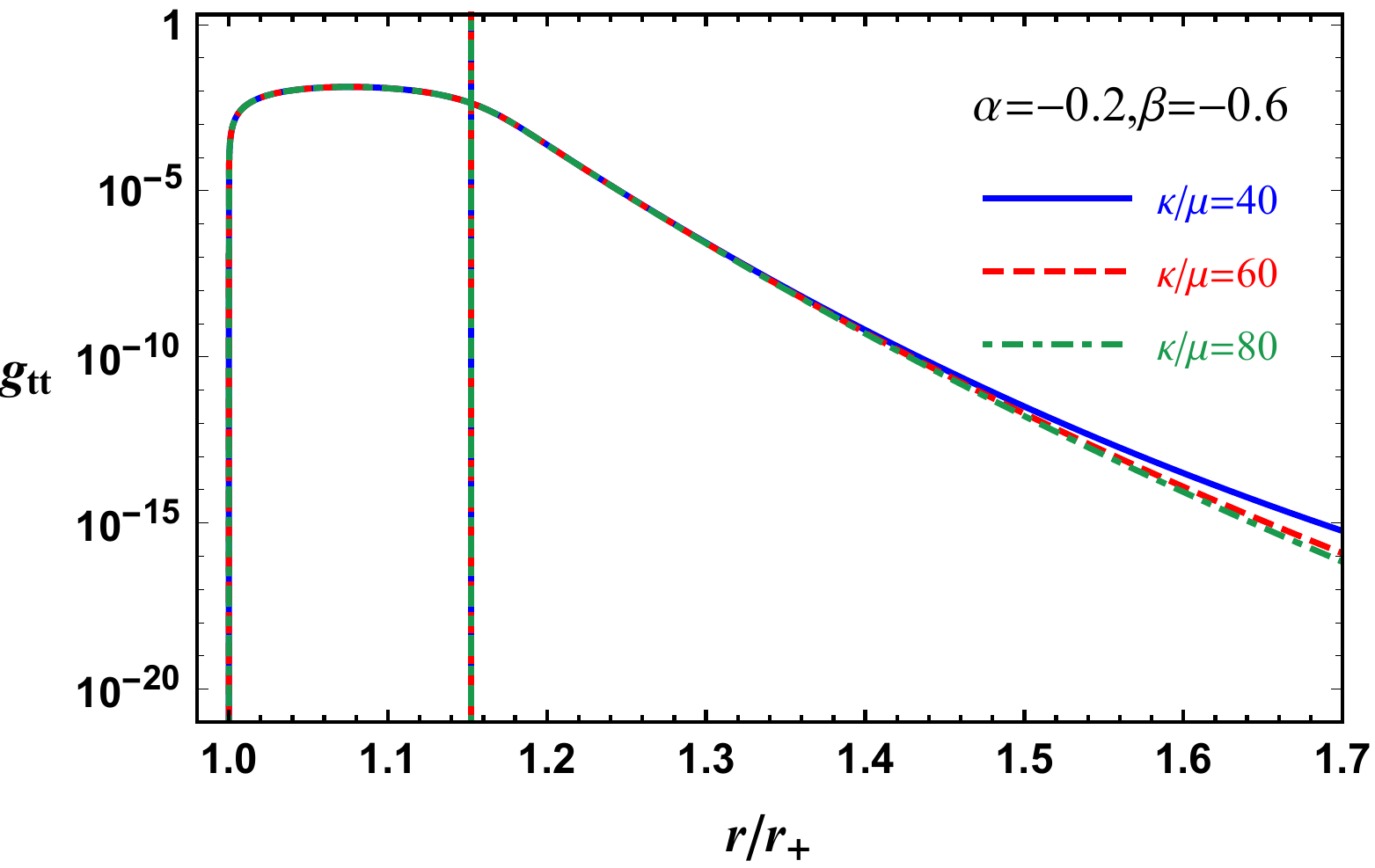}
 \begin{center}
\includegraphics[width=0.45\linewidth]{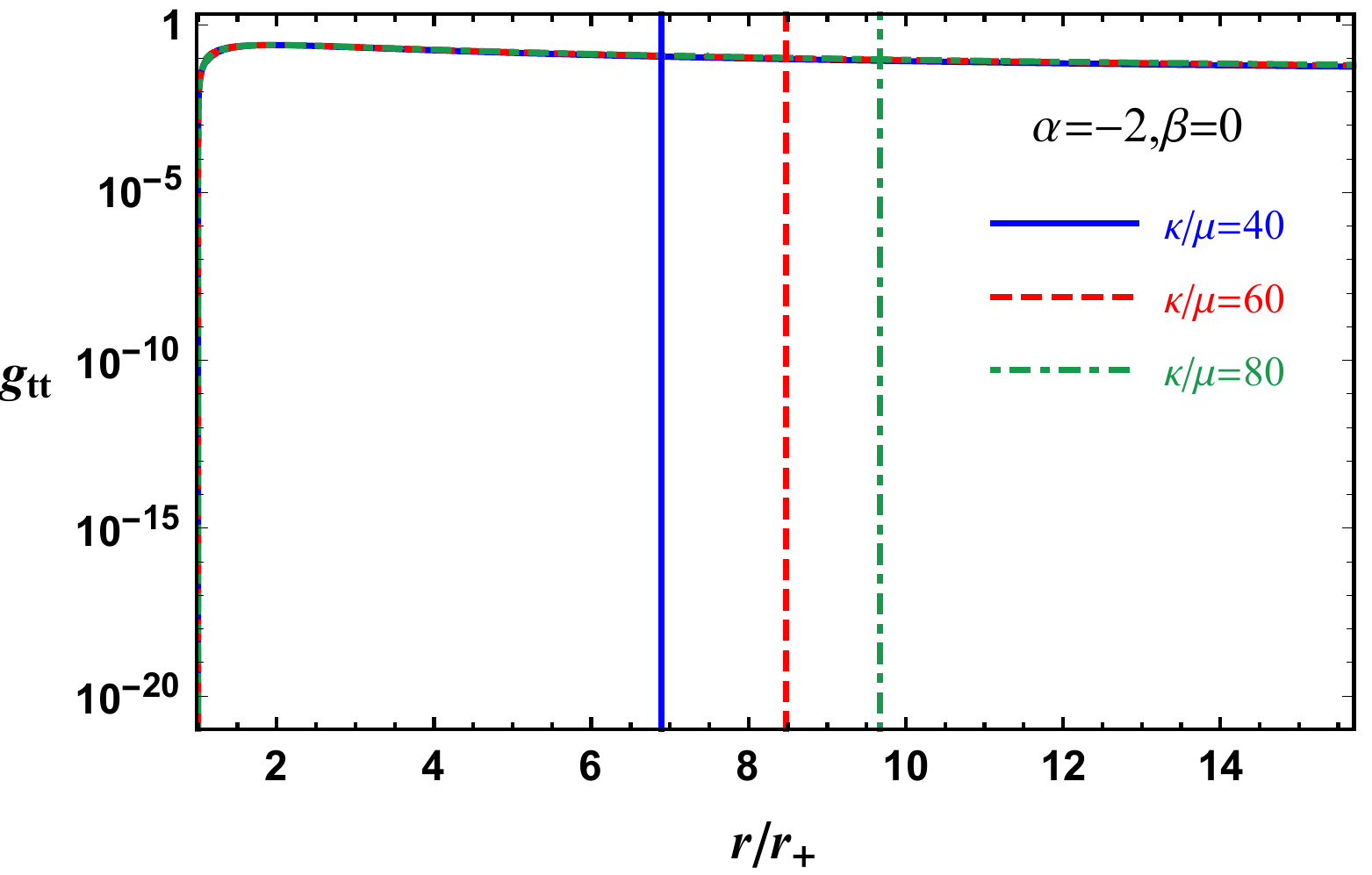}
 \end{center}
 \caption{ The metric component $g_{tt}$ as a function of $r$ near the would-be inner horizon (the vertical dashed lines) for large values of $\kappa/\mu$. Here we take $T/T_{c}=0.995$.  
 \label{gttlfig}}
\end{figure}

\begin{figure}
\includegraphics[width=0.45\linewidth]{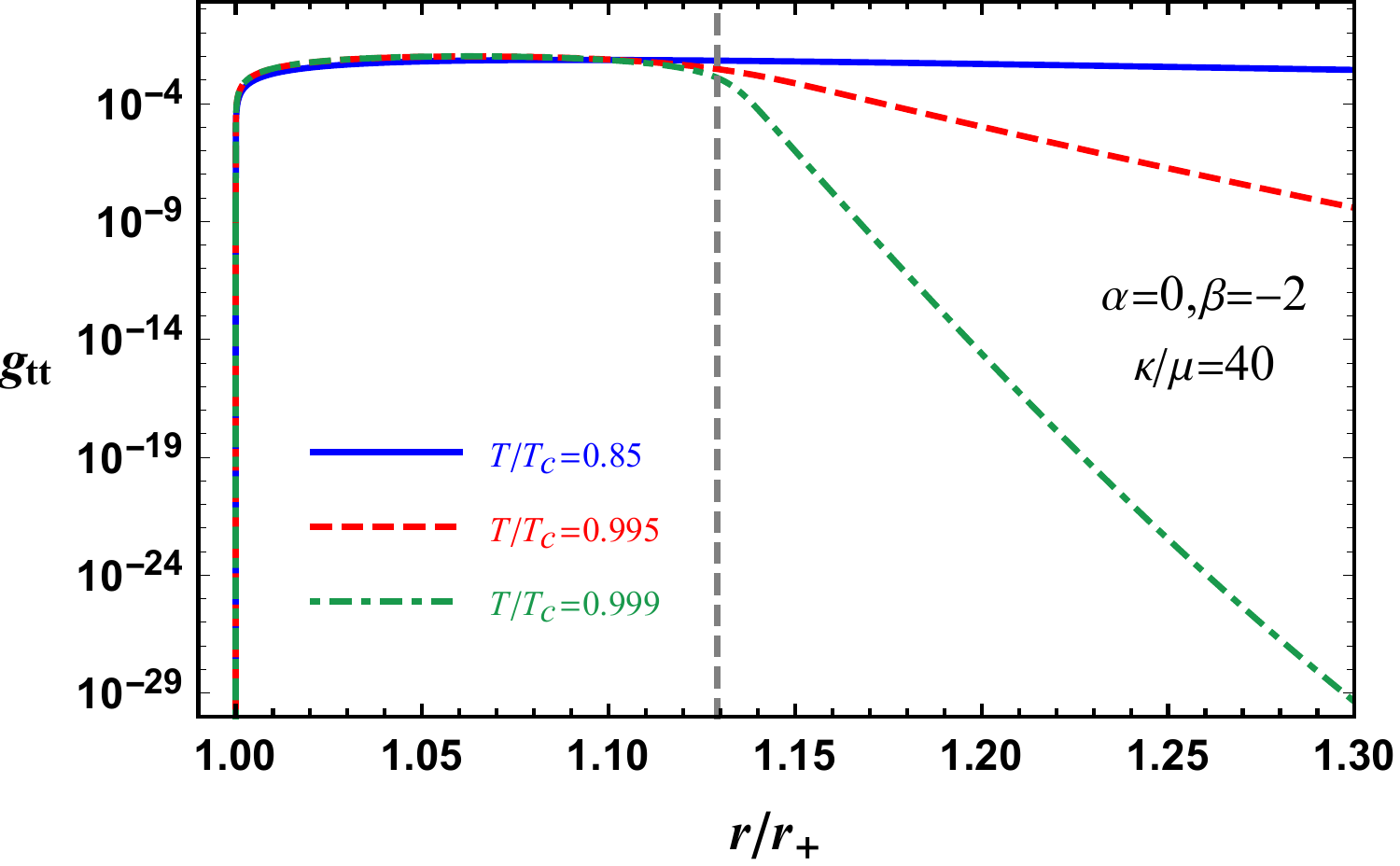}  \qquad
  \includegraphics[width=0.45\linewidth]{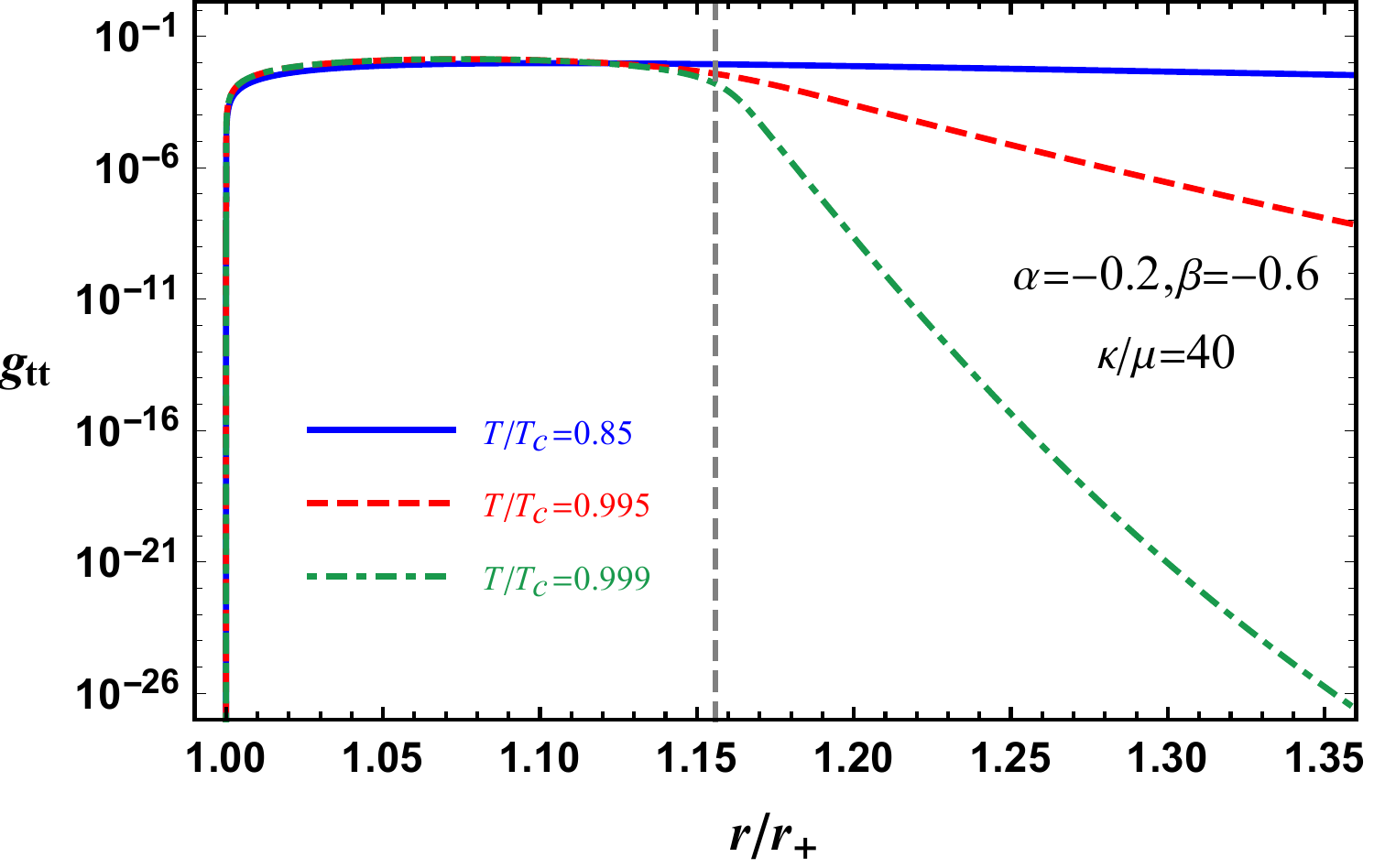}
 \begin{center}
\includegraphics[width=0.45\linewidth]{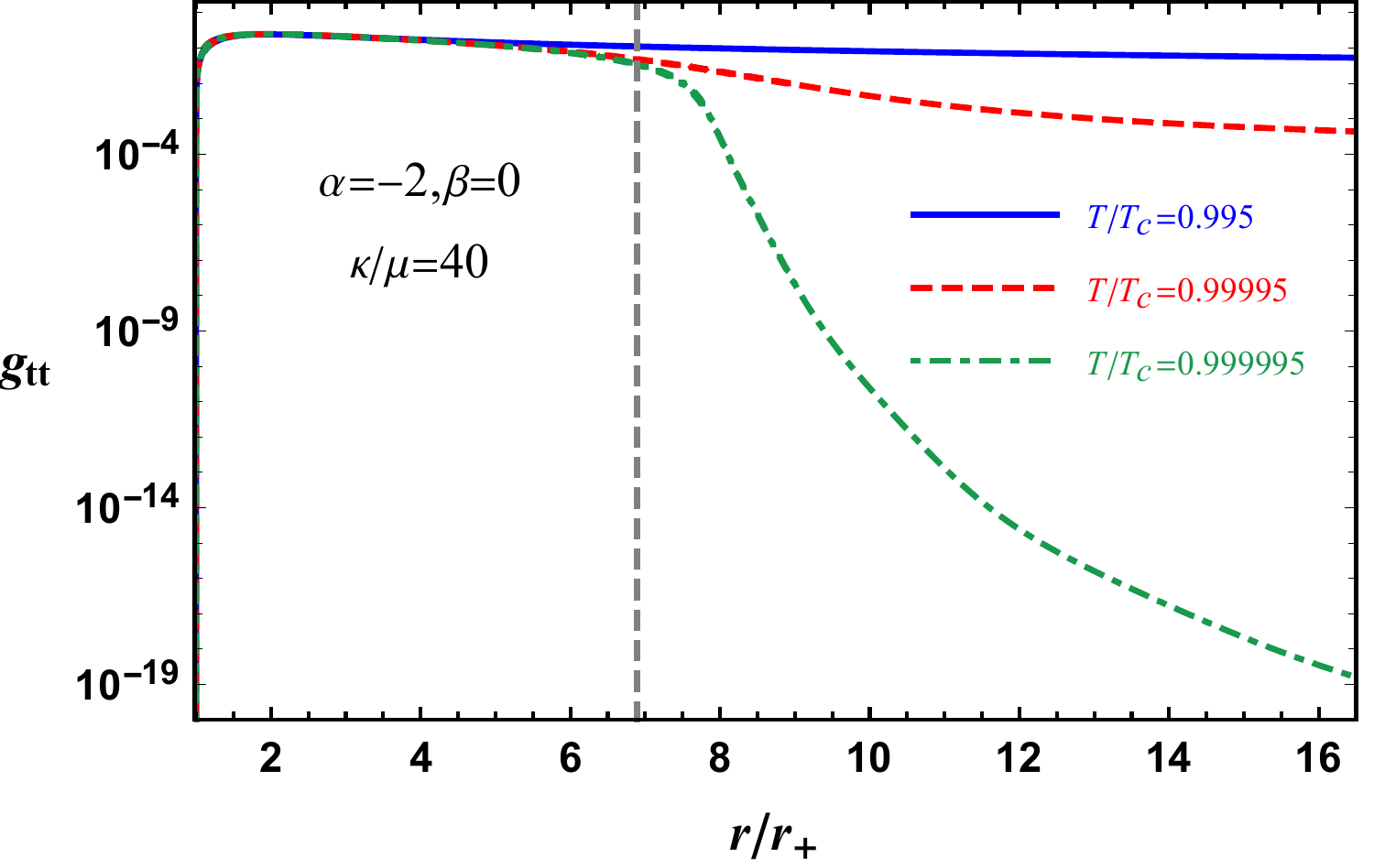}
 \end{center}
 \caption{ The metric component $g_{tt}$ as a function of $r$ near the would-be inner horizon (the vertical dashed line) for different temperature values. Here we take $\kappa/\mu=40$.  
 \label{gttlfig1}}
\end{figure}

As a concrete example, in Fig.\,\ref{gttlfig} we have depicted $g_{tt}$ behavior for large $\kappa/\mu=\{40,60,80\}$ at $T/T_{c}=0.995$. As can be seen, in the pure non-linear configuration, the ER bridge collapse is completely removed near the would-be inner horizon (the vertical dashed line), whereas this collapse still happens and is not sensitive to the change of the massive parameter in the other configurations. In addition, by keeping the massive gravity parameter fixed, we examine the impact of the temperature on the ER bridge collapse. It can be seen in Fig.\,\ref{gttlfig1} that the ER bridge collapse can be vanished at a given temperature in all massive configurations. For an example, in pure non-linear massive configuration, the accuracy of Eq.\,\eqref{gttexpand} breaks down for so small values of the scalar field at the horizon, i.e., as $T \to T_{c}$.

 \begin{figure}
\includegraphics[width=0.45\linewidth]{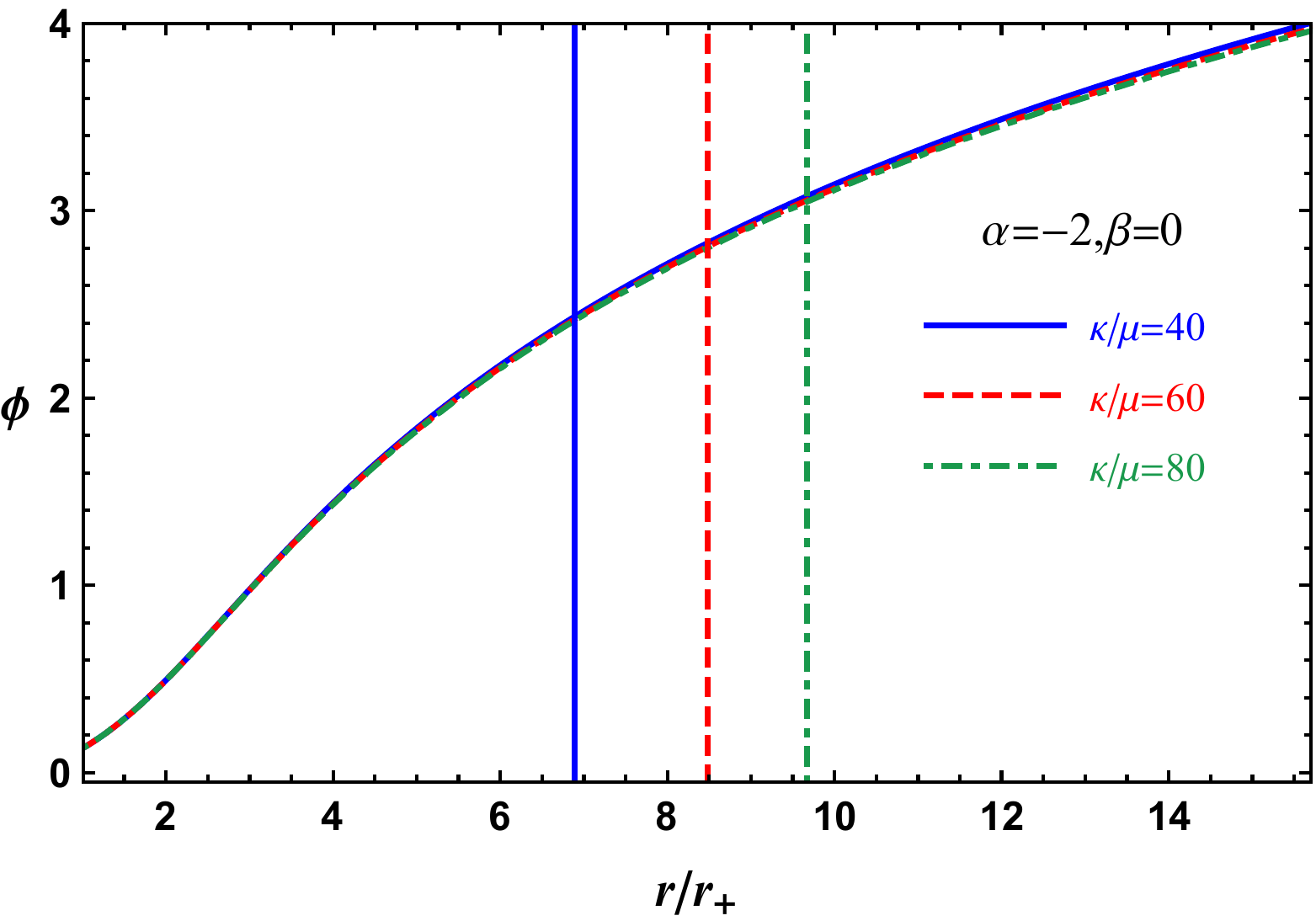}  \qquad
  \includegraphics[width=0.45\linewidth]{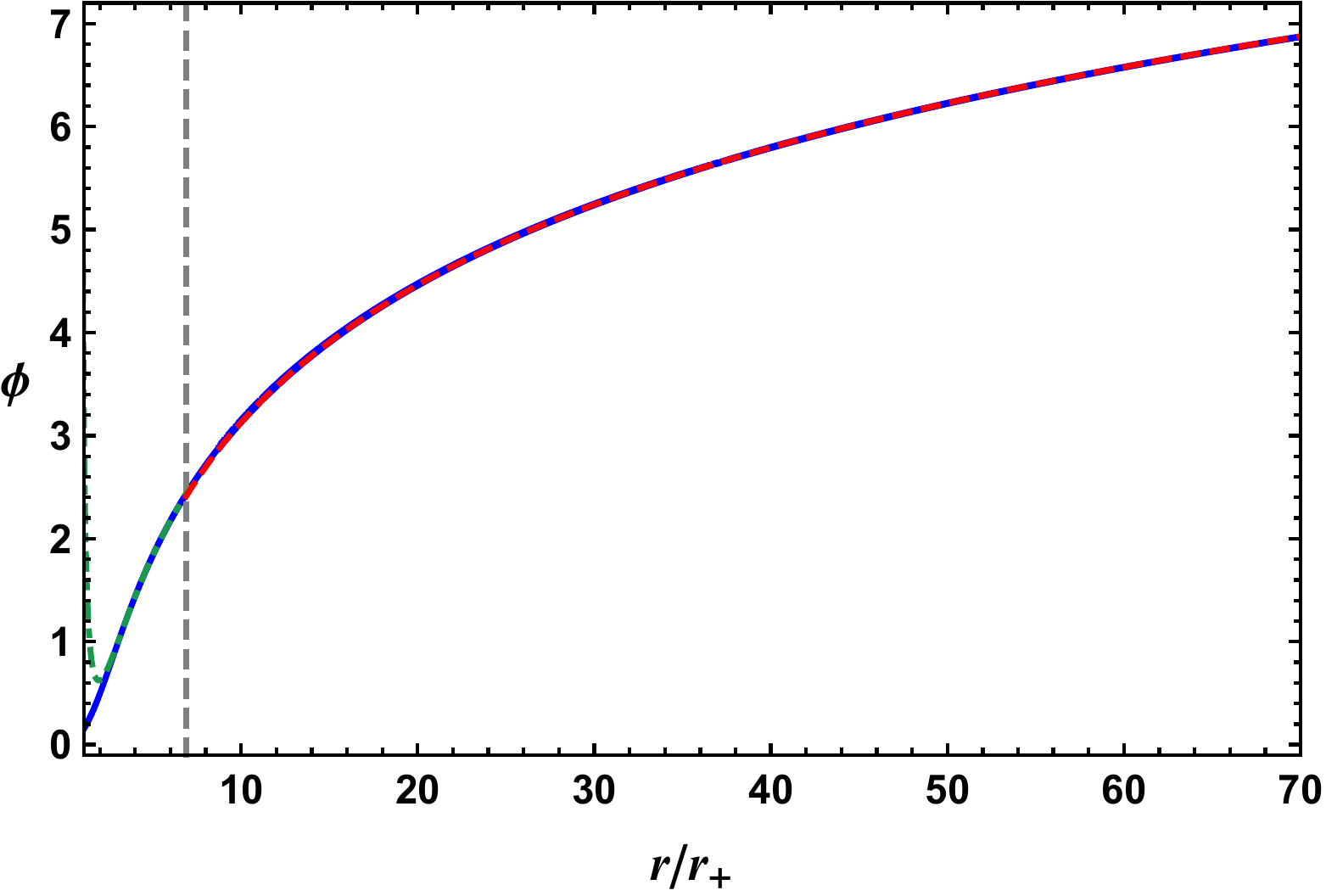}
 \caption{ Left: The evolution of $\phi$ for large  amounts of the massive parameter $\kappa/\mu$ in the pure non-linear massive configuration case. Right: A comparison of the numerical solution (solid blue curve) of $\phi$ for $\kappa/\mu=40$ in the pure non-linear massive configuration case and fits to the relation \eqref{phiER} (green dashed curve) describing ER collapse before would-be inner horizon and analytic relation  \eqref{anphi} (red dashed curve) after the would-be inner horizon. Here we have chosen  $T/T_{c}=0.995$.
 \label{philfig}}
\end{figure}
 
\begin{figure}
 \begin{center}
\includegraphics[width=0.5\linewidth]{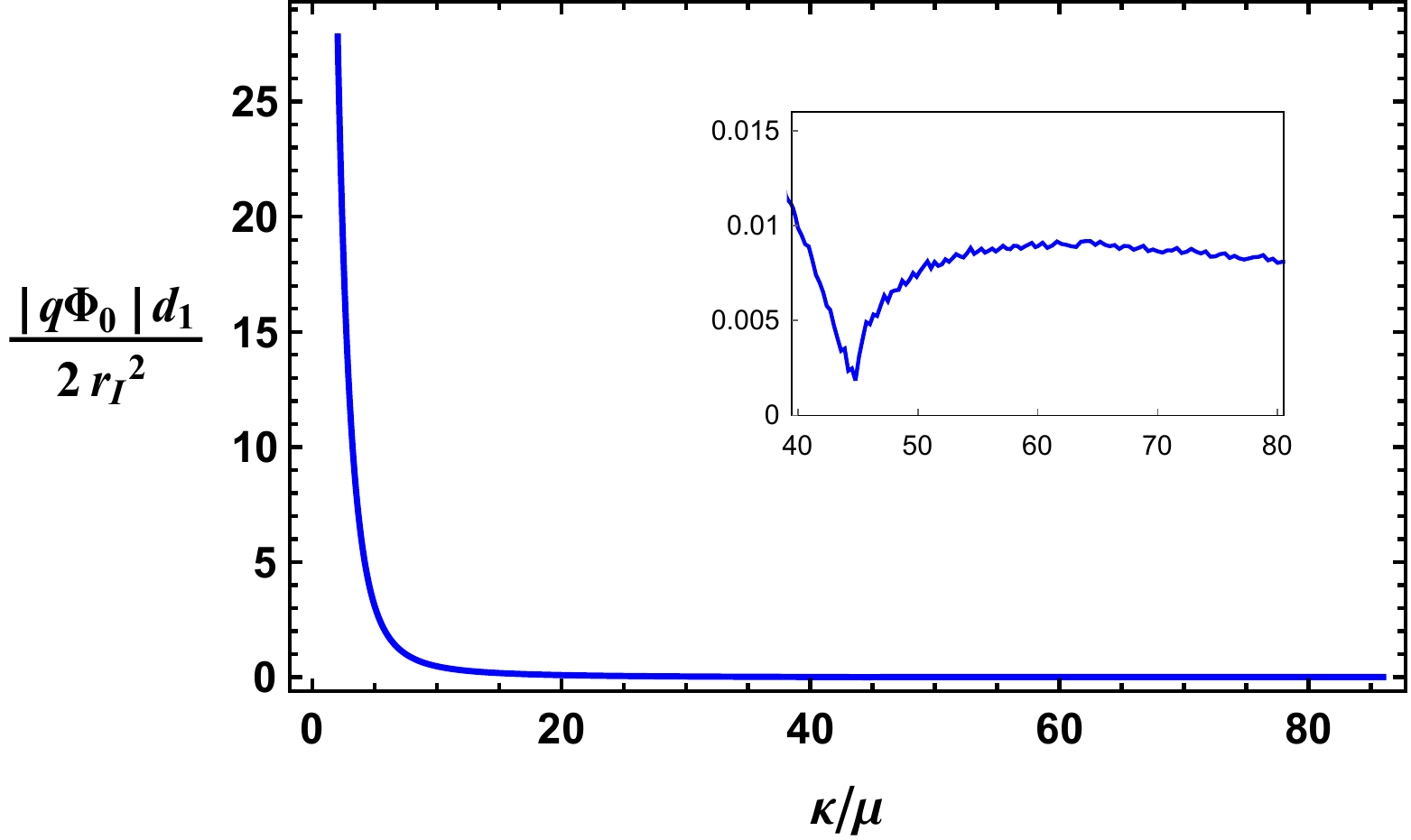}
 \end{center}
 \caption{ The behavior of the relation \eqref{argumen} with respect to $\kappa/\mu$.  
 \label{argu}}
\end{figure}

 On the other hand, what happened in the ER collapse affects directly the subsequent Josephson oscillations epoch. As illustrated in Fig.\,\ref{philfig}, the Josephson oscillation epoch is completely disappeared for large amounts of $\kappa/\mu$ in the pure non-linear massive configuration case. Now we attempt to find the analytic relation describing the dynamics of $\phi$ around the would-be inner horizon.  From Eqs.\,\eqref{c3} and \eqref{eq:gtt}, one immediately obtains the below relation for the argument inside the Bessel functions in \eqref{Joseph:psi}.   
\begin{equation}\label{argumen}
\frac{ |q \Phi_{0}|d_{1}}{2 r^2} \approx \frac{\sqrt{2}r_{\mathcal{I}}^{5/2}}{2 \phi_{0} r^2} \Big(\frac{c_{2}}{c_{1}}\Big).
\end{equation}  
It is obvious that this relation is proportional to the ratio $c_{2}/c_{1}$. Therefore, it is expected to be very small for large values of the massive parameter in the pure non-linear massive gravity. We have checked numerically this function at the would-be inner horizon for large massive gravity parameter in Fig.\,\ref{argu}. Clearly, the amount of this expression is very small at large $\kappa/\mu$ values. Note that after the would-be inner horizon ($r> r_{\mathcal{I}}$), its values becomes smaller. 

\begin{figure}
\begin{center}
\includegraphics[width=1\linewidth]{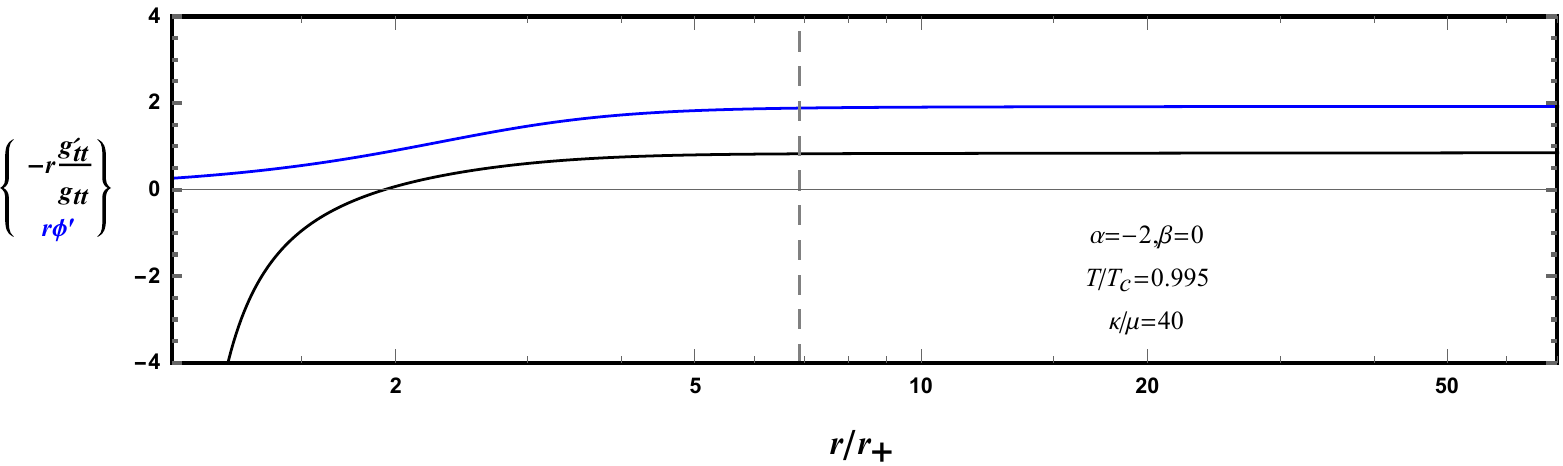}
\\
  \includegraphics[width=1\linewidth]{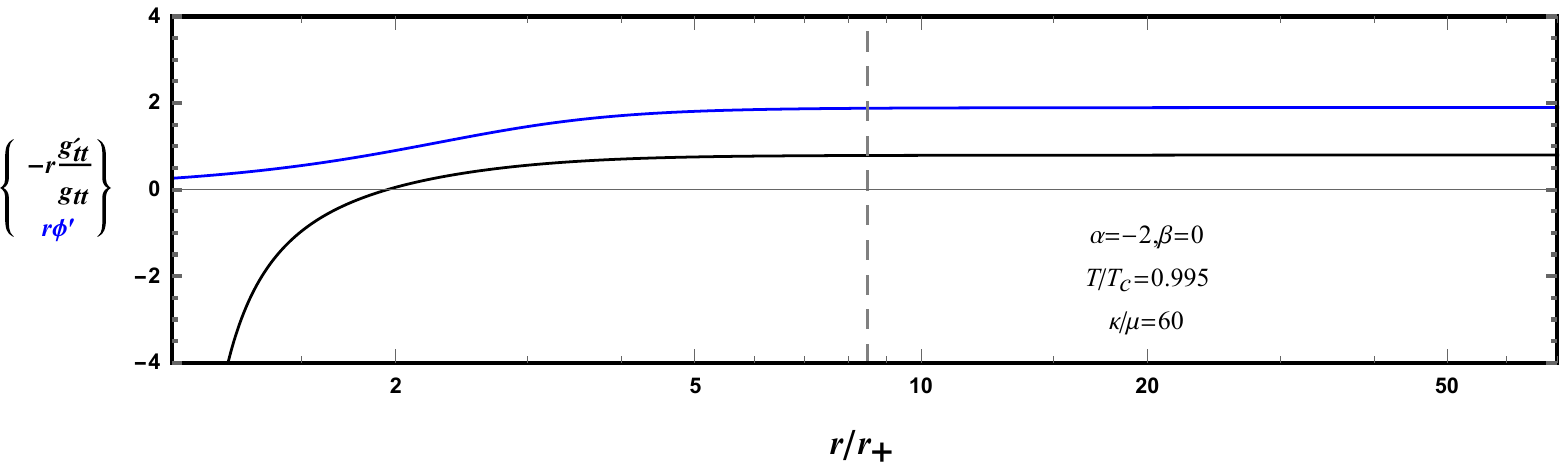}
\\
\includegraphics[width=1\linewidth]{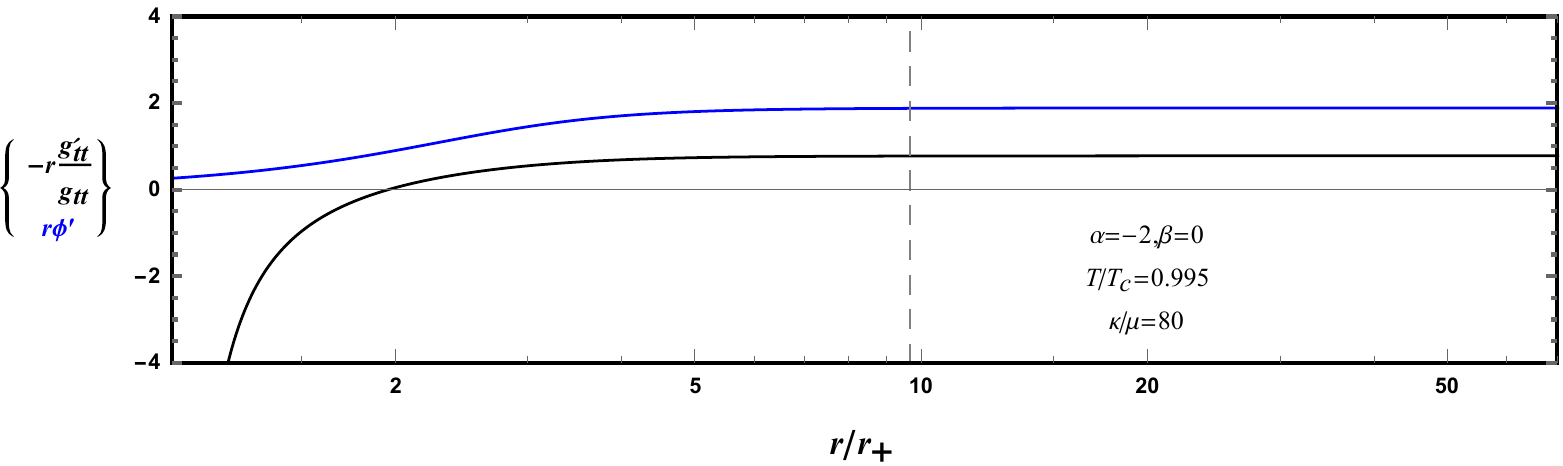}
 \end{center}
 \caption{ An example of an interior evolution with no ER bridge collapse and Josephson oscillations epochs in the pure non-linear massive configuration. Clearly, the system enters the Kasner geometry immediately after the would-be inner horizon. Numerics are at  $T/T_{c}=0.995$ and $\kappa/\mu=40,~60,~80$ from top to bottom.   
 \label{kasnerhi}}
\end{figure}

As a result of the above discussion, the expression for the Josephson oscillations in Eq.\,\eqref{Joseph:psi} for $\phi$ at the large massive parameter is written approximately as 
\begin{equation}\label{anphi}
\phi|_{r \sim r_{\mathcal{I}}}\simeq \frac{2 d_{3}}{\pi} \log \Big(\frac{c_{2} e^{\gamma_{E}}}{2 \sqrt{2}c_{1} \phi_{0}}\frac{r_{\mathcal{I}}^{5/2}}{r^2} \Big)+ d_{2}+... \simeq -\frac{4 d_{3}}{\pi} \log(r)+ s_{1}.
\end{equation}
The constant $d_{3}$ can be found through matching the above solution and its derivative for the small $r$ with the ER bridge collapse solution \eqref{phiER} and its derivative in the overlapping region at $r_\mathcal{I}$. Thus one finds
\begin{equation}\label{Bigphi}
d_{3} \approx \Big(\frac{c_{2}}{c_{1}}\Big) \sqrt{\frac{r_{\mathcal{I}}}{2}}\frac{\pi}{2}\frac{g'_{tt}(r_{\mathcal I})}{g_{tt}(r_{\mathcal{I}})} \sin(\varphi_{0}) = -\frac{\pi}{4} r_{ \mathcal{I}} \phi'(r_{\mathcal{I}}),
\end{equation}
where, we have used Eq.\,\eqref{phiER} in the last expression. As illustrated in the right hand side of Fig.\,\ref{philfig}, the analytic relation \eqref{anphi} agrees well with the numerics. More precisely, we find that the red dashed curve describes  Eq.\,\eqref{anphi} with $d_{3}=-1.51161$ and $s_{1}=-1.30245$ which is in excellent agreement with the numerical data from Eq.\,\eqref{Bigphi}.
 Interestingly, the logarithmic form of $\phi$ implies that the system moves to Kasner cosmology epoch exactly after the would-be inner horizon without experiencing neither the ER bridge collapse nor the Josephson oscillation epoch as presented in Fig.\,\ref{kasnerhi}. 
 
\begin{figure}
\begin{center}
\includegraphics[width=1\linewidth]{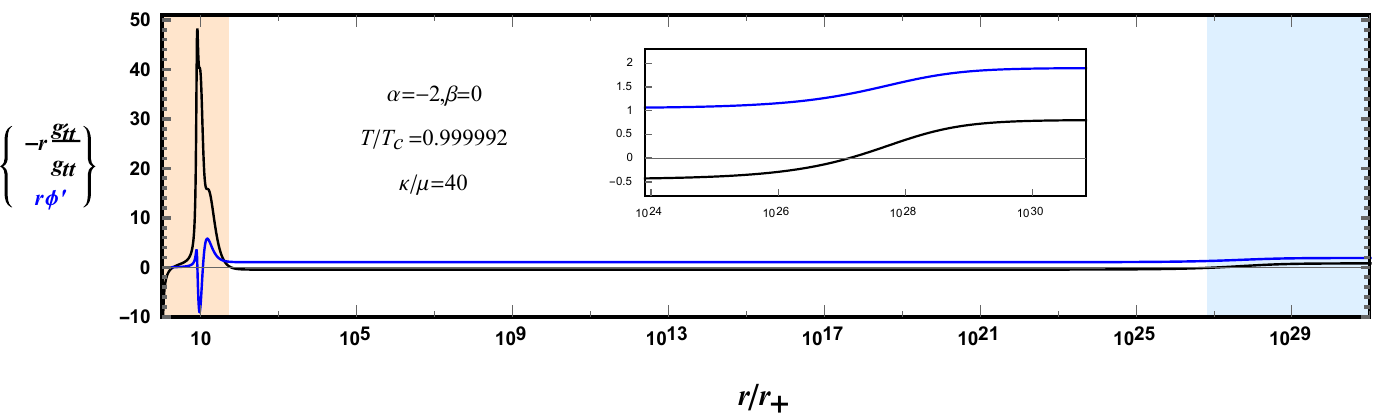}
\\
 \end{center}
 \caption{ Journey through the inside of the pure non-linear massive gravity black hole.   
 \label{kasnerhiom}}
\end{figure}

 In addition,  using Eqs.\,\eqref{Bigphi} and \eqref{kasner1}, one can find
 \begin{equation}
 c=-\frac{\sqrt{8}}{\pi} d_{3}, \hspace{0.5cm} \text{or} \hspace{0.5cm} p_{t}=\frac{8 d_{3}^2-\pi^2}{8 d_{3}^2+3 \pi^2}.
 \end{equation}
 It means that the Kasner exponent value is determined from the amount of the derivative of the scalar field $\phi$ at the would-be inner horizon. For an example, from Fig.\,\ref{kasnerhi}, one finds $c=1.3603294$ which is in agreement with the fitted graph in Fig.\,\ref{philfig} with $d_{3}=-1.51161$ (or equivalently $c=1.36093$). Notice that since the constant $c$ is larger than one, the Kasner inversion does not happen in such cases. Note that at temperatures very close to the critical temperature, all epochs can be observed in the pure non-linear massive configuration (see Fig.\,\ref{kasnerhiom} where we have shown a case with a Kasner inversion).  
 
\begin{figure}
\begin{center}
\includegraphics[width=1\linewidth]{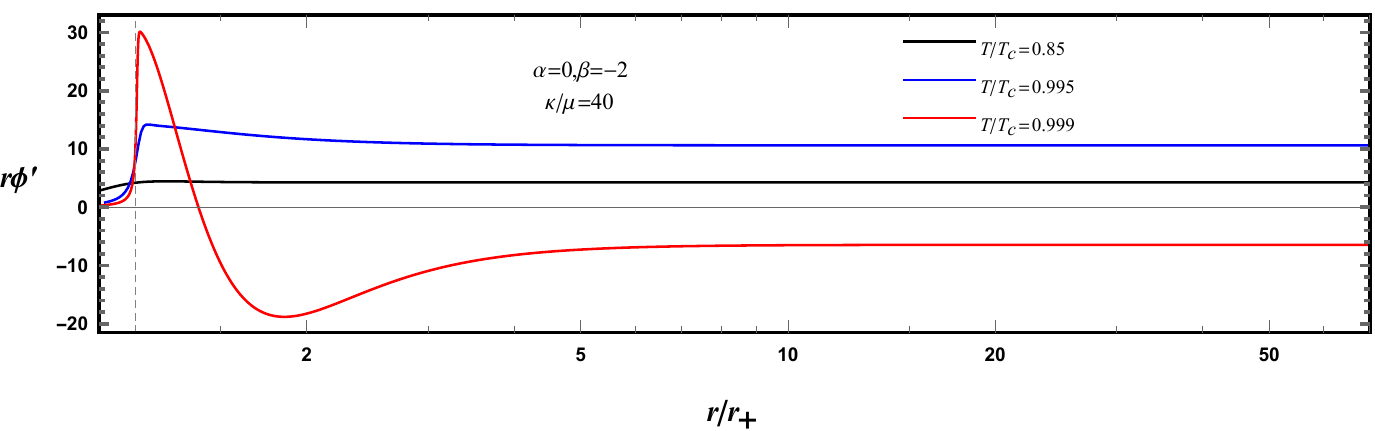}
\\
  \includegraphics[width=1\linewidth]{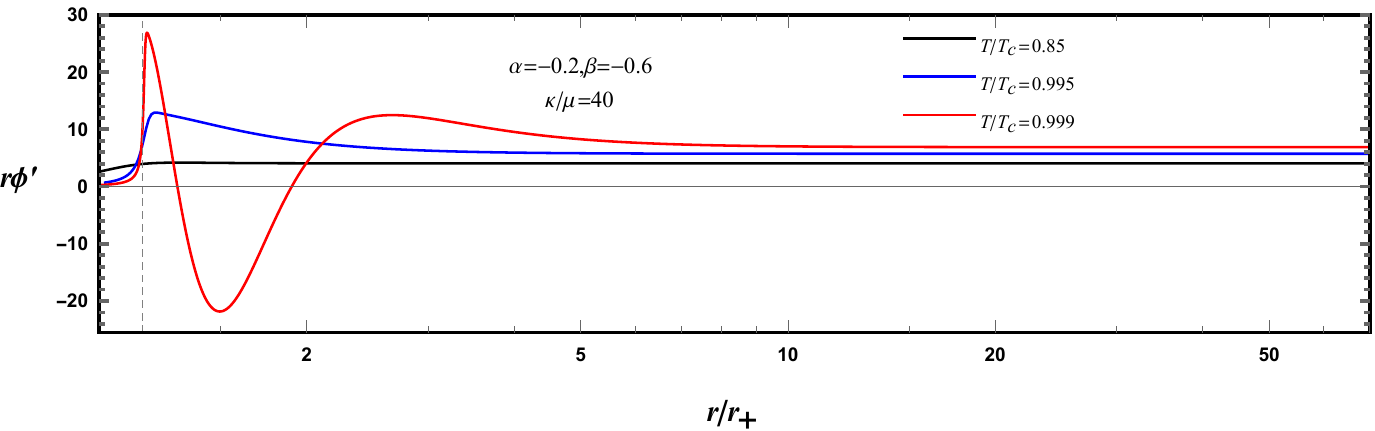}
\\
 \end{center}
 \caption{Examples of the interior evolution. Black curves show the cases with no  Josephson oscillations epochs.   
 \label{kasnerhio1}}
\end{figure}

Interestingly, as shown in Fig.\,\ref{kasnerhio1}, the relation \eqref{anphi} justifies for other massive gravity configurations at small values of the temperature. 

\section{Conclusions}\label{Sec6}

The purpose of the current study was to investigate the interior of the black hole solution of the massive gravity in the presence of the charged scalar hair. In this regard, we found a variety of dynamical epochs, with significant features for the absence of the Cauchy (inner) horizon in such a solution. 

In analogy with the intricate structure previously found inside a holographic superconductor model \cite{Hartnoll:2020fhc}, considering various configurations of the massive gravity, we have seen that just below the critical temperature where the charged massive black hole solution becomes unstable to form scalar hair, the interior dynamics passes through three epochs. More precisely, one can observe a collapse of the Einstein-Rosen (ER) bridge, the Josephson oscillations of the scalar field, and then a Kasner epoch  indicating that the Cauchy horizon is not present. 

While studying the ER collapse and Josephson oscillations epochs, we have found that the instability of the inner horizon results in the collapse of the ER bridge, for which $g_{tt}$ suddenly undergoes a very rapid  collapse and becomes exponentially small over a shot proper time. In addition, for a small value of the dimensionless massive parameter $\kappa/\mu$,
the ER bridge collapse happens sooner while increasing this parameter causes the Josephson oscillations epoch to be removed before entering the final epoch, i.e., Kasner cosmology. Moreover, like the standard holographic superconductor case \cite{Hartnoll:2020fhc}, at temperatures which are far enough from $T_{c}$, the number of Josephson oscillations tends to decrease and finally disappear.  

Finally, we have shown that the interior evolution does not deviate from the Kasner cosmology when the intermediate Kasner exponent is positive, i.e., $|c|>1$ ($p_{t}>0$) at the beginning of the Kasner cosmology epoch, while for $|c|<1$ the Maxwell field growth (\ref{kasnerin}) triggers a transition to a new Kasner regime
with the exponent $c \to c_{new}=1/c$, or equivalently 
$p_{t} \to p_{t}^{new}=-p_{t}/(2p_{t}+1)$ at a certain radial position (called the Kasner inversion). In addition, we have proved that the effect of the massive gravity contribution is not significant in the final Kasner regime near the spacelike singularity.

Interestingly, in the large massive gravity parameter regime for a given temperature, the Einstein-Rosen bridge collapse and subsequent Josephson oscillations epochs completely are removed from the interior dynamics. In addition, the system moves to the final Kasner cosmology epoch exactly after the would-be inner horizon and does not experience any Kasner inversion.
\section*{Acknowledgements}
We would like to thank Matteo Baggioli and Li Li for many useful comments and discussions.

\bibliographystyle{JHEP}
\bibliography{kasner}

\end{document}